\documentclass[journal]{IEEEtran}
\usepackage{amsmath,amssymb,amsthm,lipsum} %������ѧģʽ���%
\usepackage{cite}
\usepackage{algorithmic}
\usepackage{algorithm}
\usepackage{graphicx}
\usepackage{textcomp}
\usepackage{stfloats}
\usepackage{booktabs}
\usepackage{bm}
\usepackage{subfigure}
\usepackage{setspace}
\usepackage{color}
\usepackage[T1]{fontenc}
\usepackage{cases}
\usepackage{indentfirst}
\allowdisplaybreaks[4]

\ifCLASSINFOpdf
\else
\fi
\theoremstyle{Theorem}
\newtheorem{theo}{Theorem}
\newtheorem{theoremmaindescription1}[theo]{Theorem}

\newtheorem{theoremmaindescription6}[theo]{Theorem}
\newtheorem{theoremmaindescription7}[theo]{Theorem}

\newtheorem{theoremmaindescription10}[theo]{Theorem}
\newtheorem{theoremmaindescription11}[theo]{Theorem}
\newtheorem{theoremmaindescription12}[theo]{Theorem}
\newtheorem{theoremmaindescription13}[theo]{Theorem}

\theoremstyle{remark}
\newtheorem{rmk}{Remark}
\newtheorem{rmk2}[rmk]{Remark}

\newtheorem{rmk4}[rmk]{Remark}
\newtheorem{rmk5}[rmk]{Remark}
\newtheorem{rmk6}[rmk]{Remark}
\newtheorem{rmk7}[rmk]{Remark}
\newtheorem{rmk8}[rmk]{Remark}

\newtheorem{rmk11}[rmk]{Remark}
\newtheorem{rmk12}[rmk]{Remark}

\theoremstyle{Definition}
\newtheorem{def1}{Definition}
\newtheorem{def2}[def1]{Definition}

\newtheorem{def4}[def1]{Definition}

\theoremstyle{Lemma}
\newtheorem{lemma1}{Lemma}
\newtheorem{lemma2}[lemma1]{Lemma}
\newtheorem{lemma3}[lemma1]{Lemma}
\newtheorem{lemma4}[lemma1]{Lemma}
\newtheorem{lemma5}[lemma1]{Lemma}
\newtheorem{lemma6}[lemma1]{Lemma}

\newtheorem{lemma8}[lemma1]{Lemma}

\newtheorem{lemma13}[lemma1]{Lemma}

\theoremstyle{Corollary}
\newtheorem{Corollary1}{Corollary}
\newtheorem{Corollary2}[Corollary1]{Corollary}

\newtheorem{Corollary4}[Corollary1]{Corollary}

\newtheorem{Corollary7}[Corollary1]{Corollary}
\newtheorem{Corollary8}[Corollary1]{Corollary}

\theoremstyle{Proposition}
\newtheorem{prop}{Proposition}

\newtheorem{proposition2}[prop]{Proposition}
\newtheorem{proposition3}[prop]{Proposition}

\begin{document}


\title{Hierarchically Block-Sparse Recovery \\With Prior Support Information}

\author{Liyang~Lu,~\IEEEmembership{Member,~IEEE,}
	Haochen~Wu,~\IEEEmembership{Student Member,~IEEE,}
	Wenbo~Xu,~\IEEEmembership{Senior Member,~IEEE,}\\
	    Zhaocheng~Wang,~\IEEEmembership{Fellow,~IEEE,} 
	    H. Vincent~Poor,~\IEEEmembership{Life Fellow,~IEEE} %%
	    \thanks{This work was supported in part by
	    	the National Natural Science Foundation of China under Grant 62401319. The work of H. V. Poor was supported in part by an Innovation Grant from Princeton NextG. (\textit{Corresponding author: Zhaocheng Wang}.)}
	    
\thanks{L.~Lu is with the State Key Laboratory of Networking and Switching Technology, Beijing University of Posts and Telecommunications, Beijing 100876, China (e-mail: luliyang25@bupt.edu.cn).}

\thanks{H. Wu and Z.~Wang are with the Beijing National Research Center for Information Science and Technology, Department of Electronic Engineering, Tsinghua University, Beijing 100084, China (e-mails: wuhc23@mails.tsinghua.edu.cn, zcwang@tsinghua.edu.cn). Z.~Wang is also with the Tsinghua Shenzhen International Graduate School, Shenzhen 518055, China.}

\thanks{W. Xu is with the Key Laboratory of
	Universal Wireless Communication, School of Artificial Intelligence, Beijing
	University of Posts and Telecommunications, Beijing, China (e-mail: xuwb@bupt.edu.cn).} %

\thanks{H. Vincent Poor is with the Department of Electrical and Computer
	Engineering, Princeton University, Princeton, NJ 08544 USA (e-mail:
	poor@princeton.edu).}

\thanks{L. Lu and H. Wu contributed equally to the project and should be considered
	co-first authors.} %
}


\markboth{}%
{Shell \MakeLowercase{\textit{et al.}}: Bare Demo of IEEEtran.cls for IEEE Journals}



\maketitle

\begin{abstract}
We provide new recovery bounds for hierarchical compressed sensing (HCS) based on prior support information (PSI). A detailed PSI-enabled reconstruction model is formulated using various forms of PSI. The hierarchical block orthogonal matching pursuit with PSI (HiBOMP-P) algorithm is designed in a recursive form to reliably recover hierarchically block-sparse signals. We derive exact recovery conditions (ERCs) measured by the mutual incoherence property (MIP), wherein hierarchical MIP concepts are proposed, and further develop reconstructible sparsity levels to reveal sufficient conditions for ERCs. Leveraging these MIP analyses, we present several extended insights, including reliable recovery conditions in noisy scenarios and the optimal hierarchical structure for cases where sparsity is not equal to zero. Our results further confirm that HCS offers improved recovery performance even when the prior information does not overlap with the true support set, whereas existing methods heavily rely on this overlap, thereby compromising performance if it is absent.
\end{abstract}

\begin{IEEEkeywords}
Exact recovery conditions, hierarchical compressed sensing, mutual incoherence property, orthogonal matching pursuit, prior support information.
\end{IEEEkeywords}

\IEEEpeerreviewmaketitle

\section{Introduction}\label{Introduction} % S1
\IEEEPARstart{R}{ecent} years have witnessed widespread use of compressed sensing (CS) in efficient signal acquisition \cite{Donoho2006tit,Candes2008,Yonina2012press,greed2004,Aeron2010,liyang2024}. It assumes that if the signals of interest are compressible, CS enables signal acquisition by employing far fewer measurements than conventional acquisition approaches \cite{Candes2006tit,ttcai2011,Khanna2022,Wan2020,Matsumoto2024}. Due to the compressibility of signals, CS has been applied in many applications, involving wireless communications, image processing, signal classification, and sensor networks \cite{Cuiy2021,Zhang2020,WangRen2020,Huangyao}. 

In this paper, we focus on enhanced recovery conditions of hierarchical CS (HCS), where signals to be recovered exhibit a hierarchically sparse structure. We show that whenever prior or side information about the actual signal to be recovered is available, the recovery restriction of HCS can be reduced even further \cite{Scarlett2013,Ge2021}. 

\emph{Compressed sensing (CS):} Consider the problem of representing a vector $\mathbf{y}\in\mathbb{C}^M$ in a given measurement matrix $\mathbf{D}\in\mathbb{C}^{M\times N}$, such that 
\begin{align}\label{system}
	\mathbf{y} = \mathbf{D}\mathbf{x} ,
\end{align}
where $\mathbf{x}\in\mathbb{C}^{N}$ is the underlying signal. It has been shown that if $\mathbf{x}$ is assumed to be sparse, i.e., $\mathbf{x}$ only has a few nonzero atoms relative to its dimension, then the uniqueness of the representation in the underdetermined system of equations (\ref{system}) can be guaranteed \cite{greed2004}. Consider a more general case where the signal $\mathbf{x}$ exhibits block sparsity, i.e., the nonzero supports of $\mathbf{x}$ occur in blocks \cite{Eldar2010,liyangtccn2022,liyang2022}. Letting $d$ denote the block length, the block-sparse signal $\mathbf{x}$ is formulated as
\begin{align}
	\mathbf{x}=[\underbrace{\mathbf{x}_1\cdots  \mathbf{x}_d }_{\mathbf{x}^{\rm T}[1]} \underbrace{\mathbf{x}_{d+1}\cdots \mathbf{x}_{2d}}_{\mathbf{x}^{\rm T}[2]}\cdots \underbrace{\mathbf{x}_{N-d+1}\cdots \mathbf{x}_{N}}_{\mathbf{x}^{\rm T}[N_B]}]^{\rm T},\nonumber
\end{align}
where $N=N_Bd$, $(\cdot)^{\rm T}$ denotes the transposition of its objective, and $\mathbf{x}[i]\in \mathbb{C}^{d}$ is the $i$th block of $\mathbf{x}$ $(i\in\{1,2,\cdots,N_B\})$. A signal $\mathbf{x}$ is called $k$ block-sparse if $\mathbf{x}$ has $k$ nonzero $\ell_2$-norm blocks. The recovery of $\mathbf{x}$ has been extensively studied in the literature via  convex optimization and greedy algorithms; see for instance \cite{greed2004,Eldar2010,Soussen2013,liyang2022,Kim2020,liyang2024}. In general, the exact or reliable recovery guarantees in noiseless or noisy scenarios of block-sparse recovery are less restricted than those of conventional sparse recovery due to the exploitation of the block structure. For example, the mutual incoherence property (MIP), one of the widely adopted frameworks to characterize sparse signal recovery has been introduced (See \textbf{Definition~\ref{defconventionalmip}}) for computationally friendly theoretical interpretability \cite{greed2004,Eldar2010,liyang2024}.
\begin{def4}\label{defconventionalmip} % def.6
	The MIP of a given measurement matrix $\mathbf{D}$ contains two aspects. The first is the conventional matrix coherence, which is defined as $\mu=\max\limits_{i,j\neq i}|\mathbf{D}^{\rm H}_i\mathbf{D}_j|$, where $\mathbf{D}_i$ and $\mathbf{D}_j$ are the $i$th and $j$th columns of $\mathbf{D}$. The second part consists of block coherence and sub-coherence. The block coherence is defined as $\mu_B=\max\limits_{i,j\neq i}\frac{\|\mathbf{M}_{[i\times j]}\|_2}{d}$, where $\mathbf{M}_{[i \times j]}=\mathbf{D}^{\rm H}_{[i]}\mathbf{D}_{[j]}$, and $\mathbf{D}_{[i]}$ and $\mathbf{D}_{[j]}$ are the $i$th and $j$th column-block submatrices of $\mathbf{D}$. The sub-coherence is given by $\nu=\max\limits_l\max\limits_{i,j\neq i}|\mathbf{D}^{\rm H}_i\mathbf{D}_j|$, where $\mathbf{D}_i,\mathbf{D}_j\in \mathbf{D}[l]$, and $\mathbf{D}[l]$ denotes the $l$th column-block submatrix of $\mathbf{D}$.
\end{def4}

Among various MIP conditions studied, we only mention three sharp conditions obtained in \cite{greed2004,Eldar2010,Herzet2016}. In \cite{greed2004}, Tropp showed that if $K<\frac{1}{2}(\frac{1}{\mu}+1)$, then algorithms such as orthogonal matching pursuit (OMP) exactly recover all $K$-sparse signals in noiseless scenarios, which is called the exact recovery condition (ERC). For exactly reconstructing $k$ block-sparse signals with block length $d$, Eldar \emph{et al.} demonstrated that the ERC becomes $kd<\frac{1}{2}(\frac{1}{\mu_B}+d-(d-1)\frac{\nu}{\mu_B})$ in \cite{Eldar2010}. This result shows a less restrictive condition for exact recovery of block-sparse signals than recovering conventional sparse signals. Further, in \cite{Herzet2016}, Herzet \emph{et al.} developed the tightest achievable MIP-based guarantees by assuming a decaying structure of $k$ sparse signals, leading to the bound $k<\frac{1}{\mu}$. Since ERCs consider the worst-case scenarios, the presented results are usually more pessimistic than practical outcomes \cite{liyang2022}. So far, a significant gap remains between theoretical and practical results, presenting an opportunity for further analysis.

\emph{Hierarchical compressed sensing (HCS):} Assume that $\mathbf{x}$ exhibits an $n$-mode hierarchical block sparsity (See \textbf{Definition \ref{Hierarchicallyblocksparsesignal}}), i.e., a natural setting that possesses a more intricate structure than mere sparsity and mere block sparsity, for which even superior reconstruction performance can be anticipated \cite{Rothtsp2020,Eisert2022}. That is, for each mode of the hierarchical structure, $\mathbf{x}$ has only a few \emph{significant blocks}, wherein not all elements in a significant block are nonzero. This significant block is also referred to as a support block.

\begin{def1}\label{Hierarchicallyblocksparsesignal} % def.1
	A vector $\mathbf{x}\in\mathbb{C}^{N_0N_1 N_2\cdots N_{n} d}$ is $n$-mode $(k_0,k_1,k_2,\cdots,k_n)$ $(n\geq0)$ hierarchically block-sparse if $\mathbf{x}$ consists of $N_0N_1 N_2\cdots N_i$ blocks sized as $N_{i+1} N_{i+2}\cdots N_n d$ for the $i$th mode $(0\leq i< n)$ and sized as $d$ for the $n$th mode, where $N_0=1$, $k_0=1$, and at most $k_i$ blocks have non-vanishing atoms and additionally each nonzero block itself is $(n-i)$-mode $(k_{i+1},k_{i+2},\cdots,k_n)$ hierarchically block-sparse. Specially, for $n=1$, the vector $\mathbf{x}\in\mathbb{C}^{N_{1} d}$ converges to a conventional block-sparse signal that consists of $N_1$ blocks sized as $d$, $k_n$ is the true number of nonzero blocks with block length $d$ in each block of the $n$th hierarchical mode, and $k_0k_1\cdots k_n$ denotes the true block sparsity of $\mathbf{x}$.
\end{def1}

This structure is a general version of conventional hierarchically sparse structure, wherein the latter one assumes that $d=1$. More specifically, when setting $d=1$, hierarchically block-sparse signal $\mathbf{x}$ in \textbf{Definition \ref{Hierarchicallyblocksparsesignal}} degenerates into conventional hierarchically sparse signals as presented in \cite{Rothtsp2020}, wherein the minimum unit support of $\mathbf{x}$ is a single atom. Hierarchically block-sparse structure can be further seen as an integration of block sparsity and level sparsity.
It is evident that the concept of hierarchical block
	sparsity is more general than conventional block sparsity and draws much attention in recent applications \cite{Wunder2019MIMOCE}. It offers the improved performance when the signal to be recovered exhibits an additional hierarchical structure, 
	such as channel estimation in holographic multiple-input multiple-output (HMIMO) \cite{du2025wcl,Pizzo2020}, 
	or when the signal exhibits only a hierarchical structure without the conventional block structure, such as channel estimation in massive MIMO \cite{Wunder2019MIMOCE} and sporadic traffic scenarios in massive machine-type
	communications (mMTC) \cite{Rothtsp2020}.
 Related studies on recovery guarantees of HCS mainly focus on the restricted isometry property (RIP) metric, which has made certain progress compared to the conventional CS methods \cite{zhang2019}. However, calculating the RIP of a given matrix is NP-hard and therefore cannot be intuitively compared with MIP. 

\emph{HCS with prior information:} Assume that, in addition to the set of measurements $\mathbf{y}$ in (\ref{system}) and the hierarchical structure of $\mathbf{x}$, we also have access to \emph{prior information} \cite{Scarlett2013,Gamma2019,sideinformation2020,sideinformationren2024,sideinformationren2025}. That is, we have a prior support information (PSI) index set $\mathbf{\Theta}$, whose atoms indicate positions of the nonzero coefficients of the sparse signal $\mathbf{x}$. This occurs in numerous applications, such as video acquisition, channel estimation, and medical imaging, where past signals or certain available signals similar to the original signal of interest can be leveraged to generate an estimate of the desired signal \cite{Mota2017,liyanghaochen2024}. As few studies have considered recovery conditions for HCS with PSI, we instead present some intuitive results for conventional CS to illustrate the benefit of exploiting PSI. In \cite{Herzet2013}, Herzet \emph{et al.} showed that some algorithms, e.g., OMP and orthogonal least squares (OLS), exactly recover $K$ sparse signals if $K<\frac{1}{2}(\frac{1}{\mu}+g-b+1)$, where $g$ and $b$ denote respectively the numbers of good and bad atoms corresponding to the PSI. More clearly, denote the true support index set of $\mathbf{x}$ and the support set of the prior information vector by $\mathbf{\Xi}$ and $\mathbf{\Theta}$, respectively, and we have $g=|\mathbf{\Xi}\cap\mathbf{\Theta}|$ and $b=|\mathbf{\Theta}\backslash\mathbf{\Xi}|$. This bound becomes better than the classic one, i.e., $K<\frac{1}{2}(\frac{1}{\mu}+1)$, derived by Tropp \cite{greed2004}, when there exist more than $50\%$ good atoms in the prior information, which means that $g>b$. Moreover, recent studies such as \cite{Ge2020,Ge2021} also focus on the scenario where the accuracy of prior information on the support, i.e., the proportion of overlapping between the PSI and the true support index set, is at least $50\%$, which indeed improves the theoretical guarantees to less restrictive ones. These improvements align with the intuition that a larger number of good atoms in PSI leads to better recovery conditions, and once this number surpasses $50\%$, it significantly improves the current sharp bounds that do not rely on any additional prior information.

Since hierarchically block-sparse signals generalize block-sparse $(d>1)$ and conventional sparse $(d=1)$ signals, the existing theoretical guarantees based on MIP still apply to hierarchically block-sparse recovery when the signal is considered non-hierarchically sparse. Thus, \emph{a natural question to ask is whether the recovery conditions can be further improved for hierarchically block-sparse recovery based on the hierarchical structure, and even with the assistance of PSI.} Consider a support block, marked as the $i$th block, of a hierarchically block-sparse signal $\mathbf{x}$ in the $t$th hierarchical mode. This block contains several true support indices where the corresponding blocks of $\mathbf{x}$ are nonzero, while the remaining parts are non support blocks, which are zero in $\mathbf{x}$. It is worth mentioning that, regarding the recovery condition in terms of ERC, if the power of the $i$th support block exceeds that of the non support blocks to a greater extent, then the recovery condition will be improved. This process is referred to as correct support selection \cite{liyang2022}. It is evident that multiple true supports are included in the $i$th hierarchical block, jointly contributing to more significant power. The support selection procedure based on this enhanced power would improve the accuracy of support selection. Furthermore, for subsequent hierarchical modes, the number of non support blocks is reduced due to the constraints of the hierarchical structure. This reduction weakens the selection competition between support blocks and non support blocks. Moreover, numerous studies have confirmed the use of PSI in enhancing recovery performance, which is also applicable to hierarchically block-sparse recovery. These factors provide opportunities for improved recovery condition analysis in hierarchically block-sparse recovery. 

Previous studies have concentrated on the overlap between the PSI and the true support index set. If this overlap proportion is equal to $50\%$, then the theoretical recovery conditions for the ERC of PSI-assisted CS typically align with those of conventional CS \cite{Herzet2013,Ge2021}.
Therefore, \emph{a practical question to be answered is whether overlap lower than $50\%$ can lead to a better recovery guarantee for hierarchically block-sparse reconstruction.} This is indeed a practical consideration, as mismatches between the PSI and the true support index set often occur, presenting a challenge that must be carefully addressed in PSI-based CS. From the perspective of beneficial information for recovery, reducing the proportion below $50\%$ while maintaining recovery conditions that are equal to or even better than those of conventional CS seems impossible. This is because the useful information must be substantial enough to counteract harmful interference, and $50\%$ represents the equilibrium point. In the hierarchically block-sparse recovery formulation, a promising way forward is to take the structural properties into account for enhanced PSI efficiency. This is because, for the $t$th hierarchical mode, the $i$th support block contains zero elements that are not included in the true support, as mentioned above. These zero elements may contribute to the correct selection of the $i$th support block if certain available weights are assigned to them. 
Nevertheless, this hierarchical structure-based method is still in its infancy, with even the basic model foundation requiring thorough examination and exploration.

The goal of this paper is to shed light on the recovery performance of HCS aided by PSI, while providing   affirmative answers to the above two questions. Our main contributions are summarized as follows.

\begin{enumerate}
	\item A hierarchically block-sparse recovery model assisted by PSI is formulated. Hierarchical MIP concepts, including hierarchical block coherence and sub-coherence, are introduced, and various forms of PSI are constructed. Within this framework, a recursive form of hierarchical block OMP with PSI (HiBOMP-P) is developed to ensure reliable reconstruction. A computational complexity analysis is presented, indicating that accurate PSI can significantly reduce the number of iterations. In actual scenarios with low-order hierarchical structures, the complexity of HiBOMP-P is comparable to its traditional counterparts.  It is evident that the proposed HiBOMP-P serves as a representative methodology and acts as a foundational component for uncovering the recovery conditions of HCS with PSI. 
	
	\item We show that if the measurement matrix $\mathbf{D}$ satisfies that ${G}_{*}+{G}_{\circ}<1$ (See \textbf{Theorem \ref{theorem1}}) with the parameters ${G}_{*}$ and ${G}_{\circ}$ depending on mixed norm properties of $\mathbf{D}$, the PSI support index sets, and the support power ratio of $\mathbf{x}$ in different support blocks, then HiBOMP-P exactly recovers $\mathbf{x}$ from (\ref{system}), thereby constituting the ERC of HiBOMP-P. Moreover, sufficient conditions of the ERC have been derived, indicating that if $\overline{G}_{*'}+\overline{G}_{\circ'}<1$ (See \textbf{Theorem \ref{theo6}}), the ERC is established. Furthermore, a combined result derived from \textbf{Theorem \ref{theorem1}} and \textbf{Theorem \ref{theo6}} establishes the complete recovery assurance of HiBOMP-P (See \textbf{Theorem \ref{theo12}}).
	
	\item Based on the results related to ERC, this paper further provides reliable recovery conditions in noisy scenarios (see \textbf{Section \ref{noisyrecoverycondition}}). Additionally, we emphasize the benefits of an optimal hierarchical structure for hierarchically block-sparse recovery, effectively eliminating the influence from outside support blocks (see \textbf{Section~\ref{secoptimalhier}}), where the definition of outside support blocks is presented in \textbf{Section~\ref{sectionMathematical}}. Further analysis demonstrates that the bounds derived under certain assumptions outperform the existing results, indicating a more assured recoveryability of HiBOMP-P.
	
	\item The analytical results confirm that HiBOMP-P is capable of achieving improved recovery performance, even in cases where the PSI does not overlap with the true support set (See \emph{Remark~\ref{rmk11}}). This enhancement can be attributed to the hierarchical structure, as evidenced by simulation results. Furthermore, asymptotic analysis provides meaningful comparisons between existing bounds and those developed in this paper, considering parameters such as reconstructible block sparsity, number of measurements, block length of signal supports, support power, and coherence levels. Notably, our proposed methodology could be easily generalized and extended to various scenarios without block structures or PSI.
\end{enumerate}

The remainder of this paper is organized as follows. \textbf{Section~II} presents the notation, mathematical formulation, and the proposed recovery algorithm. The main results, in terms of ERCs, are detailed in \textbf{Section III}. In \textbf{Section IV}, we discuss further extensions based on ERCs. \textbf{Section V} contains the technical proofs. Simulation results are provided in \textbf{Section~VI}, and conclusions are drawn in \textbf{Section VII}.

\section{Preliminaries}\label{problemFormulation} % S1

In this section, we first present the symbolic representation of the hierarchically block-sparse recovery formulation. Then, we propose two hierarchically block-sparse recovery algorithms. Finally, several useful lemmas are presented that facilitate the derivation of the main theorems provided.

\subsection{Notation}\label{Notations}

\begin{figure*}[!tp]
	\begin{center}
		\includegraphics[width=.7\textwidth]{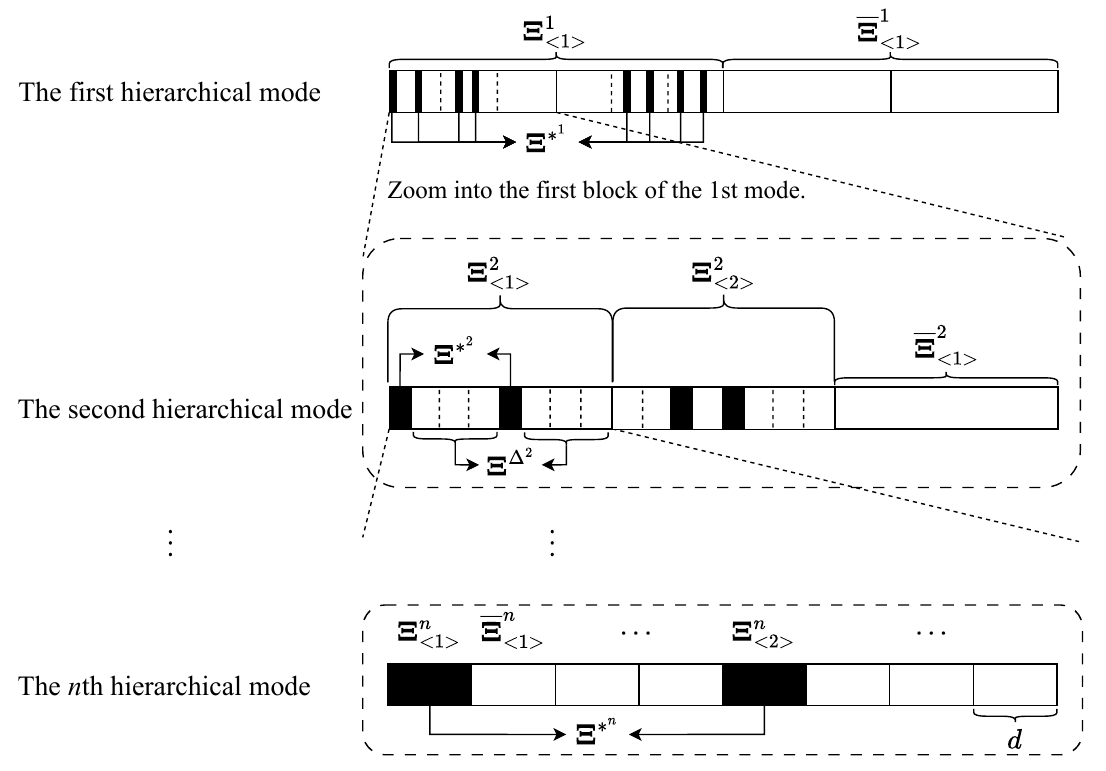}
	\end{center}
	\caption{{Illustration of the $n$-mode hierarchically block-sparse signal.}}
	\label{blocksparsemodel} % Fig.1
\end{figure*}

We briefly summarize the notation used in this paper. Vectors are denoted by boldface lowercase letters, e.g., $\mathbf{x}$, and matrices are denoted by boldface uppercase letters, e.g., $\mathbf{D}$. $\mathbf{D}_{i}$ denotes the $i$th column of $\mathbf{D}$. $\mathbf{D}_{\mathbf{\Theta}}$ is a submatrix of $\mathbf{D}$ that contains the columns indexed by set $\mathbf{\Theta}$, and $\mathbf{x}_{\mathbf{\Theta}}$ is a subvector containing the elements of $\mathbf{x}$ indexed by $\mathbf{\Theta}$. $\mathbf{D}^{\rm H}$ denotes the conjugate transpose of $\mathbf{D}$. Given a block length $d$, the $i$th block of $\mathbf{x}$ is denoted as $\mathbf{x}_{[i]}$, and the $i$th column-block submatrix of $\mathbf{D}$ is denoted by $\mathbf{D}_{[i]}$. $|\mathbf{\Theta}|$ stands for the cardinality of set $\mathbf{\Theta}$, and $|c|$ is the absolute value of constant $c$. The disjoint union of two sets $\mathbf{\Theta}$ and $\mathbf{\Lambda}$ is denoted by $\mathbf{\Theta}\sqcup\mathbf{\Lambda}$. Given an integer $n$, $\{\mathring{n}\}$ denotes the set $\{1,2,\cdots,n\}$. $\mathbf{I}$ stands for the identity matrix. If $\mathbf{D}_{\mathbf{\Theta}}$ has the full column rank, $\mathbf{D}_{\mathbf{\Theta}}^\dag\! =\! \big(\mathbf{D}_{\mathbf{\Theta}}^{\rm H}\mathbf{D}_{\mathbf{\Theta}}\big)^{-1}\mathbf{D}^{\rm H}_{\mathbf{\Theta}}$ is the pseudoinverse of $\mathbf{D}_{\mathbf{\Theta}}$. $\text{span}(\mathbf{D}_{\mathbf{\Theta}})$ denotes the space spanned by the columns of $\mathbf{D}_{\mathbf{\Theta}}$, and $\mathbf{P}_{\mathbf{D}_{\mathbf{\Theta}}}\! =\! \mathbf{D}_{\mathbf{\Theta}}\mathbf{D}_{\mathbf{\Theta}}^\dag$ is the projection onto $\text{span}(\mathbf{D}_{\mathbf{\Theta}})$, while $\mathbf{P}_{\mathbf{D}_{\mathbf{\Theta}}}^\bot\! =\! \mathbf{I}-\mathbf{P}_{\mathbf{D}_{\mathbf{\Theta}}}$ is the projection onto the orthogonal complement of span$(\mathbf{D}_{\mathbf{\Theta}})$. Throughout the paper, the columns in the measurement matrix are normalized to have the unit $\ell_2$-norm.

To provide an intuitive understanding of the recursive form of the $n$-mode hierarchically block-sparse signal defined in \textbf{Definition \ref{Hierarchicallyblocksparsesignal}}, we present Fig.~\ref{blocksparsemodel}, which illustrates an example of such a signal.
 Given an $n$-mode hierarchically block-sparse signal $\mathbf{x}\in\mathbb{C}^{N_1 N_2\cdots N_{n} d}$, we emphasize that the minimum block length unit used in this paper is $d$, which is important for the clarity. As the hierarchically block-sparse signal has $n$ hierarchical modes, the block in the $t$th hierarchical mode may come from the $i$th $(i\in\{\mathring{N_{t-1}}\})$ block of the $(t-1)$th hierarchical mode. The support set in the $t$th $(t\in\{\mathring{n}\})$ hierarchical mode coming from the $i$th block in the $(t-1)$th hierarchical mode is denoted by $\mathbf{\Xi}^t_{<i>}$, where the signal block corresponding to the index in $\mathbf{\Xi}^t_{<i>}$ has a block length of $d$. As illustrated in Fig.~\ref{blocksparsemodel}, we zoom into the first block of the first mode, from which the corresponding block in the second mode is derived. The support sets in the second hierarchical mode, denoted by $\mathbf{\Xi}^2_{<1>}$ and $\mathbf{\Xi}^2_{<2>}$, are obtained from the first block in the first hierarchical mode. This hierarchical structure continues recursively up to the $n$th mode.
 Specially, the index set corresponding to the nonzero blocks in $\mathbf{\Xi}^t_{<i>}$ is called the true support set denoted by $\mathbf{\Xi}^{*^t}$, while the index set that corresponds to all zero blocks in $\mathbf{\Xi}^t_{<i>}$ is called the additional support set denoted by $\mathbf{\Xi}^{\Delta^t}$. Meanwhile, the index set that corresponds to the zero blocks in the $t$th hierarchical mode coming from the $i$th block of the $(t-1)$th hierarchical mode is called the non support index set denoted by $\overline{\mathbf{\Xi}}^t_{<i>}$. Note that we have dropped the dependence on the block number $<\!\!\!i\!\!\!>$ in $\mathbf{\Xi}^{*^t}$ and $\mathbf{\Xi}^{\Delta^t}$ for symbol simplification.
  Correspondingly, the measurement matrix exhibits hierarchical structure from the perspective of matrix columns as similar to the definition in \textbf{Definition~\ref{Hierarchicallyblocksparsesignal}}. 

\subsection{Mathematical Formulation}\label{sectionMathematical}
\begin{figure}[!tp]
	\begin{center}
		\includegraphics[width=.38\textwidth]{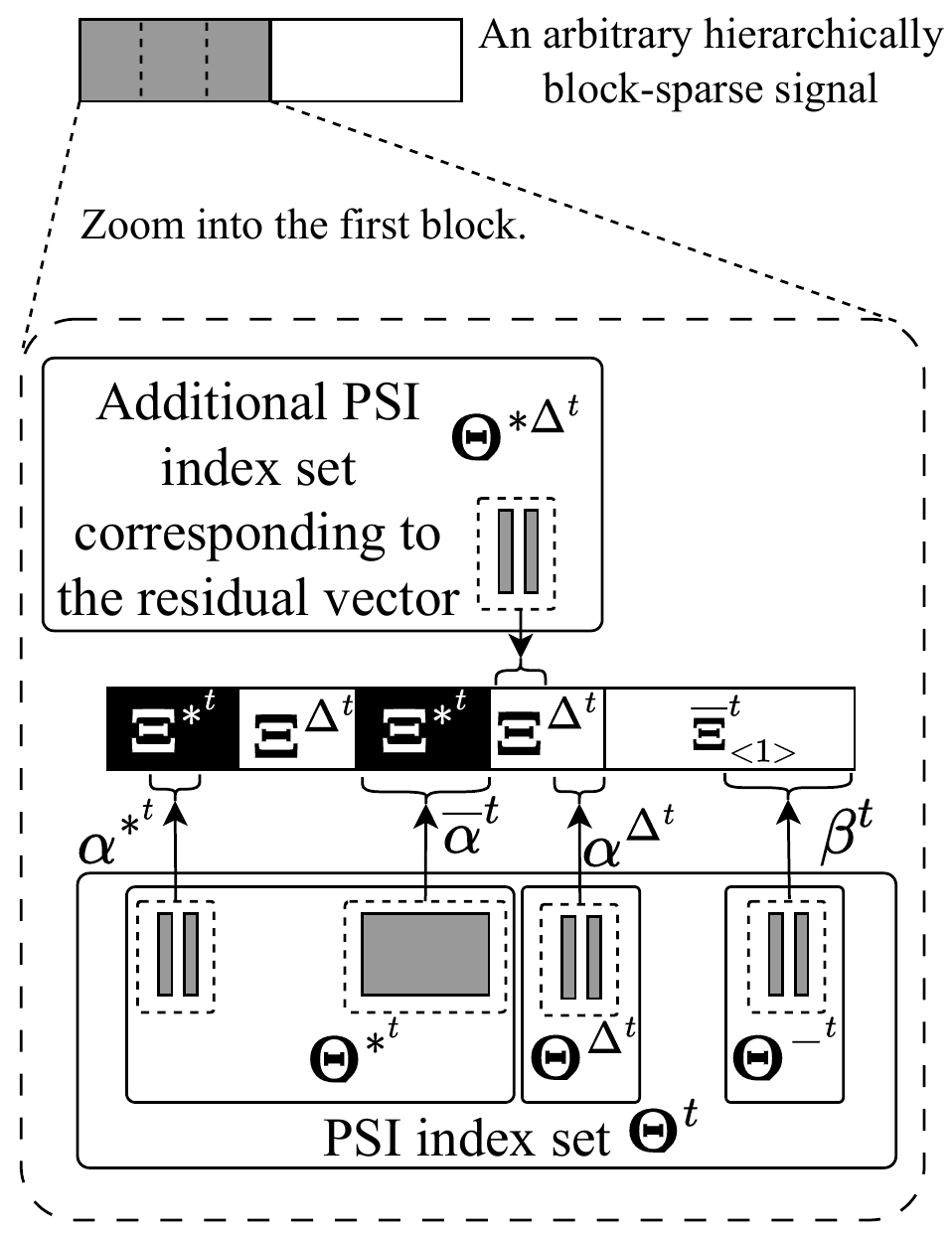}
	\end{center}
	\caption{Illustration of the PSI index set in the $t$th hierarchical mode.}
	\label{blocksparsemodel2} % Fig.1
\end{figure}

We consider recovering a hierarchically block-sparse signal $\mathbf{x}$ from its compressed embodiment $\mathbf{y}$ with the assistance of PSI, formulated by the index sets $\mathbf{\Theta}$ and $\mathbf{\Theta}^{*\Delta}$. As shown in Fig. \ref{blocksparsemodel2}, for the PSI in the $t$th $(t\in\{\mathring{n}\})$ hierarchical mode, the corresponding index sets can be further denoted by $\mathbf{\Theta}^t$ and $\mathbf{\Theta}^{*\Delta^ t}$.  It is evident that there is overlap between $\mathbf{\Theta}^t$, $\mathbf{\Theta}^{*\Delta^ t}$ and $\mathbf{\Xi}^{*^t}$, $\mathbf{\Xi}^{\Delta^t}$, $\overline{\mathbf{\Xi}}^{t}_{<i>}$.
To this end, we define the following parameters to better quantify the overlap of the corresponding indices for the $t$th hierarchical mode.

(1) The overlapping index number of PSI and the true support index set $\mathbf{\Xi}^{*^t}$ consists of two parts. The first part indicates the overlap between $\mathbf{\Theta}^t$ and $\mathbf{\Xi}^{*^t}$ with a block length of $d$, denoted by $\alpha^{*^t}$. The second part indicates the overlap between $\mathbf{\Theta}^t$ and $\mathbf{\Xi}^{*^t}$ with a block length of $N_{t+1}N_{t+2}\cdots N_n d$, denoted by $\overline{\alpha}^{t}$. Thus, we have $|\mathbf{\Xi}^{*^t}\cap\mathbf{\Theta}^{t}|=\alpha^{*^t}+N_{t+1}N_{t+2}\cdots N_n\overline{\alpha}^t$. The symbol corresponding to the PSI set is denoted by $\mathbf{\Theta}^{*^t}$.

(2) The overlapping index number of PSI and the additional support index set $\mathbf{\Xi}^{\Delta^t}$ consists of two parts. The first part is contained within the PSI set $\mathbf{\Theta}^t$. Denote this overlapping number, with block length $d$, as $\alpha^{\Delta^t}$, such that $|\mathbf{\Xi}^{\Delta^t}\cap\mathbf{\Theta}^t|=\alpha^{\Delta^t}$. The corresponding notation for the PSI set is given by $\mathbf{\Theta}^{\Delta^t}$. The second part is denoted by $\mathbf{\Theta}^{*\Delta^t}$, with block length $d$, corresponding to the additional augmented PSI to the residual vector. The overlapping number is given by $|\mathbf{\Xi}^{\Delta^t}\cap\mathbf{\Theta}^{*\Delta^t}|=\alpha^{*\Delta^{t}}$. Note that $\mathbf{\Theta}^t\cap\mathbf{\Theta}^{*\Delta^t}=\mathbf{\emptyset}$.

(3) The overlapping index number of PSI and the non support index set $\overline{\mathbf{\Xi}}^{t}$ is denoted by $\beta^t$ with a block length of $d$, i.e., $|\overline{\mathbf{\Xi}}^{t}\cap\mathbf{\Theta}^t|=\beta^{t}$. The symbol corresponding to the PSI set is represented by $\mathbf{\Theta}^{-}$.

For the sake of concise notation throughout the paper, we assume that the indices of $\mathbf{\Theta}^t$ are uniformly distributed within each true support block, each additional support block, and each non support block.

As MIP provides computable theoretical guarantees for sparse recovery algorithms \cite{liyang2022,Miandji2017,liyang2023}, this paper also considers the theoretical aspects of MIP. However, conventional MIP, including matrix coherence, and block MIP, which consists of block coherence and sub-coherence as defined in \textbf{Definition~\ref{defconventionalmip}}, cannot capture the uncertain relations between measurement matrix blocks in the hierarchically block-sparse recovery formulation.
  This limitation arises because conventional CS assumes a fixed and regular signal structure, where different signal blocks are continuously connected with the same block length. However, in HCS, discontinuous signal blocks may form a desired support that needs to be identified due to the hierarchical structure, as illustrated in Fig. \ref{blocksparsemodel}. To address this issue, we introduce the concept of hierarchical block MIP, which enables the representation of uncertain relations within the hierarchical measurement matrix. This concept encompasses hierarchical block coherence and hierarchical sub-coherence.

\begin{def2}\label{hierarchicallyMIP} % def.1
	Given a matrix $\mathbf{D}\in\mathbb{C}^{M\times N}$ with minimum block length $d$, the hierarchical block coherence of $\mathbf{D}$, which represents the similarity of its column-block submatrices, is defined as
	\begin{equation}
		\mu_{d^*} = \max\limits_{\forall i,j\neq i} \frac{\big\|\overline{\mathbf{M}}_{[i\times j]}\big\|_2}{d^*},\nonumber
	\end{equation}
	where $\overline{\mathbf{M}}_{[i\times j]}=\mathbf{D}_{\mathbf{\Xi}_{(i)}}^{\rm H}\mathbf{D}_{\mathbf{\Xi}_{(j)}}$, $d^*$ is an integer multiple of $d$, and $\mathbf{\Xi}_{(i)}$ and $\mathbf{\Xi}_{(j)}$ are arbitrary valid index sets satisfying that  $|\mathbf{\Xi}_{(i)}|=|\mathbf{\Xi}_{(j)}|=\frac{d^*}{d}$ and $\mathbf{\Xi}_{(i)}\cap\mathbf{\Xi}_{(j)}=\mathbf{\emptyset}$. Following the definition of hierarchical block coherence, the hierarchical sub-coherence of $\mathbf{D}$ is defined as
	\begin{equation}
		\nu_{d^*} = \max\limits_{\forall j}\max\limits_{\forall l,r\neq l} \big|\mathbf{D}_{\mathbf{\Xi}_l}^{\rm H}\mathbf{D}_{\mathbf{\Xi}_r}\big| ,\nonumber
	\end{equation}
	where $\mathbf{\Xi}_l$ and $\mathbf{\Xi}_r$ is the $l$th and $r$th indices in the set $\mathbf{\Xi}$, $\mathbf{\Xi}$ satisfies that $|\mathbf{\Xi}|=\frac{d^*}{d}$, and $\mathbf{\Xi}$ contains the indices in the $j$th block from the $(t-1)$th hierarchical mode.
\end{def2}

As we can see, a special case is when $d=1$, leading to the MIP measured in conventional hierarchically sparse recovery without considering the block structure. The conventional matrix coherence $\mu$ defined in \textbf{Definition \ref{defconventionalmip}} considers the similarity between arbitrary matrix columns, without accounting for block or hierarchical structures. The block coherence and sub-coherence defined in \textbf{Definition \ref{defconventionalmip}} take the block structure into account, but they only calculate the relationships between continuous matrix blocks, neglecting the possibility of hierarchical structures that may include separated matrix blocks within a single support matrix block. In contrast, the hierarchical block MIP defined in \textbf{Definition \ref{hierarchicallyMIP}} not only reveals the block structure of the measurement matrix, but also illustrates its potential hierarchical properties. When $d=1$, the newly defined hierarchical block coherence measures the aggregation of matrix columns within a possible support block. Mathematically, hierarchical block coherence captures the uncertain relations among matrix blocks of length $d^*$, consisting of $d^*$ matrix columns. In the following, we present the bounds for hierarchical block coherence and hierarchical sub-coherence, with \textbf{Lemma \ref{lemma4}} serving as the basis.

\begin{lemma4}\label{lemma4} % L2
	(\cite[Lemma 2]{liyang2024}) Given a matrix $\mathbf{D}\in\mathbb{C}^{M\times N}$ with $M=md$ and $N=nd$, we have
	\begin{align} % eqs.43,44
		\|\mathbf{D}\|_2\leq\sqrt{\rho_c(\mathbf{D})\rho_r(\mathbf{D})},\nonumber
	\end{align}
	where $\max\limits_{i}\sum\limits_{j}\|\mathbf{D}_{[i,j]}\|_2\triangleq\rho_{r}(\mathbf{D})$ and $\max\limits_{j}\sum\limits_{i}\|\mathbf{D}_{[i,j]}\|_2\triangleq\rho_{c}(\mathbf{D})$ are mixed matrix norm, and $\mathbf{D}_{[i,j]}$ is the $(i,j)$th $d\times d$ block of $\mathbf{D}$.
\end{lemma4}

\begin{rmk8}\label{rmk8}
	\emph{Assume that $d^*=cd$. Using the symbols in \textbf{Definition \ref{hierarchicallyMIP}}, we have
	\begin{equation}
		\mu_{d^*} = \max\limits_{\forall i,j\neq i} \frac{\big\|\overline{\mathbf{M}}_{[i\times j]}\big\|_2}{d^*}\leq\max\limits_{\forall i,j\neq i} \frac{\sqrt{\rho_c(\overline{\mathbf{M}}_{[i\times j]})\rho_r(\overline{\mathbf{M}}_{[i\times j]})}}{d^*},\nonumber
	\end{equation}
where the inequality is from \textbf{Lemma \ref{lemma4}}. 
Note that the matrix $\overline{\mathbf{M}}_{[i\times j]}$ consists of $c^2$ matrix blocks sized as $d\times d$. Based on \textbf{Definition \ref{defconventionalmip}}, we have $\rho_c(\overline{\mathbf{M}}_{[i\times j]})\leq cd\mu_B$ and $\rho_r(\overline{\mathbf{M}}_{[i\times j]})\leq cd\mu_B$. Thus, the following inequality holds:
\begin{align}
	\mu_{d^*}\leq\frac{\sqrt{cd\mu_{B}\times cd\mu_{B}}}{cd}=\mu_{B}\leq\mu,\nonumber
\end{align}
where the second inequality is because $\mu_B\leq\mu$ as given by \cite[Propoisition 2]{Eldar2010}. Meanwhile, for the hierarchical sub-coherence, we have
\begin{align}
	\nu\leq\nu_{d^*} = \max\limits_{\forall j}\max\limits_{\forall l,r\neq l} \big|\mathbf{D}_{\mathbf{\Xi}_l}^{\rm H}\mathbf{D}_{\mathbf{\Xi}_r}\big|\leq\mu,\nonumber
\end{align}
where $\nu$ and $\mu$ represent the sub-coherence and conventional matrix coherence, respectively. }
\end{rmk8}

\begin{rmk12}\label{rmk1212}
	\emph{In this remark, we present certain results related to the hierarchical block orthogonality of the measurement matrix. Conventional block orthogonality indicates that the matrix consists of orthogonal blocks with isometric block lengths, i.e., $\nu=0$. As for hierarchical block MIP, it reveals the uncertain relations within a matrix block that is the cascade of discontinuous matrix blocks. For the $t$th $(t\in\{\mathring{n}\})$ hierarchical mode, denote the block length of each hierarchical block as $\hat{d}$; then the measurement matrix satisfies hierarchical block orthogonality with $\nu_{\hat{d}}=0$. In this case, if we assume that the minimum block length is equal to $\hat{d}$, then we have
	\begin{align}
		\mu_{\hat{d}}=&\max\limits_{\forall i,j\neq i} \frac{\big\|\mathbf{M}_{[i\times j]}\big\|_2}{\hat{d}}\nonumber\\
		=&\max\limits_{\forall i,j\neq i} \frac{\big\|\mathbf{D}_{\mathbf{\Xi}_{(i)}}^{\rm H}\mathbf{D}_{\mathbf{\Xi}_{(j)}}\big\|_2}{\hat{d}}\nonumber\\
		\leq&\max\limits_{\forall i,j\neq i} \frac{\big\|\mathbf{D}_{\mathbf{\Xi}_{(i)}}^{\rm H}\big\|_2\|\mathbf{D}_{\mathbf{\Xi}_{(j)}}\big\|_2}{\hat{d}}\nonumber\\
		=&\frac{1}{\hat{d}},\nonumber
\end{align}  
where the block length of the block corresponding to the index in $\mathbf{\Xi}_{(i)}$ and $\mathbf{\Xi}_{(j)}$ is $\hat{d}$.}
\end{rmk12}

\subsection{Recovery Algorithms}\label{Recoveryalgorithm}

At present, there exist two categories of sparse recovery algorithms: those utilizing convex optimization techniques and those exploiting greedy iterative mechanisms. Optimization-based algorithms, such as the well-known basis pursuit (BP) \cite{Chen2001BP,Mota2017,Goldstein2018}, provide desirable recovery performance but come with high computational complexity, making practical implementation challenging. On the other hand, greedy algorithms, e.g., OMP and OLS \cite{greed2004,liyang2023,boyangchen2023,liyangsensorjournal}, offer simple and fast implementations, which are beneficial for practical use and have been widely adopted. The main difference between OMP and OLS lies in their greedy rules for selecting the support in each iteration. OMP selects a column in the measurement matrix that is most strongly correlated with the current residual vector, while OLS searches for a candidate that provides the most significant decrease in the residual power. It is worth mentioning that algorithms based on the OLS framework exhibit preferable convergence properties but have higher computational complexity compared to OMP framework-based algorithms.
This paper focuses on OMP-based framework for hierarchically block-sparse recovery. To achieve this, we propose the HiBOMP-P algorithm in a recursive form, which is illustrated in Algorithm~\ref{alg:HiBOXX}.

	The HiBOMP-P algorithm requires as input the measurement matrix, measurement vector, the hierarchically block-sparse structure of the signal to be recovered, the PSI, weight vectors for leveraging the additional PSI index set $\mathbf{\Theta}^{*\Delta}$, and a residual tolerance. Firstly, it defines global variable sets $\mathbf{\Xi}^l_{1}=\mathbf{\emptyset},\mathbf{\Xi}^l_{2}=\mathbf{\emptyset},\cdots,\mathbf{\Xi}^l_{n}=\mathbf{\emptyset}$ $(l\geq0)$, which are used throughout the recursive procedures of HiBOMP-P, and initializes the residual vector as $\mathbf{r}^0=\mathbf{y}$. Secondly, the algorithm performs support selection recursively to accurately identify the true support set. Specifically, for the $t$th hierarchical mode in the $l$th iteration, the residual vector is updated as $\mathbf{r}^l=\mathbf{r}^l+\mathbf{D}_{\mathbf{\Theta}^{*\Delta^t}}\mathbf{x}^l_{*\Delta^t}$ with the assistance of the additional PSI index set $\mathbf{\Theta}^{{*\Delta}^t}$ and the corresponding weight vector $\mathbf{x}^l_{*\Delta^t}$. The support selection rule for HiBOMP-P is $\arg\max\limits_{i\in\{\mathring{\mathbf{N}_1}\}\backslash\mathbf{\Xi}^{t^l}}\big\|(\mathbf{P}^{\bot}_{\mathbf{D}_{\mathbf{\Xi}^{t^l}\cup\mathbf{\Theta}^t}}\mathbf{D}_{[i]})^{\rm H}\mathbf{r}^l\big\|^2_2$, which selects the block most correlated with the residual, aided by the PSI-related index set, where $\mathbf{\Xi}^{t^l}$ denotes the selected index set for the $t$th hierarchical mode in the $l$th iteration. After support selection, the residual is restored to its previous state by subtracting the additional PSI component from the current residual. This recursive procedure continues until HiBOMP-P has selected candidate blocks corresponding to the true support set in the $n$th hierarchical mode. The recursion is repeated to help HiBOMP-P search for other true support sets. Finally, the hierarchically block-sparse signal is recovered by applying least squares (LS) on the estimated support set obtained through the support selection process.

	We briefly analyze the computational complexity of HiBOMP-P. Assume that the signal to be recovered is an $n$-mode hierarchically block-sparse signal with sparsity pattern $(k_0, k_1, k_2, \dots, k_n)$ $(n \geq 0)$, as defined in \textbf{Definition \ref{Hierarchicallyblocksparsesignal}}. For clarity, we consider the PSI index set $\mathbf{\Theta}^t$ $(t \in \{\mathring{n}\})$ to include only the term $\overline{\alpha}^t$. The primary computational cost of HiBOMP-P lies in the support selection function in Algorithm 1, whose complexity can be further broken down into the following components:
	\begin{itemize}
		\item Additional PSI calculations in steps 5 and 7: Since the matrix $\mathbf{D}_{\mathbf{\Theta}^{*\Delta^t}}$ has dimensions $M \times \overline{\alpha}^{t}d$, the computational complexity for each PSI calculation is $\mathcal{O}(M\overline{\alpha}^td)$ for each $t \in \mathring{n}$.
		Given the hierarchically block-sparse structure $(k_1, k_2, \dots, k_n)$ and the fact that the PSI index set $\mathbf{\Theta}^t$ helps reduce complexity, the total computational cost of steps 4 and 6 is $\mathcal{O}\bigg(\sum\limits_{i=1}^{n}M\overline{\alpha}^{i}d\prod\limits_{j=1}^{i}(k_j-\overline{\alpha}^j)\bigg)$.	
		\item Support selection in step 6: The matrix $\mathbf{D}_{\mathbf{\Xi}^{t^l} \cup \mathbf{\Theta}^t}$ has dimensions $M\times \Big(\overline{\alpha}^td+jd\prod\limits_{c=t+1}^{n}N_c\Big)$, where $j$ denotes the selected number of support blocks, and $j\leq k_t-\overline{\alpha}^t$. The complexity of computing the orthogonal projection $\mathbf{P}^{\bot}_{\mathbf{D}_{\mathbf{\Xi}^{t^l} \cup \mathbf{\Theta}^t}}$ is at most $\mathcal{O}\bigg(M^2\Big(\overline{\alpha}^td+(k_t-\overline{\alpha}^t)d\prod\limits_{c=t+1}^{n}N_c\Big)\bigg)$, and the complexity of computing $\mathbf{P}^{\bot}_{\mathbf{D}_{\mathbf{\Xi}^{t^l} \cup \mathbf{\Theta}^t}} \mathbf{D}_{[i]}$ for valid indices $i$ is $\mathcal{O}\bigg(M^2\Big(-\overline{\alpha}^td+d\prod\limits_{c=t-1}^{n}N_c\Big)\bigg)$. Therefore, the total complexity of step 5 is $\mathcal{O}\Bigg(\sum\limits_{i=1}^n\bigg(M^2\Big(\overline{\alpha}^id+(k_i-\overline{\alpha}^i)d\prod\limits_{c=i+1}^{n}N_c\Big)+M^2\Big(-\overline{\alpha}^id+d\prod\limits_{c=i-1}^{n}N_c\Big)\bigg)\prod\limits_{j=1}^i\Big(k_j-\overline{\alpha}^j\Big)\Bigg)$.
		\item LS computation in step 12: The computational complexity of the LS algorithm is on the order of $\mathcal{O}((k_nd)^2M)$. Considering the hierarchical block structure, the total computational complexity is upper bounded by $\mathcal{O}\bigg((k_nd)^2M\prod\limits_{i=0}^{n-1}N_i\bigg)$.
		\item Residual vector update in step 13: The total computational complexity of the residual update step is $\mathcal{O}\bigg(k_ndM\prod\limits_{i=0}^{n-1}N_i\bigg)$.
	\end{itemize}	  
	
	The total computational complexity of HiBOMP-P is the sum of the complexities of its constituent steps. As observed, when the PSI index set includes correct support positions, the complexity is reduced because there is no need to compute support selection results for the indices already contained in $\mathbf{\Theta}^t$.
	However, as the hierarchical mode $n$ increases, the overall complexity also grows due to the additional operations required in both the PSI calculations and the support selection steps.
	
	To provide a more intuitive understanding, we consider the case where the hierarchical mode is set to $n = 2$. In this scenario, the complexity of HiBOMP-P is given by: $\mathcal{O}\Big((k_1-\overline{\alpha}^1)M\overline{\alpha}^1d+(k_1-\overline{\alpha}^1)(k_2-\overline{\alpha}^2)M\overline{\alpha}^2d+(k_1-\overline{\alpha}^1)(M^2(N_{2}(k_1-\overline{\alpha}^{1})d+\overline{\alpha}^1d)+M^2(N_{1}N_{2}d-\overline{\alpha}^1d))+(k_1-\overline{\alpha}^1)(k_2-\overline{\alpha}^2)(M^2(N_{2}(k_2-\overline{\alpha}^2)d+\overline{\alpha}^2d)+M^2(N_{1} N_{2}d-\overline{\alpha}^2d))+(k_2d)^2M N_{1} + k_2dM N_{1}\Big)$. Assuming that $\overline{\alpha}^1$ and $\overline{\alpha}^2$ are of the same order, the expression can be simplified to $\mathcal{O}\Big((k_1-\overline{\alpha}^1)(k_2-\overline{\alpha}^2)(M^2(N_{2}(k_2-\overline{\alpha}^2)d+\overline{\alpha}^2d)+M^2(N_{1} N_{2}d-\overline{\alpha}^2d))+(k_2d)^2M N_{1} \Big)$. In contrast, the computational complexity of conventional OMP and BOMP with sparsity level $k_1 k_2$ is given by $\mathcal{O}(k_1k_2d^2MN_1N_2+M(k_1k_2)^3d^3)$ and $\mathcal{O}(k_1k_2MN_1N_2d+M(k_1k_2)^3d^2)$ \cite{liyang2023tcom}, respectively. It can be seen that BOMP generally runs faster than OMP due to the exploitation of block structures. Compared to BOMP, the first term in the complexity of HiBOMP-P includes an extra factor of $M$, which implies higher cost in certain steps. However, the second term of HiBOMP-P is smaller than that of BOMP. Overall, this suggests that the complexity of HiBOMP-P can be comparable to that of BOMP when $n = 2$.

\begin{algorithm}
	%\doublespacing
	\renewcommand{\algorithmicrequire}{\textbf{Input:}}
	\renewcommand{\algorithmicensure}{\textbf{Output:}}
	\caption{HiBOMP-P}
	\label{alg:HiBOXX} % Alg.1
	\begin{algorithmic}[1]
		\REQUIRE The measurement matrix $\mathbf{D}$, the measurement vector $\mathbf{y}$, hierarchical block sparsity set $\mathbf{k}=$ $\{k_0,k_{1},k_{2},\cdots,k_{n}\}$, hierarchical dimension set $\mathbf{N}=\{N_0,N_1,N_2,\cdots,N_n,d\}$, index sets $\mathbf{\Theta}^t$ and $\mathbf{\Theta}^{*\Delta^t}$  corresponding to the PSI, weight vector $\mathbf{x}_{*\Delta^t}$ $(t\in\{\mathring{n}\})$,
		and residual tolerant $\epsilon$.
		\ENSURE $\hat{\mathbf{\Xi}}$, $\hat{\mathbf{x}}$.\\
		Global variable sets $\mathbf{\Xi}^l_{1}=\mathbf{\emptyset},\mathbf{\Xi}^l_{2}=\mathbf{\emptyset},\cdots,\mathbf{\Xi}^l_{n}=\mathbf{\emptyset}$ $(l\geq0)$.\\
		$\mathbf{r}^0=\mathbf{y}$.\\
		$(\hat{\mathbf{\Xi}},\: \hat{\mathbf{x}})=\text{Support-selection}(\mathbf{D},0,\mathbf{r}^0,\mathbf{k},\mathbf{N},\epsilon,\mathbf{\emptyset},\mathbf{\Theta}^1,1)$.\\
		\vspace{-3mm}
		\hspace{-1.5em}\rule[-10pt]{8.8cm}{0.1em}\\
		\vspace{1mm}
		\hspace{-1.5em} $\text{Support-selection}(\mathbf{D},c,\mathbf{r}^0,\mathbf{k},\mathbf{N},\epsilon,\mathbf{\Lambda},\mathbf{\Theta}^t,t)$\\
		\STATE $\mathbf{Initialization}$: $l=0$, $\mathbf{\Xi}^{t^0}=\mathbf{\Lambda}$, $\mathbf{x}^0=\mathbf{0}$.
		\WHILE {$l< \mathbf{k}_1$ and $\|\mathbf{r}^l\|_2>\epsilon$} 
		\STATE Block length $\tilde{d}=\mathbf{N}_2\mathbf{N}_3\cdots\mathbf{N}_nd$.			
		\IF {$t<n$}
		\STATE $\mathbf{r}^l=\mathbf{r}^l+\mathbf{D}_{\mathbf{\Theta}^{*\Delta^t}}\mathbf{x}^l_{*\Delta^t}$.		
		\STATE $i_l= 
		\arg\max\limits_{i\in\{\mathring{\mathbf{N}_1}\}\backslash\mathbf{\Xi}^{t^l}}\big\|(\mathbf{P}^{\bot}_{\mathbf{D}_{\mathbf{\Xi}^{t^l}\cup\mathbf{\Theta}^t}}\mathbf{D}_{[i]})^{\rm H}\mathbf{r}^l\big\|_2$.
		\STATE	$\mathbf{r}^l=\mathbf{r}^l-\mathbf{D}_{\mathbf{\Theta}^{*\Delta^t}}\mathbf{x}^l_{*\Delta^t}$.
		\STATE $c=i_l$.
		\STATE  $\mathbf{\Xi}^{t^{l+1}}=\mathbf{\Xi}^{t^l}\cup \text{Support-selection}(\mathbf{D}_{[i_l]},c,\mathbf{r}^l,\mathbf{k}\backslash\mathbf{k}_1,$\\
		$\mathbf{N}\backslash\mathbf{N}_1,\epsilon,\mathbf{\Xi}^l,\mathbf{\Theta}^{t+1},t+1)$.
		\ELSE 
		\STATE  $\mathbf{\Xi}^{l+1}=\mathbf{\Xi}^{l}\cup i_l$.
		\STATE $\mathbf{x}^{l+1}=\arg \min\limits_{\mathbf{x}:\;{\rm supp}(\mathbf{x})=\mathbf{\Xi}^{l+1}}\|\mathbf{y}-\mathbf{D}\mathbf{x}\|_2^2$.
		\STATE  $\mathbf{r}^{l+1}=\mathbf{y}-\mathbf{D}\mathbf{x}^{l+1}$.
		\STATE  $l=l+1$.
		\ENDIF
				\ENDWHILE
		\STATE \textbf{return} $\hat{\mathbf{\Xi}}=\mathbf{\Xi}^l$, $\hat{\mathbf{x}}=\mathbf{x}^l$.
	\end{algorithmic}
\end{algorithm}

 It can be observed that the primary support selection mechanism of HiBOMP-P focuses on choosing a matrix block rather than a single column, similar to conventional block-sparse recovery algorithms, e.g., BOMP and BOLS \cite{Kim2021,liyang2023tcom,liyang2023tvt}. For some column-wise index set of $\mathbf{D}$ denoted by $\mathbf{\Delta}$, we define 
\begin{align}
	\ddot{\mathbf{A}}_{i}=\mathbf{P}^{\bot}_{\mathbf{D}_{\mathbf{\Delta}}}\mathbf{D}_{i},\nonumber
\end{align}
and thus we have
\begin{align}
	\ddot{\mathbf{A}}_{[i]}=\mathbf{P}^{\bot}_{\mathbf{D}_{\mathbf{\Delta}}}\mathbf{D}_{[i]}.\nonumber
\end{align}
Based on these descriptions, the selection rule of HiBOMP-P in the $l$th iteration for the hierarchical $t$th mode can be reformulated as $i_l= 
\arg\max\limits_{i\in\{\mathring{\mathbf{N}}_1\}\backslash\mathbf{\Xi}^{t^l}}\big\|\ddot{\mathbf{A}}_{[i]}^{\rm H}\mathbf{r}^l\big\|_2$. It worth mentioning that if $\mathbf{\Theta}=\mathbf{\emptyset}$, then HiBOMP-P converges to the algorithm without prior information, which we refer to as HiBOMP. Furthermore, when the minimum block length $d=1$, then HiBOMP converges to HiOMP, the algorithm that does not consider block structure.

In the following, we usually drop the dependence on the hierarchical mode $t$ for simplified notation without causing confusion.

\section{Main Results}\label{ERCsresults} % S1

In the first subsection, we present several theoretical foundations that facilitate the derivation process. Following that, in the second subsection, we outline our main results regarding the ERCs of HiBOMP-P.

\subsection{Useful Foundations}

To establish recovery bounds related to MIP, determining the tight boundary of the matrix norm is crucial. Therefore, in this subsection, we primarily focus on presenting the tight bounds of the mixed matrix norm. The following definitions and results are provided first, which are useful in the sequel.

As defined in \cite{Eldar2010}, for $\mathbf{x}\in\mathbb{C}^N$, the general mixed $\ell_2/\ell_p$ $(p=1,2,\infty)$ norm is given by
\begin{align}
	\|\mathbf{x}\|_{(d)2,p}=\|\mathbf{r}\|_p,\nonumber
\end{align}
where $\mathbf{r}_i=\|\mathbf{x}_{[i]}\|_2$, and $\mathbf{x}_{[i]}$ are consecutive blocks sized as $d$.
For an $M\times N$ matrix $\mathbf{D}$ with $M=mq$ and $N=nd$, where $m$ and $n$ are integers, we define the following matrix operation:
\begin{align}
	\|\mathbf{D}\|_{(q,d)2,p} = \max\limits_{\mathbf{x}\neq\mathbf{0}}\frac{\|\mathbf{D}\mathbf{x}\|_{(q)2,p}}{\|\mathbf{x}\|_{(d)2,p}}.\label{matrixope}
\end{align}

The matrix operation in (\ref{matrixope}) differs from \cite[Eqn. (45)]{Eldar2010}, as the latter assumes that the mixed norms in the numerator and the denominator are related to the same block length. For the operation in (\ref{matrixope}), we present the subsequent proposition.

\begin{proposition2}\label{prop2}
	$\|\mathbf{D}\|_{(q,d)2,p}$ as defined in (\ref{matrixope}) is a matrix norm, which satisfies the following properties:
	
	(1) Non-negative: $\|\mathbf{D}\|_{(q,d)2,p}\geq0$, and $\|\mathbf{D}\|_{(q,d)2,p}=0$ if and only if $\mathbf{D}=0$;

	(2) Homogeneous: $\|\beta\mathbf{D}\|_{(q,d)2,p}=|\beta|\|\mathbf{D}\|_{(q,d)2,p}$ for all $\beta\in\mathbb{C}$;
	
	(3) Triangle inequality: $\|\mathbf{D}+\mathbf{A}\|_{(q,d)2,p}\leq\|\mathbf{D}\|_{(q,d)2,p}+\|\mathbf{A}\|_{(q,d)2,p}$, where $\mathbf{A}\in\mathbb{C}^{M\times N}$ with $M=mq$ and $N=nd$;
	
	(4) Submultiplicative: $\!\!\|\mathbf{D}\mathbf{B}\|_{\!(q,\overline{d})2,p}\!\!\leq\!\!\|\mathbf{D}\|_{\!(q,d)2,p}\|\mathbf{B}\|_{\!(d,\overline{d})2,p}$,
	where $\mathbf{B}\in\mathbb{C}^{N\times\overline{N}}$ with $\overline{N}=\overline{n}\overline{d}$.
\end{proposition2}

The proof of \textbf{Proposition \ref{prop2}} is omitted, as it can be proved based on the definition of $\|\mathbf{D}\|_{(q,d)2,p}$.
The following lemma provides several useful bounds on $\|\mathbf{D}\|_{2,p}$.

\begin{lemma5}\label{lemma5}
	Let $\mathbf{D}$ be an $M\times N$ matrix with $M=mq$ and $N=nd$. Denote by $\mathbf{D}_{[i,j]}$ the $(i,j)$th block of $\mathbf{D}$ sized as $q\times d$. Then, we have
	\begin{align}
		&\|\mathbf{D}\|_{(q,d)2,\infty}\leq\max_{i}\sum_{j}\|\mathbf{D}_{[i,j]}\|_2\triangleq\rho_{r(q,d)}(\mathbf{D}),\nonumber\\
		&\|\mathbf{D}\|_{(q,d)2,1}\leq\max_{j}\sum_{i}\|\mathbf{D}_{[i,j]}\|_2\triangleq\rho_{c(q,d)}(\mathbf{D}).\nonumber
	\end{align}
\end{lemma5}

The proof of \textbf{Lemma \ref{lemma5}} is similar to that of \cite[Lemma 1]{Eldar2010}, thus we omit it here. Similarly, $\rho_{r(q,d)}(\mathbf{D})$ and $\rho_{c(q,d)}(\mathbf{D})$ are matrix norm, as they are non-negative, homogeneous, submultiplicative, and they also satisfy the triangle inequality, which are similar to those in \textbf{Proposition \ref{prop2}}. Note that when $q=d=h$, $\rho_{r(h,h)}(\mathbf{D})$ and $\rho_{c(h,h)}(\mathbf{D})$ converge to the definitions presented by \cite[Lemma~1]{Eldar2010}. Here we provide the submultiplicativity of $\rho_{c(q,d)}(\mathbf{D})$ in the following lemma as it is formulated differently from that in \cite[Lemma~1]{Eldar2010}. 
\begin{lemma8}\label{lemma8}
	Let $\mathbf{A}$ and $\mathbf{B}$ be $M\times N$ and $N\times G$ matrices with $M=md_1$, $N=nd_2$, and $G=gd_3$. The $d_1\times d_2$ and $d_2\times d_3$ blocks of $\mathbf{A}$ and $\mathbf{B}$ are denoted by $\mathbf{A}_{[m',n']}$ and $\mathbf{B}_{[n',g']}$, respectively. Then $\rho_{c(d_1,d_3)}(\mathbf{A}\mathbf{B})$ satisfies that $\rho_{c(d_1,d_3)}(\mathbf{A}\mathbf{B})\leq\rho_{c(d_1,d_2)}(\mathbf{A})\rho_{c(d_2,d_3)}(\mathbf{B})$.
\end{lemma8}

\begin{IEEEproof}
	See Appendix \ref{profoflemma8}.
\end{IEEEproof}

\begin{rmk2}
	\emph{\textbf{Lemma \ref{lemma8}} applies to the vector version. That is, given a matrix $\mathbf{A}$ sized as $M\times N$ with $M=md_1$ and $N=nd_2$, for any vector $\mathbf{x}$, we have $\|\mathbf{A}\mathbf{x}\|_{(d_1)2,\infty}\leq\rho_{c(d_1,d_2)}(\mathbf{A})\|\mathbf{x}\|_{(d_2)2,\infty}$. The proof is similar to that of \textbf{Lemma~\ref{lemma8}}.}
\end{rmk2}

\begin{rmk5}
	\emph{As for matrix $\mathbf{A}$ given in \textbf{Lemma \ref{lemma8}}, we have $\rho_{c(d_1,d_2)}(\mathbf{A})\leq\big\lceil\frac{d_2}{d_1}\big\rceil\rho_{c(d_1,d_1)}(\mathbf{A})$, where $\lceil\cdot\rceil$ denotes the ceiling function. To prove this inequality, we first denote a new matrix stemming from $\mathbf{A}$ by adding all 0 matrices, which is formulated as 
		\begin{align}
			\overline{\mathbf{A}}=\begin{bmatrix}
				\overline{\mathbf{A}}_{[1]},\overline{\mathbf{A}}_{[2]},\cdots,\overline{\mathbf{A}}_{[n]}
			\end{bmatrix},\nonumber
		\end{align}
	where 
		\begin{align}
		\overline{\mathbf{A}}_{[j]}=\bigg[
			& \mathbf{B}_{(md_1,d_1)[1]} \quad \mathbf{B}_{(md_1,d_1)[2]} \quad\cdots\nonumber\\
			&\mathbf{B}_{\big(md_1,d_2-\big\lfloor\frac{d_2}{d_1}\big\rfloor d_1\big)\big[\big\lceil\frac{d_2}{d_1}\big\rceil\big]} \quad\mathbf{0}_{\big(md_1,d_1-d_2+\big\lfloor\frac{d_2}{d_1}\big\rfloor d_1\big)}
		\bigg],\nonumber
	\end{align}
$\mathbf{B}=\mathbf{A}_{(md_1,d_2)[j]}$, and $\mathbf{B}_{(a,b)[i]}$ denotes the $i$th column-block submatrix of $\mathbf{B}$ sized as $a\times b$. It can be observed that $\overline{\mathbf{A}}$ consists of regular matrix blocks sized as $d_1\times d_1$. Denote by $\overline{\mathbf{A}}_{[j]_{(a,b)[i,r]}}$ the $(i,r)$th $a\times b$ matrix block of $\overline{\mathbf{A}}_{[j]}$.
		Then, for any $\frac{d_2}{d_1}$ is not an integer, we have 
	\begin{align}
		\rho_{c(d_1,d_2)}(\mathbf{A})&=\max_{j}\sum_{i}\|\mathbf{A}_{(d_1,d_2)[i,j]}\|_2\nonumber\\
		&\leq\max_j\sum_{i}\Bigg(\bigg(\sum_{r=1}^{\big\lfloor\frac{d_2}{d_1}\big\rfloor}\|\overline{\mathbf{A}}_{[j]_{(d_1,d_1)[i,r]}}\|_2\bigg)\nonumber\\
		&\quad+\Bigg\|\overline{\mathbf{A}}_{[j]_{\big(d_1,d_2-\big\lfloor\frac{d_2}{d_1}\big\rfloor d_1\big)\big[i,\big\lceil\frac{d_2}{d_1}\big\rceil\big]}}\Bigg\|_2\Bigg)\nonumber\\
		&=\max_j\sum_{i}\Bigg(\bigg(\sum_{r=1}^{\big\lfloor\frac{d_2}{d_1}\big\rfloor}\|\overline{\mathbf{A}}_{[j]_{(d_1,d_1)[i,r]}}\|_2\bigg)\nonumber\\
		&\quad+\Bigg\|\overline{\mathbf{A}}_{[j]_{(d_1,d_1)\big[i,\big\lceil\frac{d_2}{d_1}\big\rceil\big]}}\Bigg\|_2\Bigg)\nonumber\\
		&\leq\bigg\lfloor\frac{d_2}{d_1}\bigg\rfloor\rho_{c(d_1,d_1)}(\overline{\mathbf{A}})+\rho_{c(d_1,d_1)}(\overline{\mathbf{A}})\nonumber\\
		&=\bigg\lceil\frac{d_2}{d_1}\bigg\rceil\rho_{c(d_1,d_1)}(\overline{\mathbf{A}}),\nonumber
	\end{align}
where $\lfloor\cdot\rfloor$ denotes the floor function, and the first inequality follows from \textbf{Lemma \ref{lemma13}}. Note that for the sake of symbol simplicity, we do not distinguish between $\mathbf{A}$ and its transformation $\overline{\mathbf{A}}$ in the sequel without causing confusion.}
\end{rmk5}

The following lemma and corollary provide useful bounds of the mixed $\ell_{2}/\ell_{\infty}$ norm.

\begin{lemma6}\label{lemma6}
	For any vector set $\mathbf{\Gamma}=\{\mathbf{x}^1, \mathbf{x}^2, \cdots, \mathbf{x}^n\}$ with isometric length, denote the formulation consisting of their any two partitions as $\mathbf{x}^i=[(\mathbf{x}^i_{[1]})^{\rm T},(\mathbf{x}^i_{[2]})^{\rm T}]^{\rm T}$, where $1\leq i\leq n$. Then, we have 
	\begin{align}
		&\max\Big(\max_{1\leq p\leq n}\|\mathbf{x}^p_{[1]}\|^2_2+\min_{1\leq q\leq n}\|\mathbf{x}^q_{[2]}\|^2_2,\nonumber\\
		&\qquad\quad\max_{1\leq q\leq n}\|\mathbf{x}^q_{[2]}\|^2_2+\min_{1\leq p\leq n}\|\mathbf{x}^p_{[1]}\|^2_2\Big)\nonumber\\
		&\leq \max\limits_{1\leq t\leq n} \|\mathbf{x}^t\|^2_2
		\leq\max_{1\leq p\leq n}\|\mathbf{x}^p_{[1]}\|^{2}_2 + \max_{1\leq q\leq n}\|\mathbf{x}^q_{[2]}\|_2^2,\nonumber
	\end{align}
where $\max(\cdot,\cdot)$ returns the maximum value of its targets.
\end{lemma6}
\begin{IEEEproof}
	See Appendix \ref{profoflemma6}.
\end{IEEEproof}

The following corollary follows from \textbf{Lemma \ref{lemma6}}.

\begin{Corollary1}\label{corollary1}
	For any vector $\mathbf{x}\in\mathbb{C}^N$ with block length $d$, denote its index set corresponding to block length $d$ as $\mathbf{\Xi}$, and thus $|\mathbf{\Xi}|=\frac{N}{d}$. Letting each of the $d$ size block partitioned as two sets with equal lengths denoted by $\mathbf{\Xi}^*_{(i)}$ and $\mathbf{\Xi}^{\Delta}_{(i)}$ $(i\in\{1,2,\cdots,\frac{N}{d}\})$, where $|\mathbf{\Xi}^*_{(i)}|=d_{(i)}^1$, $|\mathbf{\Xi}^{\Delta}_{(i)}|=d_{(i)}^2$, and $d_{(i)}^1+d_{(i)}^2=d$, we have $\mathbf{\Xi} = [\mathbf{\Xi}^*_{(1)},\mathbf{\Xi}^{\Delta}_{(1)},\mathbf{\Xi}^*_{(2)},\mathbf{\Xi}^{\Delta}_{(2)},\cdots,\mathbf{\Xi}^*_{(\frac{N}{d})},\mathbf{\Xi}^{\Delta}_{(\frac{N}{d})}]$. Based on \textbf{Lemma \ref{lemma6}}, the following inequality holds:
	\begin{align}
		&\max\Big(\max\limits_{i\in\{1,2,\cdots,\frac{N}{d}\}}\|\mathbf{x}_{\mathbf{\Xi}^*_{(i)}}\|^2_2+\min\limits_{j\in\{1,2,\cdots,\frac{N}{d}\}}\|\mathbf{x}_{\mathbf{\Xi}^{\Delta}_{(j)}}\|^2_2,\nonumber\\
		&\qquad\quad\max\limits_{j\in\{1,2,\cdots,\frac{N}{d}\}}\|\mathbf{x}_{\mathbf{\Xi}^{\Delta}_{(j)}}\|^2_2+\min\limits_{i\in\{1,2,\cdots,\frac{N}{d}\}}\|\mathbf{x}_{\mathbf{\Xi}^*_{(i)}}\|^2_2\Big)\nonumber\\
		&\leq\!\|\mathbf{x}\|^2_{(d)2,\infty}\leq\!\max\limits_{i\in\{1,2,\cdots,\frac{N}{d}\}}\!\|\mathbf{x}_{\mathbf{\Xi}^*_{(i)}}\|_2^2 + \max\limits_{j\in\{1,2,\cdots,\frac{N}{d}\}}\!\|\mathbf{x}_{\mathbf{\Xi}^{\Delta}_{(j)}}\|_2^2.\nonumber
	\end{align} 
\end{Corollary1}

\subsection{Exact Recovery Conditions}\label{ercssss}

After completing these preparations, we now illustrate the ERCs of HiBOMP-P. It can be observed that for an $n$-mode hierarchically block-sparse signal, HiBOMP-P should perform block selection at least $n$ times, as the algorithm performs at least once per hierarchical mode. Additionally, the HiBOMP-P algorithm performs block selection exactly $n$ times when the block sparsity of the $n$th mode is equal to 1, with HiBOMP-P selecting the correct block in each iteration. Considering this, we gradually derive the ERCs of HiBOMP-P by allowing the algorithm to select the correct block in each iteration within each hierarchical mode. We then integrate all $n$-mode conditions to obtain the overall reconstruction guarantees. 

To clarify, we start with the following notation. $\mathbf{\Xi}_{<i>}$ and $\overline{\mathbf{\Xi}}_{<i>}$denotes the support and non support index sets from the $i$th block in the $(t-1)$th hierarchical mode, respectively, $\mathbf{\Xi}^*$ is the true support index set, $\mathbf{\Xi}^{\Delta}$ is the additional support index set, $\mathbf{\Xi}^{\circ}$ is the index set outside the $i$th block in the $(t-1)$th hierarchical mode, $\mathbf{\Theta}$ is the PSI index set, $\mathbf{\Theta}^*$, and $\mathbf{\Theta}^{-}$ are the PSI index sets corresponding to the true support, and the non support indices, $\mathbf{\Theta}^{\Delta}$ is the PSI index sets corresponding to the additional support indices, and $\mathbf{\Theta}^{\circ}$ denotes the PSI index set corresponding to the indices outside the $i$th block. $\mathbf{\Theta}^{*\Delta}$ is the PSI index set corresponding to the additional augmented indices with $\mathbf{\Theta}\cap\mathbf{\Theta}^{*\Delta}=\mathbf{\emptyset}$.
Based on these preparations, the following theorem presents an ERC of HiBOMP-P in each hierarchical mode.

\begin{theoremmaindescription1}\label{theorem1}
	Let $\mathbf{x}\in\mathbb{C}^{N_1N_2\cdots N_{n} d}$ be an $n$-mode hierarchically block-sparse vector, and let $\mathbf{y}=\mathbf{D}\mathbf{x}$ for a given matrix $\mathbf{D}\in\mathbb{C}^{M\times N_1N_2\cdots N_{n} d}$. For the $t$th $(t\in\{\mathring{n}\})$ hierarchical mode, suppose that $\mathbf{D}_{\mathbf{\Xi}_{<i>}\cup\mathbf{\Theta}}$ is full rank with $|\mathbf{\Xi}_{<i>}|=k_t$,  $|\mathbf{\Xi}^*\cap\mathbf{\Theta}|=|\mathbf{\Xi}^*\cap\mathbf{\Theta}^*|=\alpha^*$, $|\mathbf{\Xi}^{\Delta}\cap\mathbf{\Theta}|=|\mathbf{\Xi}^{\Delta}\cap\mathbf{\Theta}^{\Delta}|=\alpha^{\Delta}$, $\mathbf{\Xi}^*\cap\mathbf{\Xi}^{\Delta}=\mathbf{\emptyset}$, $|\overline{\mathbf{\Xi}}_{<i>}\cap\mathbf{\Theta}|=|\overline{\mathbf{\Xi}}_{<i>}\cap\mathbf{\Theta}^{-}|=\beta$, $|{\mathbf{\Xi}}^{\circ}\cap\mathbf{\Theta}|=|\mathbf{\Xi}^{\circ}\cap\mathbf{\Theta}^{\circ}|=\gamma$, and $\mathbf{\Theta}^{*}\sqcup\mathbf{\Theta}^{\Delta}\sqcup\mathbf{\Theta}^{-}\sqcup\mathbf{\Theta}^{\circ}=\mathbf{\Theta}$.
	For block selection within the $i$th hierarchical block of the $(t-1)$th hierarchical mode, a sufficient condition for HiBOMP-P with PSI to correctly identify the support blocks of $\mathbf{x}$ in the $t$th hierarchical mode at the $l$th iteration $(t \in \{1, 2, \dots, n\})$ is that
\begin{align}\label{mainmain}
	{G}_{*}+{G}_{\circ}<1,
\end{align}
where  
\begin{align}
	{G}_{*}\triangleq&\rho_{c(d^*+d^{*\Delta},\overline{d})}(\ddot{\mathbf{A}}^{\dagger}_{[\mathbf{\Xi}^*\backslash\mathbf{\Theta},\mathbf{\Theta}^{*\Delta}]}\ddot{\mathbf{A}}_{\overline{\mathbf{\Xi}}_{<i>}\backslash\mathbf{\Theta}})\nonumber\\
	&\times\bigg(1-\big(\rho_{c(d^{\circ},d^*+d^{*\Delta})}(\ddot{\mathbf{A}}^{\rm H}_{\mathbf{\Xi}^{\circ}\backslash\mathbf{\Theta}^{\circ}}\ddot{\mathbf{A}}_{[\mathbf{\Xi}^*\backslash\mathbf{\Theta},\mathbf{\Theta}^{*\Delta}]})\nonumber\\
	&\qquad\qquad+\rho_{c(d^{\circ},d^{\Delta})}(\ddot{\mathbf{A}}^{\rm H}_{\mathbf{\Xi}^{\circ}\backslash\mathbf{\Theta}^{\circ}}\ddot{\mathbf{A}}_{\mathbf{\Xi}^{\Delta}\backslash\mathbf{\Theta}})\big)^{\frac{1}{2}}\nonumber\\
	&\qquad\qquad\!\!\!\times\big(\sigma_{\min}(\ddot{\mathbf{A}}^{\rm H}_{[\mathbf{\Xi}^*\backslash\mathbf{\Theta},\mathbf{\Theta}^{*\Delta}]}\ddot{\mathbf{A}}_{[\mathbf{\Xi}^*\backslash\mathbf{\Theta},\mathbf{\Theta}^{*\Delta}]})\big)^{-1}\nonumber\\
	&\qquad\qquad\!\!\!\times\frac{\|\mathbf{x}^l_{\mathbf{\Xi}^{\circ}\backslash\mathbf{\Theta}^{\circ}}\|_{(d^{\circ})2,\infty}}{\|\mathbf{x}^l_{[\mathbf{\Xi}^*\backslash\mathbf{\Theta},\mathbf{\Theta}^{*\Delta}]}\|_{(d^*+d^{*\Delta})2,\infty}}\bigg)^{-1},\nonumber\\
	{G}_{\circ}\triangleq&\rho_{c(d^{\circ},\overline{d})}(\ddot{\mathbf{A}}^{\rm H}_{\mathbf{\Xi}^{\circ}\backslash\mathbf{\Theta}^{\circ}}\ddot{\mathbf{A}}_{\overline{\mathbf{\Xi}}_{<i>}\backslash\mathbf{\Theta}})\nonumber\\
	&\times\bigg(\sigma_{\min}(\ddot{\mathbf{A}}^{\rm H}_{[\mathbf{\Xi}^*\backslash\mathbf{\Theta},\mathbf{\Theta}^{*\Delta}]}\ddot{\mathbf{A}}_{[\mathbf{\Xi}^*\backslash\mathbf{\Theta},\mathbf{\Theta}^{*\Delta}]})\nonumber\\
	&\qquad\times\frac{\|\mathbf{x}^l_{[\mathbf{\Xi}^*\backslash\mathbf{\Theta},\mathbf{\Theta}^{*\Delta}]}\|_{(d^*+d^{*\Delta})2,\infty}}{\|\mathbf{x}^l_{\mathbf{\Xi}^{\circ}\backslash\mathbf{\Theta}^{\circ}}\|_{(d^{\circ})2,\infty}}\nonumber\\
	&\qquad-\big(\rho_{c(d^{\circ},d^*+d^{*\Delta})}(\ddot{\mathbf{A}}^{\rm H}_{\mathbf{\Xi}^{\circ}\backslash\mathbf{\Theta}^{\circ}}\ddot{\mathbf{A}}_{[\mathbf{\Xi}^*\backslash\mathbf{\Theta},\mathbf{\Theta}^{*\Delta}]})\nonumber\\
	&\qquad+\rho_{c(d^{\circ},d^{\Delta})}(\ddot{\mathbf{A}}^{\rm H}_{\mathbf{\Xi}^{\circ}\backslash\mathbf{\Theta}^{\circ}}\ddot{\mathbf{A}}_{\mathbf{\Xi}^{\Delta}\backslash\mathbf{\Theta}})\big)^{\frac{1}{2}}\bigg)^{-1},\nonumber
\end{align}
$d^*$, $d^{\Delta}$, $d^{*\Delta}$, $\overline{d}$, and $d^{\circ}$ are column-wise block lengths related to submatrices consisting of the indices in the index sets $\mathbf{\Xi}^*\backslash\mathbf{\Theta}$,  $\mathbf{\Xi}^{\Delta}\backslash\mathbf{\Theta}$, $\mathbf{\Theta}^{*\Delta}$,  $\overline{\mathbf{\Xi}}_{<i>}\backslash\mathbf{\Theta}$, and $\mathbf{\Xi}^{\circ}\backslash\mathbf{\Theta}^{\circ}$, and $\sigma_{\min}(\cdot)$ returns the minimum eigenvalue or singular value of its objective.
\end{theoremmaindescription1}

In the derived condition (\ref{mainmain}), the index set outside the $i$th block in the $(t-1)$th hierarchical mode plays a crucial role in tightening the bound. Specifically, when $\mathbf{\Xi}^{\circ} = \mathbf{\emptyset}$ and $\mathbf{\Theta}^{\circ} = \mathbf{\emptyset}$, meaning that the support is entirely contained within the $i$th block of the $(t-1)$th hierarchical mode, the sufficient condition in (\ref{mainmain}) simplifies to ${G}_{*} < 1$, where ${G}_{*}=\rho_{c(d^*+d^{*\Delta},\overline{d})}(\ddot{\mathbf{A}}^{\dagger}_{[\mathbf{\Xi}^*\backslash\mathbf{\Theta},\mathbf{\Theta}^{*\Delta}]}\ddot{\mathbf{A}}_{\overline{\mathbf{\Xi}}_{<i>}\backslash\mathbf{\Theta}})$. This condition is less restrictive than the original bound, suggesting that signal recovery is easier when the support is concentrated within a single block. Moreover, condition (\ref{mainmain}) depends on the amplitude of the signal to be recovered. In particular, it includes the term $\frac{\|\mathbf{x}^l_{\mathbf{\Xi}^{\circ}\backslash\mathbf{\Theta}^{\circ}}\|_{(d^{\circ})2,\infty}}{\|\mathbf{x}^l_{[\mathbf{\Xi}^*\backslash\mathbf{\Theta},\mathbf{\Theta}^{*\Delta}]}\|_{(d^*+d^{*\Delta})2,\infty}}$. It can be seen that $\frac{\partial G_*}{\|\mathbf{x}^l_{\mathbf{\Xi}^{\circ}\backslash\mathbf{\Theta}^{\circ}}\|_{(d^{\circ})2,\infty}}>0$,  $\frac{\partial G_{*}}{\|\mathbf{x}^l_{[\mathbf{\Xi}^*\backslash\mathbf{\Theta},\mathbf{\Theta}^{*\Delta}]}\|_{(d^*+d^{*\Delta})2,\infty}}<0$, and similarly, $\frac{\partial G_{\circ}}{\|\mathbf{x}^l_{\mathbf{\Xi}^{\circ}\backslash\mathbf{\Theta}^{\circ}}\|_{(d^{\circ})2,\infty}}>0$, $\frac{\partial G_{\circ}}{\|\mathbf{x}^l_{[\mathbf{\Xi}^*\backslash\mathbf{\Theta},\mathbf{\Theta}^{*\Delta}]}\|_{(d^*+d^{*\Delta})2,\infty}}<0$. These results reveal that the bound in condition (\ref{mainmain}) is more likely to hold when the amplitudes of the signal components corresponding to the support and the additional PSI index set $\mathbf{\Theta}^{*\Delta}$ are large. In contrast, if the amplitudes of $\mathbf{x}^l$ outside the $i$th block in the $(t-1)$th hierarchical mode are large, it becomes more difficult for condition (\ref{mainmain}) to be satisfied. This leads to degraded performance in recovering the support blocks of $\mathbf{x}$ in the $t$th hierarchical mode at the $l$th iteration. In other words, during the identification of the current support block, signal components outside the target block act as interfering factors. Enhancing $\|\mathbf{x}^l_{[\mathbf{\Xi}^*\backslash\mathbf{\Theta},\mathbf{\Theta}^{*\Delta}]}\|_{(d^*+d^{*\Delta})2,\infty}$ or reducing $\|\mathbf{x}^l_{\mathbf{\Xi}^{\circ}\backslash\mathbf{\Theta}^{\circ}}\|_{(d^{\circ})2,\infty}$ can help mitigate this negative impact.

\begin{rmk4}\label{rmk4}
	\emph{In choosing the support block of HiBOMP-P in the $i$th hierarchical block, two important terms should be considered, i.e., $\|\ddot{\mathbf{A}}^{\rm H}_{\mathbf{\Xi}^*\backslash\mathbf{\Theta}}\mathbf{r}^l\|_{(d^*)2,\infty}$ and $\|\ddot{\mathbf{A}}^{\rm H}_{\mathbf{\Xi}^{\Delta}\backslash\mathbf{\Theta}}\mathbf{r}^l\|_{(d^{\Delta})2,\infty}$. It can be observed that $\|\ddot{\mathbf{A}}^{\rm H}_{\mathbf{\Xi}^{\Delta}\backslash\mathbf{\Theta}}\mathbf{r}^l\|_{(d^{\Delta})2,\infty}$ is a term that could cause adverse effects to support selection since the index set $\mathbf{\Xi}^{\Delta}\backslash\mathbf{\Theta}$ does not contain the true support indices. Thus, it is crucial to categorize and discuss the two cases where $\|\ddot{\mathbf{A}}^{\rm H}_{\mathbf{\Xi}^*\backslash\mathbf{\Theta}}\mathbf{r}^l\|_{(d^*)2,\infty}\geq\|\ddot{\mathbf{A}}^{\rm H}_{\mathbf{\Xi}^{\Delta}\backslash\mathbf{\Theta}}\mathbf{r}^l\|_{(d^{\Delta})2,\infty}$ and $\|\ddot{\mathbf{A}}^{\rm H}_{\mathbf{\Xi}^*\backslash\mathbf{\Theta}}\mathbf{r}^l\|_{(d^*)2,\infty}<\|\ddot{\mathbf{A}}^{\rm H}_{\mathbf{\Xi}^{\Delta}\backslash\mathbf{\Theta}}\mathbf{r}^l\|_{(d^{\Delta})2,\infty}$, respectively. This phenomenon also suggests a potential operation of a hierarchical structure. By enlarging the term $\|\ddot{\mathbf{A}}^{\rm H}_{\mathbf{\Xi}^{\Delta}\backslash\mathbf{\Theta}}\mathbf{r}^l\|_{(d^{\Delta})2,\infty}$, one might achieve better atom selection. However, without additional intervention, $\|\ddot{\mathbf{A}}^{\rm H}_{\mathbf{\Xi}^*\backslash\mathbf{\Theta}}\mathbf{r}^l\|_{(d^*)2,\infty}<\|\ddot{\mathbf{A}}^{\rm H}_{\mathbf{\Xi}^{\Delta}\backslash\mathbf{\Theta}}\mathbf{r}^l\|_{(d^{\Delta})2,\infty}$ 
		indicates that the power of the true support is too small, leading to an inferior recovery formulation and resulting in unreliable recovery.    In practice, we can see that $\|\ddot{\mathbf{A}}^{\rm H}_{\mathbf{\Xi}^*\backslash\mathbf{\Theta}}\mathbf{r}^l\|_{(d^*)2,\infty}\geq\|\ddot{\mathbf{A}}^{\rm H}_{\mathbf{\Xi}^{\Delta}\backslash\mathbf{\Theta}}\mathbf{r}^l\|_{(d^{\Delta})2,\infty}$ is easier to establish compared to $\|\ddot{\mathbf{A}}^{\rm H}_{\mathbf{\Xi}^*\backslash\mathbf{\Theta}}\mathbf{r}^l\|_{(d^*)2,\infty}<\|\ddot{\mathbf{A}}^{\rm H}_{\mathbf{\Xi}^{\Delta}\backslash\mathbf{\Theta}}\mathbf{r}^l\|_{(d^{\Delta})2,\infty}$, since residual $\mathbf{r}^l$ may contain the components in $\ddot{\mathbf{A}}_{\mathbf{\Xi}^*\backslash\mathbf{\Theta}}$, leading to larger mixed norm of the vector $\ddot{\mathbf{A}}^{\rm H}_{\mathbf{\Xi}^*\backslash\mathbf{\Theta}}\mathbf{r}^l$. For a more intuitive display, consider the case where $\mathbf{\Xi}^{\circ}=\mathbf{\emptyset}$. When $(d^*-1)\nu_{d^*}+(k_t-\overline{\alpha}-1)d^*\mu_{d^*}<1$ and $(d^*-1)\nu_{d^*}+(\lceil\frac{rd}{d^*}\rceil-1)d^*\mu_{d^*}<1$,  define the following variables:
	\begin{align}
		&\delta^* \triangleq (1-(d^*-1)\nu_{d^*}-(k_t-\overline{\alpha}-1)d^*\mu_{d^*}),\nonumber\\
		&\delta^{\Delta}\triangleq
		\bigg\lceil\frac{d^{\Delta}}{d^*}\bigg\rceil (k_t-\overline{\alpha})d^*\mu_{d^*}\nonumber\\
		&\qquad+\frac{\lceil\frac{rd}{d^*}\rceil^2\lceil\frac{(k_t-\overline{\alpha})d^{\Delta}}{d^*}\rceil (k_t-\overline{\alpha})d^{*^2}\mu^2_{d^*}}{1-(d^*-1)\nu_{d^*}-(\lceil\frac{rd}{d^*}\rceil-1)d^*\mu_{d^*}}.\nonumber
	\end{align}
	If $\delta^*\geq\delta^{\Delta}$, then $\|\ddot{\mathbf{A}}^{\rm H}_{\mathbf{\Xi}^*\backslash\mathbf{\Theta}}\mathbf{r}^l\|_{(d^*)2,\infty}\geq\|\ddot{\mathbf{A}}^{\rm H}_{\mathbf{\Xi}^{\Delta}\backslash\mathbf{\Theta}}\mathbf{r}^l\|_{(d^{\Delta})2,\infty}$ holds. Consider that $\mathbf{\Theta}=\mathbf{\emptyset}$, the condition $\delta^*\geq\delta^{\Delta}$ in terms of reconstructible sparsity for the $t$th hierarchical mode becomes that
\begin{align}
	kd^*\leq k_0k_1\cdots k_{t-1}\frac{\frac{1}{\mu_{d^*}}(1-(d^*-1)\nu_{d^*})+d^*}{1+\lceil\frac{d^{\Delta}}{d^*}\rceil}\triangleq K^*,\label{kxing}
\end{align}
where $k$ denotes the reconstructible block sparsity, and thus $kd^*$ represents the reconstructible sparsity. From (\ref{kxing}), we directly obtain that $\frac{\partial K^*}{\partial \mu_{d^*}}<0$, $\frac{\partial K^*}{\partial \nu_{d^*}}<0$, and $\frac{\partial K^*}{\partial d^{\Delta}}<0$. Meanwhile, assuming that $\nu_{d^*}=\frac{1}{d^*-1}$, we have $\frac{\partial K^*}{\partial d^*}>0$. These observations indicate that a lower coherence, in terms of hierarchical block coherence and hierarchical sub-coherence of the measurement matrix, and a smaller block length of the non support block in each hierarchical signal block make it easier for $\delta^*\geq\delta^{\Delta}$ to hold. Meanwhile, the larger the block length of the support signal block in each hierarchical signal block, the easier it is for the condition $\delta^*\geq\delta^{\Delta}$ to hold. 
}
\end{rmk4}

\begin{IEEEproof}
	See Appendix \ref{profofrmk4}.
\end{IEEEproof}

It is evident that the recovery condition derived in \textbf{Theorem~\ref{theorem1}} is specific to the particular support index set, making it neither general nor practical for real-world applications. A fundamental question is to test whether the ERC holds as studied in many works \cite{greed2004,liyang2024}. However, the result in revealing the sufficient condition of the condition in \textbf{Theorem \ref{theorem1}} becomes quite challenging, as the introduction of hierarchical structure induces certain complex terms related to the mixed norm $\rho_{c(\cdot,\cdot)}(\cdot)$ and the minimum eigenvalue or singular value of complex block matrices. To address this, we consider the relationship between the norm properties of block matrices and hierarchical block MIP, having the potential to unveil tight bounds for eigenvalues or singular values and revealing the uncertain relationships of matrix blocks within hierarchical measurement matrices. Based on this, the following theorem is presented, providing a sufficient condition for establishing the ERC of HiBOMP-P.
\begin{theoremmaindescription6}\label{theo6}
	Let $\mathbf{x}\in\mathbb{C}^{N_1N_2\cdots N_{n} d}$ be an $n$-mode hierarchically block-sparse vector, and let $\mathbf{y}=\mathbf{D}\mathbf{x}$ for a given matrix $\mathbf{D}\in\mathbb{C}^{M\times N_1N_2\cdots N_{n} d}$. Denote by $\mu_{d^{*}+d^{*\Delta}}$ and $\mu_{d^{\circ}}$ the hierarchical block coherence with respect to block length $d^{*}+d^{*\Delta}$ and $d^{\circ}$, respectively. Denote by $\nu_{d^{*}+d^{*\Delta}}$ and $\nu_{d^{\circ}}$ the hierarchical sub-coherence with respect to block length $d^{*}+d^{*\Delta}$ and $d^{\circ}$, respectively. Denote by $k_t^{\circ}$ the outside block sparsity with block length $d^{\circ}$ for the $t$th hierarchical mode. Suppose that $(d^*+d^{*\Delta}-1)\nu_{d^*+d^{*\Delta}}+(\lceil\frac{rd}{d^*+d^{*\Delta}}\rceil-1)(d^*+d^{*\Delta})\mu_{d^*+d^{*\Delta}}<1$, $(d^*+d^{*\Delta}-1)\nu_{d^*+d^{*\Delta}}  +(k_t-\overline{\alpha}-1)(d^*+d^{*\Delta})\mu_{d^*+d^{*\Delta}}<1$, and $(d^{\circ}-1)\nu_{d^{\circ}}+(\lceil\frac{rd}{d^{\circ}}\rceil-1)d^{\circ}\mu_{d^{\circ}}<1$, where $r=\alpha^{*}+\alpha^{\Delta}+\overline{\alpha}+\beta+\gamma$.
	For block selection in the $i$th hierarchical block from the $(t-1)$th hierarchical mode, a sufficient condition for (\ref{mainmain}) in \textbf{Theorem \ref{theorem1}} is that
	\begin{align}
			\overline{G}_{*}+\overline{G}_{\circ}<1,\label{theo6main}
	\end{align}
where $\overline{G}_{*}$ and $\overline{G}_{\circ}$ are formulated in (\ref{somanyparameters}).
\begin{figure*}[hb]
	\rule[0pt]{18.2cm}{0.05em}
	\begin{align}
		\overline{G}_{*}\triangleq&\frac{\delta_{d^*+d^{*\Delta},\overline{d}}}{1-\frac{(\delta_{d^{\circ},d^*+d^{*\Delta}}+\delta_{d^{\circ},d^{\Delta}})^{\frac{1}{2}}}{\delta_{\sigma_{\min}}}\times\frac{\|\mathbf{x}^l_{\mathbf{\Xi}^{\circ}\backslash\mathbf{\Theta}^{\circ}}\|_{(d^{\circ})2,\infty}}{\|\mathbf{x}^l_{[\mathbf{\Xi}^*\backslash\mathbf{\Theta},\mathbf{\Theta}^{*\Delta}]}\|_{(d^*+d^{*\Delta})2,\infty}}},\nonumber\\
		\overline{G}_{\circ}\triangleq&\frac{\delta_{d^{\circ},\overline{d}}}{\delta_{\sigma_{\min}}\times\frac{\|\mathbf{x}^l_{[\mathbf{\Xi}^*\backslash\mathbf{\Theta},\mathbf{\Theta}^{*\Delta}]}\|_{(d^*+d^{*\Delta})2,\infty}}{\|\mathbf{x}^l_{\mathbf{\Xi}^{\circ}\backslash\mathbf{\Theta}^{\circ}}\|_{(d^{\circ})2,\infty}}-(\delta_{d^{\circ},d^*+d^{*\Delta}}+\delta_{d^{\circ},d^{\Delta}})^{\frac{1}{2}}},\nonumber\\
		\delta_{d^*+d^{*\Delta},\overline{d}}\triangleq&\bigg(\bigg\lceil\frac{\overline{d}}{d^*+d^{*\Delta}}\bigg\rceil(k_t-\overline{\alpha})(d^*+d^{*\Delta})\mu_{d^*+d^{*\Delta}}\nonumber\\
		&+\frac{(k_t-\overline{\alpha})(d^*+d^{*\Delta})^2\mu^2_{d^*+d^{*\Delta}}\lceil\frac{\overline{d}}{d^*+d^{*\Delta}}\rceil \lceil\frac{rd}{d^*+d^{*\Delta}}\rceil}{1-(d^*+d^{*\Delta}-1)\nu_{d^*+d^{*\Delta}}-(\lceil\frac{rd}{d^*+d^{*\Delta}}\rceil-1)(d^*+d^{*\Delta})\mu_{d^*+d^{*\Delta}}}\bigg)\nonumber\\
		&
		\times
		\bigg((1-(d^*+d^{*\Delta}-1)\nu_{d^*+d^{*\Delta}}  -(k_t-\overline{\alpha}-1)(d^*+d^{*\Delta})\mu_{d^*+d^{*\Delta}})\nonumber\\
		&-\frac{(d^*+d^{*\Delta})^2\mu_{d^*+d^{*\Delta}}^2\lceil\frac{rd}{d^*+d^{*\Delta}} \rceil (k_t-\overline{\alpha})}{1-(d^*+d^{*\Delta}-1)\nu_{d^*+d^{*\Delta}}-(\lceil\frac{rd}{d^*+d^{*\Delta}}\rceil-1)(d^*+d^{*\Delta})\mu_{d^*+d^{*\Delta}}}\bigg)^{-1},\nonumber\\
		\delta_{d^{\circ},d^*+d^{*\Delta}}\triangleq&\bigg\lceil\frac{d^{*}+d^{*\Delta}}{d^{\circ}}\bigg\rceil\bigg(k^{\circ}_{t}d^{\circ}\mu_{d^{\circ}}+\frac{(k_t^{\circ}-\gamma)(d^{\circ})^2\mu^2_{d^{\circ}} \lceil\frac{rd}{d^{\circ}}\rceil}{1-(d^{\circ}-1)\nu_{d^{\circ}}-(\lceil\frac{rd}{d^{\circ}}\rceil-1)d^{\circ}\mu_{d^{\circ}}}\bigg),\nonumber\\
		\delta_{d^{\circ},d^{\Delta}}\triangleq&\bigg\lceil\frac{d^{\Delta}}{d^{\circ}}\bigg\rceil\bigg(k_t^{\circ}d^{\circ}\mu_{d^{\circ}}+\frac{(k_t^{\circ}-\gamma)(d^{\circ})^2\mu^2_{d^{\circ}} \lceil\frac{rd}{d^{\circ}}\rceil}{1-(d^{\circ}-1)\nu_{d^{\circ}}-(\lceil\frac{rd}{d^{\circ}}\rceil-1)d^{\circ}\mu_{d^{\circ}}}\bigg),\nonumber\\
		\delta_{d^{\circ},\overline{d}}\triangleq&\bigg\lceil\frac{\overline{d}}{d^{\circ}}\bigg\rceil\bigg(k_t^{\circ}d^{\circ}\mu_{d^{\circ}}+\frac{(k_t^{\circ}-\gamma)(d^{\circ})^2\mu^2_{d^{\circ}} \lceil\frac{rd}{d^{\circ}}\rceil}{1-(d^{\circ}-1)\nu_{d^{\circ}}-(\lceil\frac{rd}{d^{\circ}}\rceil-1)d^{\circ}\mu_{d^{\circ}}}\bigg),\nonumber\\
		\delta_{\sigma_{\min}}\triangleq&(1-(d^*+d^{*\Delta}-1)\nu_{d^*+d^{*\Delta}}  -(k_t-\overline{\alpha}-1)(d^*+d^{*\Delta})\mu_{d^*+d^{*\Delta}})\nonumber\\
		&-\frac{(d^*+d^{*\Delta})^2\mu_{d^*+d^{*\Delta}}^2\lceil\frac{rd}{d^*+d^{*\Delta}} \rceil (k_t-\overline{\alpha})}{1-(d^*+d^{*\Delta}-1)\nu_{d^*+d^{*\Delta}}-(\lceil\frac{rd}{d^*+d^{*\Delta}}\rceil-1)(d^*+d^{*\Delta})\mu_{d^*+d^{*\Delta}}}.\label{somanyparameters}
	\end{align}
\end{figure*}
\end{theoremmaindescription6}

Notice that there exists a term $\frac{\|\mathbf{x}^l_{[\mathbf{\Xi}^*\backslash\mathbf{\Theta},\mathbf{\Theta}^{*\Delta}]}\|_{(d^*+d^{*\Delta})2,\infty}}{\|\mathbf{x}^l_{\mathbf{\Xi}^{\circ}\backslash\mathbf{\Theta}^{\circ}}\|_{(d^{\circ})2,\infty}}$ in the main condition of \textbf{Theorem \ref{theo6}}. This term indicates the power 
ratio of the support being identified in the current $i$th block to the support outside the current block. As we can see, when this term becomes larger, then the left-hand side of (\ref{theo6main}) becomes smaller, and the sufficient condition becomes less restrictive. This reveals that a larger power of the support in the current block leads to a better recovery performance. It is worth mentioning that if $\|\mathbf{x}^l_{\mathbf{\Xi}^{\circ}\backslash\mathbf{\Theta}^{\circ}}\|_{(d^{\circ})2,\infty}$ refers to the power of the non support outside the current block, then we have $\|\mathbf{x}^l_{\mathbf{\Xi}^{\circ}\backslash\mathbf{\Theta}^{\circ}}\|_{(d^{\circ})2,\infty}=0$. This indicates that all of the supports outside the current block have been selected, and we obtain $\overline{G}_{*}=\delta_{d^*+d^{*\Delta},\overline{d}}$ and $\overline{G}_{\circ}=0$. The sufficient condition in (\ref{theo6main}) thus becomes $\delta_{d^*+d^{*\Delta},\overline{d}}<1$, which is less restrictive than the condition when $\|\mathbf{x}^l_{\mathbf{\Xi}^{\circ}\backslash\mathbf{\Theta}^{\circ}}\|_{(d^{\circ})2,\infty}\neq0$. This result also provides insight that if there are no supports to be identified outside the current block in hierarchically block-sparse recovery, the recovery conditions improve.  

As for the cardinality of the input PSI index set, since $r=\alpha^{*}+\alpha^{\Delta}+\overline{\alpha}+\beta+\gamma$, we have the following observations: 1) $\frac{\partial (\overline{G}_{*}+\overline{G}_{\circ})}{\alpha^{*}}>0$; 2) $\frac{\partial (\overline{G}_{*}+\overline{G}_{\circ})}{\alpha^{\Delta}}>0$; 3) $\frac{\partial (\overline{G}_{*}+\overline{G}_{\circ})}{\beta}>0$. It worth mentioning that if the left-hand side of (\ref{theo6main}) becomes larger, then the sufficient condition becomes more restricted. Thus, these partial derivative results reveal that the PSI input corresponding to $\alpha^{*}$, $\alpha^{\Delta}$, and $\beta$ cause adverse effects to the recovery of HiBOMP-P. The reasons of $\alpha^{\Delta}$ and $\beta$ are evident, since the related PSI index sets do not overlap with the true support set, and cannot provide beneficial information for recovery. Conversely, as they wrongly change the residual projection in the current iteration, the mismatching between the residual and the measurement matrix blocks occur more frequently, which lead to significant performance degradation. The result related with $\alpha^{*}$ that the partial derivative is larger than 0 is unexpected, as the corresponding index set consists of support indices. This is result is also different from the existing studies \cite{Scarlett2013,Herzet2013,Ge2020}, wherein the results indicates that the PSI related to true support indices would significantly improve the performance of the algorithm. We now present certain analytic results to explain the abnormal phenomenon of $\frac{\partial (\overline{G}_{*}+\overline{G}_{\circ})}{\alpha^{*}}>0$. When PSI input represents subsets of the true support, it weakens the block structure within one hierarchical block. For instance, assume that the unit block length as $d$, and there are $c$ support blocks in each hierarchical block. Then, we say that the hierarchical block length is $cd$. When PSI related to $c^*$ $(c^*<c)$ block indices is entered, the correlation between these corresponding entries with the residual is equal to 0. In this case, only $c-c^*$ blocks could contribute to the support block selection. The hierarchical block structure has been decimated, causing performance deterioration in support block selection. 

In the following corollary, we further consider the hierarchical block orthogonality of the measurement matrix as illustrated in \emph{Remark \ref{rmk1212}}. In conventional block-sparse recovery, block orthogonality leads to $\nu=0$ of the measurement matrix, wherein a block orthogonal measurement matrix usually improves the performance of the recovery algorithm \cite{liyang2024}. Similarly, for hierarchically block-sparse recovery and given a measurement matrix $\mathbf{D}\in\mathbb{C}^{M\times N_0N_1\cdots N_nd}$, the hierarchical block orthogonality indicates that $\nu_{d^\xi}=0$ for the $t$th hierarchical mode, where $d^\xi=N_{t+1}N_{t+2}\cdots N_nd$. Then, the following corollary, stemming from \textbf{Theorem \ref{theo6}}, holds for hierarchical block orthogonal measurement matrix with $\nu_{d^*+d^{*\Delta}}=0$ and $\nu_{d^{\circ}}=0$. Note that $\nu_{d^{\xi}}=0$ could induce $\nu_{d^*+d^{*\Delta}}=0$ and $\nu_{d^{\circ}}=0$, as the matrix block with length $d^{\xi}$ contains the matrix blocks with lengths $d^*+d^{*\Delta}$ and $d^{\circ}$. Moreover, consider hierarchical block orthogonality of the hierarchically block-sparse recovery formulation, it is necessary to further assume that the PSI index set $\mathbf{\Theta}$ aligns with the block lengths $d^{*}+d^{*\Delta}$ and $d^{\circ}$, such that the partitioned blocks corresponding to the indices in $\mathbf{\Theta}$ with respect to these block lengths fall within each hierarchical block regularly. The mentioned corollary is presented as follows.

\begin{Corollary7}\label{coro7}
	Let $\mathbf{x} \in \mathbb{C}^{N_1N_2\cdots N_n d}$ be an $n$-mode hierarchically block-sparse vector, and let $\mathbf{y} = \mathbf{D}\mathbf{x}$, where $\mathbf{D} \in \mathbb{C}^{M \times N_1N_2\cdots N_n d}$ is a given matrix. Suppose that the hierarchical sub-coherence satisfies $\nu_{N_{t+1}N_{t+2}\cdots N_n d} = 0$, which implies that $\nu_{d^* + d^{\Delta}} = 0$ and $\nu_{d^{\circ}} = 0$, where $\nu_{d^ + d^{*\Delta}}$ and $\nu_{d^{\circ}}$ are defined in \textbf{Theorem 2}. Denote by $\mu_{d^{*}+d^{*\Delta}}$ and $\mu_{d^{\circ}}$ the hierarchical block coherence with respect to block length $d^{*}+d^{*\Delta}$ and $d^{\circ}$, respectively.
	Suppose that $(\lceil\frac{rd}{d^*+d^{*\Delta}}\rceil-1)(d^*+d^{*\Delta})\mu_{d^*+d^{*\Delta}}<1$ and  $(\lceil\frac{rd}{d^{\circ}}\rceil-1)d^{\circ}\mu_{d^{\circ}}<1$. For block selection in the $i$th hierarchical block from the $(t-1)$th hierarchical mode, a sufficient condition for (\ref{mainmain}) in \textbf{Theorem~\ref{theorem1}} is that
	\begin{align}
		\overline{G}_{*'}+\overline{G}_{\circ'}<1,\label{coro7MAIN}
	\end{align}
	where $\overline{G}_{*'}$ and $\overline{G}_{\circ'}$ are provided in (\ref{gggggxingxing})
	\begin{figure*}[hb]
		\rule[0pt]{18.2cm}{0.05em}
	\begin{align}
		\overline{G}_{*'}\triangleq&\frac{\delta^{'}_{d^*+d^{*\Delta},\overline{d}}}{1-\frac{(\delta^{'}_{d^{\circ},d^*+d^{*\Delta}}+\delta^{'}_{d^{\circ},d^{\Delta}})^{\frac{1}{2}}}{\delta^{'}_{\sigma_{\min}}}\times\frac{\|\mathbf{x}^l_{\mathbf{\Xi}^{\circ}\backslash\mathbf{\Theta}^{\circ}}\|_{(d^{\circ})2,\infty}}{\|\mathbf{x}^l_{[\mathbf{\Xi}^*\backslash\mathbf{\Theta},\mathbf{\Theta}^{*\Delta}]}\|_{(d^*+d^{*\Delta})2,\infty}}},\nonumber\\
		\overline{G}_{\circ'}\triangleq&\frac{\delta^{'}_{d^{\circ},\overline{d}}}{\delta^{'}_{\sigma_{\min}}\times\frac{\|\mathbf{x}^l_{[\mathbf{\Xi}^*\backslash\mathbf{\Theta},\mathbf{\Theta}^{*\Delta}]}\|_{(d^*+d^{*\Delta})2,\infty}}{\|\mathbf{x}^l_{\mathbf{\Xi}^{\circ}\backslash\mathbf{\Theta}^{\circ}}\|_{(d^{\circ})2,\infty}}-(\delta^{'}_{d^{\circ},d^*+d^{*\Delta}}+\delta^{'}_{d^{\circ},d^{\Delta}})^{\frac{1}{2}}}.\label{gggggxingxing}
	\end{align}
	\end{figure*}
		with $\delta^{'}_{d^*+d^{*\Delta},\overline{d}}$, $\delta^{'}_{d^{\circ},d^*+d^{*\Delta}}$, $\delta^{'}_{d^{\circ},d^{\Delta}}$, $\delta^{'}_{d^{\circ},\overline{d}}$, and  $\delta^{'}_{\sigma_{\min}}$ corresponding to their unprimed versions defined in \textbf{Theorem 2}, under the conditions $\nu_{d^* + d^{\Delta}} = 0$ and $\nu_{d^{\circ}} = 0$.
\end{Corollary7}

As defined by PSI, $\mathbf{\Theta}^{*\Delta}$ denotes the index set that does not overlap with the true support index set. In conventional PSI-assisted sparse recovery (e.g., \cite{Scarlett2013,Mota2017,Ge2020}), this type of prior information is considered interference, as it does not offer beneficial information for sparse recovery. Instead, it projects the current residual into the wrong subspace, causing a conflict between correct support information and this interference from the so-called incorrect PSI. Even with sufficient correct prior information, algorithms struggle to maintain their original recoveryability. In contrast, this type of PSI significantly enhances the performance of HiBOMP-P, as indicated by \textbf{Corollary \ref{coro7}}. In the following remark, we provide an intuitive explanation of this statement.

\begin{rmk11}\label{rmk11}
\emph{ Note that $\mathbf{\Theta}\cap\mathbf{\Theta}^{*\Delta}=\mathbf{\emptyset}$, where $\mathbf{\Theta}^{*\Delta}$ denotes the PSI corresponding to the additional support index set that does not overlap with the true support set. Assume that $\big\lceil\frac{\overline{d}}{d^*+d^{*\Delta}}\big\rceil=\frac{\overline{d}}{d^*+d^{*\Delta}}$ and  $r=|\mathbf{\Theta}|=0$; then the parameters in \textbf{Corollary~\ref{coro7}} change into
	\begin{align}
		\delta^{'}_{d^*+d^{*\Delta},\overline{d}}=&\frac{k_t\mu_{d^*+d^{*\Delta}}\overline{d}}{1  -(k_t-1)(d^*+d^{*\Delta})\mu_{d^*+d^{*\Delta}}},\nonumber\\
				\delta^{'}_{d^{\circ},d^*+d^{*\Delta}}=&\bigg\lceil\frac{d^{*}+d^{*\Delta}}{d^{\circ}}\bigg\rceil k^{\circ}_{t}d^{\circ}\mu_{d^{\circ}},\nonumber\\
				\delta^{'}_{d^{\circ},d^{\Delta}}=&\bigg\lceil\frac{d^{\Delta}}{d^{\circ}}\bigg\rceil k_t^{\circ}d^{\circ}\mu_{d^{\circ}},\nonumber\\
				\delta^{'}_{d^{\circ},\overline{d}}=&\bigg\lceil\frac{\overline{d}}{d^{\circ}}\bigg\rceil k_t^{\circ}d^{\circ}\mu_{d^{\circ}},\nonumber\\
		\delta^{'}_{\sigma_{\min}}=&1  -(k_t-1)(d^*+d^{*\Delta})\mu_{d^*+d^{*\Delta}},\nonumber
	\end{align}
It is evident that $\delta^{'}_{d^*+d^{*\Delta},\overline{d}}=\frac{k_t\mu_{d^*+d^{*\Delta}}\overline{d}}{\delta^{'}_{\sigma_{\min}}}$. Meanwhile, $\delta^{'}_{\sigma_{\min}}=1  -(k_t-1)(d^*+d^{*\Delta})\mu_{d^*+d^{*\Delta}}$ indicates that
\begin{align}
	k_t\mu_{d^*+d^{*\Delta}}\overline{d}=\bigg(\frac{1-\delta^{'}_{\sigma_{\min}}}{d^*+d^{*\Delta}}+\mu_{d^*+d^{*\Delta}}\bigg)\overline{d}.\nonumber
\end{align}
Thus, we have $\delta^{'}_{d^*+d^{*\Delta},\overline{d}}=\frac{\big(\frac{1-\delta^{'}_{\sigma_{\min}}}{d^*+d^{*\Delta}}+\mu_{d^*+d^{*\Delta}}\big)\overline{d}}{\delta^{'}_{\sigma_{\min}}}$.
Then, by direct simplification, the condition in (\ref{coro7MAIN}) is transformed into 
\begin{align}
	A(\delta^{'}_{\sigma_{\min}})^2+B\delta^{'}_{\sigma_{\min}}+C<0,\label{ercifangcheng}
\end{align}
where 
\begin{align}
	&A = -\frac{\|\mathbf{x}^l_{[\mathbf{\Xi}^*\backslash\mathbf{\Theta},\mathbf{\Theta}^{*\Delta}]}\|_{(d^*+d^{*\Delta})2,\infty}}{\|\mathbf{x}^l_{\mathbf{\Xi}^{\circ}\backslash\mathbf{\Theta}^{\circ}}\|_{(d^{\circ})2,\infty}}\bigg(\frac{\overline{d}}{d^*+d^{*\Delta}}+1\bigg),\nonumber\\
	&B=\frac{\overline{d}}{d^*+d^{*\Delta}}(\delta^{'}_{d^{\circ},d^*+d^{*\Delta}}+\delta^{'}_{d^{\circ},d^{\Delta}})^{\frac{1}{2}}\nonumber\\
	&\qquad+\bigg(\frac{\overline{d}}{d^*+d^{*\Delta}}+\overline{d}\mu_{d^*+d^{*\Delta}}\bigg)\frac{\|\mathbf{x}^l_{[\mathbf{\Xi}^*\backslash\mathbf{\Theta},\mathbf{\Theta}^{*\Delta}]}\|_{(d^*+d^{*\Delta})2,\infty}}{\|\mathbf{x}^l_{\mathbf{\Xi}^{\circ}\backslash\mathbf{\Theta}^{\circ}}\|_{(d^{\circ})2,\infty}}\nonumber\\
	&\qquad+\delta^{'}_{d^{\circ},\overline{d}}+2(\delta^{'}_{d^{\circ},d^*+d^{*\Delta}}+\delta^{'}_{d^{\circ},d^{\Delta}})^{\frac{1}{2}},\nonumber\\
	&C=-\bigg(\frac{\overline{d}}{d^*+d^{*\Delta}}+\overline{d}\mu_{d^*+d^{*\Delta}}\bigg)(\delta^{'}_{d^{\circ},d^*+d^{*\Delta}}+\delta^{'}_{d^{\circ},d^{\Delta}})^{\frac{1}{2}}\nonumber\\
	&\qquad-\delta^{'}_{d^{\circ},\overline{d}}(\delta^{'}_{d^{\circ},d^*+d^{*\Delta}}+\delta^{'}_{d^{\circ},d^{\Delta}})^{\frac{1}{2}}\nonumber\\
	&\qquad\quad\!\times\frac{\|\mathbf{x}^l_{\mathbf{\Xi}^{\circ}\backslash\mathbf{\Theta}^{\circ}}\|_{(d^{\circ})2,\infty}}{\|\mathbf{x}^l_{[\mathbf{\Xi}^*\backslash\mathbf{\Theta},\mathbf{\Theta}^{*\Delta}]}\|_{(d^*+d^{*\Delta})2,\infty}}\nonumber\\
	&\qquad-(\delta^{'}_{d^{\circ},d^*+d^{*\Delta}}+\delta^{'}_{d^{\circ},d^{\Delta}})\times\frac{\|\mathbf{x}^l_{\mathbf{\Xi}^{\circ}\backslash\mathbf{\Theta}^{\circ}}\|_{(d^{\circ})2,\infty}}{\|\mathbf{x}^l_{[\mathbf{\Xi}^*\backslash\mathbf{\Theta},\mathbf{\Theta}^{*\Delta}]}\|_{(d^*+d^{*\Delta})2,\infty}}\nonumber.
\end{align}
Note that the formulas for $\overline{G}_{*'}$ and $\overline{G}_{\circ'}$ imply that $\delta^{'}_{\sigma_{\min}}$ should satisfy the following conditions:
\begin{align}
	&\delta^{'}_{\sigma_{\min}}-(\delta^{'}_{d^{\circ},d^*+d^{*\Delta}}+\delta^{'}_{d^{\circ},d^{\Delta}})^{\frac{1}{2}}\nonumber\\
	&\qquad\quad\times\frac{\|\mathbf{x}^l_{\mathbf{\Xi}^{\circ}\backslash\mathbf{\Theta}^{\circ}}\|_{(d^{\circ})2,\infty}}{\|\mathbf{x}^l_{[\mathbf{\Xi}^*\backslash\mathbf{\Theta},\mathbf{\Theta}^{*\Delta}]}\|_{(d^*+d^{*\Delta})2,\infty}}>0,\nonumber\\
	&\delta^{'}_{\sigma_{\min}}\times\frac{\|\mathbf{x}^l_{[\mathbf{\Xi}^*\backslash\mathbf{\Theta},\mathbf{\Theta}^{*\Delta}]}\|_{(d^*+d^{*\Delta})2,\infty}}{\|\mathbf{x}^l_{\mathbf{\Xi}^{\circ}\backslash\mathbf{\Theta}^{\circ}}\|_{(d^{\circ})2,\infty}}\nonumber\\
	&\qquad\quad\qquad\!\!-(\delta^{'}_{d^{\circ},d^*+d^{*\Delta}}+\delta^{'}_{d^{\circ},d^{\Delta}})^{\frac{1}{2}}>0.\nonumber
\end{align}
Obtaining the solution of (\ref{ercifangcheng}) with respect to $\delta^{'}_{\sigma_{\min}}$ in real-domain, we have a lower bound of $\delta^{'}_{\sigma_{\min}}$ given by
\begin{align}
	\delta^{'}_{\sigma_{\min}}>\frac{-B-\sqrt{B^2-4AC}}{2A}\triangleq\underline{\delta}^{'}_{\sigma_{\min}}.\label{sigmamin}
\end{align} 
Then, combining (\ref{sigmamin}) and the definition of $\delta^{'}_{\sigma_{\min}}$ yields
\begin{align}
	k_t(d^*+d^{*\Delta})<d^*+d^{*\Delta}+\frac{1-\underline{\delta}^{'}_{\sigma_{\min}}}{\mu_{d^*+d^{*\Delta}}},\label{upperbound12}
\end{align}
which is a sufficient condition of the establishment of $\overline{G}_{*'}+\overline{G}_{\circ'}<1$ with $\big\lceil\frac{\overline{d}}{d^*+d^{*\Delta}}\big\rceil=\frac{\overline{d}}{d^*+d^{*\Delta}}$ and $r=|\mathbf{\Theta}|=0$.
As presented in Remark \ref{rmk8}, $\mu_{d^*+d^{*\Delta}}\leq\mu_B$, where $\mu_B$ is the conventional block coherence of the measurement matrix. Thus, the following upper bound, obtained by substituting the hierarchical block coherence with $\mu_B$, is sufficient for (\ref{upperbound12}):
\begin{align}
	k_t(d^*+d^{*\Delta})<d^*+d^{*\Delta}+\frac{1-\underline{\delta}^{'}_{\sigma_{\min}}}{\mu_{B}}\triangleq K^*.\nonumber
\end{align}
 Note that $\underline{\delta}^{'}_{\sigma_{\min}}\leq\delta^{'}_{\sigma_{\min}}\leq1$ is well-bounded, and $\mu_B$ is not affected by the variation in $d^*+d^{*\Delta}$. Thus, we could obtain that $\frac{\partial K^*}{\partial (d^*+d^{*\Delta})}>0$, suggesting that a higher count of indices in the PSI index set $\mathbf{\Theta}^{*\Delta}$ will lead to a greater reconstructible sparsity level denoted by $K^*$. 
 These analytical results confirm that HiBOMP-P will deliver improved recovery performance even if the input PSI index set does not overlap with the true support set. In the iterative procedures of HiBOMP-P, the residual is updated as $\mathbf{r}^l=\mathbf{r}^l+\mathbf{D}_{\mathbf{\Theta}^{*\Delta^t}}\mathbf{x}^l_{*\Delta^t}$, where the additional PSI index set $\mathbf{\Theta}^{*\Delta^t}$ contributes the correct block selection. Notably, in this selected block, the true support set is included. Consequently, when performing support selection within these blocks, HiBOMP-P exhibits higher probability of correctly identifying the true support set, thereby substantially improving the overall recovery performance. They further demonstrate that HCS exhibits strong resilience in terms of whether PSI is accurately located.
}
\end{rmk11}

It is worth mentioning that \textbf{Theorems \ref{theorem1}} and \textbf{\ref{theo6}} are recovery conditions applicable to HiBOMP-P for the $i$th hierarchical block from the $(t-1)$th hierarchical mode. To this end, combined results are illustrated in the following theorem for theoretical generality of the derived HiBOMP-P's ERCs.  

\begin{theoremmaindescription12}\label{theo12}
	Let $\mathbf{x}\in\mathbb{C}^{N_1N_2\cdots N_{n} d}$ be an $n$-mode hierarchically block-sparse vector, and let $\mathbf{y}=\mathbf{D}\mathbf{x}$ for a given matrix $\mathbf{D}\in\mathbb{C}^{M\times N_1N_2\cdots N_{n} d}$. Denote by $\mu_{d^{*}+d^{*\Delta}}$ and $\mu_{d^{\circ}}$ the hierarchical block coherence with respect to block length $d^{*}+d^{*\Delta}$ and $d^{\circ}$, respectively. Denote by $\nu_{d^{*}+d^{*\Delta}}$ and $\nu_{d^{\circ}}$ the hierarchical sub-coherence with respect to block length $d^{*}+d^{*\Delta}$ and $d^{\circ}$, respectively. Then, the ERC for HiBOMP-P is given by 
	\begin{align}
		{G}_{*}+{G}_{\circ}<1,\forall i, t,\nonumber
	\end{align}
where $G_{*}$ and $G_{\circ}$ are defined in \textbf{Theorem \ref{theorem1}}. A sufficient condition of this ERC is 
\begin{align}
	\overline{G}_{*}+\overline{G}_{\circ}<1,\forall i, t,\nonumber
\end{align}
where $\overline{G}_{*}$ and $\overline{G}_{\circ}$ are defined in \textbf{Theorem \ref{theo6}}.
\end{theoremmaindescription12}

\textbf{Theorem \ref{theo12}} contains the conditions for all support selection in recovering the hierarchically block-sparse signal $\mathbf{x}$ from  $\mathbf{y}$ with respect to the current support block index $i$ and the hierarchical mode $t$. It can be easily checked that if the most restricted condition among the conditions in \textbf{Theorem \ref{theo12}} is established, then the whole ERC holds. Since this condition is an aggregate of those in \textbf{Theorems \ref{theorem1}} and \textbf{\ref{theo6}}, the corresponding analyses given as above also apply to \textbf{Theorem \ref{theo12}}.

\section{Further Discussions}\label{reliablerecover}

In this section, we present several insights, including noisy recovery conditions and optimal hierarchical structure, based on the MIP-related conditions derived in \textbf{Section \ref{ercssss}}.

\subsection{Reliable Recovery Conditions Under Noisy Settings}\label{noisyrecoverycondition}

By considering the noise vector, the system model given in (\ref{system}) transforms into
\begin{align}\label{systemnoisy}
	\mathbf{y} = \mathbf{D}\mathbf{x} +\mathbf{n},
\end{align}
where $\mathbf{n}\in\mathbb{C}^M$ denotes the additive noise. 
In this subsection, noisy recovery conditions for reliably reconstructing $\mathbf{x}$ from the noisy measurements in (\ref{systemnoisy}) are provided. We consider the scenario where $\|\mathbf{n}\|_2\leq\epsilon$. As illustrated in Algorithm \ref{alg:HiBOXX}, the stopping rule is set as $\|\mathbf{r}^l\|_2\leq\epsilon$, which is a reasonable rule widely used in most iterative CS algorithms \cite{Kim2020,liyang2024}. By employing this stopping rule, the following theorem presents a result that reveals the guarantee based on which HiBOMP-P would select a correct block in the current iteration.

\begin{theoremmaindescription10}\label{theo10}
	Let $\mathbf{x}\in\mathbb{C}^{N_1N_2\cdots N_{n} d}$ be an $n$-mode hierarchically block-sparse vector, and let $\mathbf{y}=\mathbf{D}\mathbf{x}+\mathbf{n}$ for a given matrix $\mathbf{D}\in\mathbb{C}^{M\times N_1N_2\cdots N_{n} d}$. Suppose that $\|\mathbf{n}\|_2\leq\epsilon$, and $\mathbf{D}_{\mathbf{\Xi}\cup\mathbf{\Theta}}$ is full rank. Suppose that $(d^*+d^{*\Delta}-1)\nu_{d^*+d^{*\Delta}}+(\lceil\frac{rd}{d^*+d^{*\Delta}}\rceil-1)(d^*+d^{*\Delta})\mu_{d^*+d^{*\Delta}}<1$, $(d^*+d^{*\Delta}-1)\nu_{d^*+d^{*\Delta}}  +(k_t-\overline{\alpha}-1)(d^*+d^{*\Delta})\mu_{d^*+d^{*\Delta}}<1$, and $(d^{\circ}-1)\nu_{d^{\circ}}+(\lceil\frac{rd}{d^{\circ}}\rceil-1)d^{\circ}\mu_{d^{\circ}}<1$. A sufficient condition for HiBOMP-P to select a correct atom in the current iteration for the $t$th hierarchical mode is 
	\begin{align}
		&\delta_{\sigma_{\min}}\|\mathbf{x}^l_{[\mathbf{\Xi}^*\backslash\mathbf{\Theta},\mathbf{\Theta}^{*\Delta}]}\|_{(d^*+d^{*\Delta})2,\infty}\nonumber\\
		&-(\delta_{d^{\circ},d^*+d^{*\Delta}}+\delta_{d^{\circ},d^{\Delta}})^{\frac{1}{2}}\|\mathbf{x}^l_{\mathbf{\Xi}^{\circ}\backslash\mathbf{\Theta}^{\circ}}\|_{(d^*+d^{*\Delta}+d^{\Delta})2,\infty}\nonumber\\
		&\qquad\qquad\qquad\qquad\qquad\quad>\frac{2\|\ddot{\mathbf{A}}^{\rm H}_{\overline{\mathbf{\Xi}}_{<i>}\backslash\mathbf{\Theta}}\mathbf{n}\|_{(\overline{d})2,\infty}}{1-(\overline{G}_{*}+\overline{G}_{\circ})},\label{theo4main}
	\end{align}
where $\delta_{\sigma_{\min}}$, $\delta_{d^{\circ},d^*+d^{*\Delta}}$, $\delta_{d^{\circ},d^{\Delta}}$, $\overline{G}_{*}$, and $\overline{G}_{\circ}$ are given in \textbf{Theorem \ref{theo6}}.
\end{theoremmaindescription10}

\textbf{Theorem \ref{theo10}} indicates that if the mixed norm of the remaining support entries are large enough, then HiBOMP-P would select a correct support index in the current iteration. Based on \textbf{Theorem \ref{theo10}}, we provide the condition under which the HiBOMP-P algorithm selects all the correct support indices.

\begin{theoremmaindescription11}\label{theo11}
	Let $\mathbf{x}\in\mathbb{C}^{N_1N_2\cdots N_{n} d}$ be an $n$-mode hierarchically block-sparse vector, and let $\mathbf{y}=\mathbf{D}\mathbf{x}+\mathbf{n}$ for a given matrix $\mathbf{D}\in\mathbb{C}^{M\times N_1N_2\cdots N_{n} d}$ with $\|\mathbf{n}\|_2\leq\epsilon$. Suppose that $(d^*+d^{*\Delta}-1)\nu_{d^*+d^{*\Delta}}+(\lceil\frac{rd}{d^*+d^{*\Delta}}\rceil-1)(d^*+d^{*\Delta})\mu_{d^*+d^{*\Delta}}<1$, $(d^*+d^{*\Delta}-1)\nu_{d^*+d^{*\Delta}}  +(k_t-\overline{\alpha}-1)(d^*+d^{*\Delta})\mu_{d^*+d^{*\Delta}}<1$, and $(d^{\circ}-1)\nu_{d^{\circ}}+(\lceil\frac{rd}{d^{\circ}}\rceil-1)d^{\circ}\mu_{d^{\circ}}<1$. A sufficient condition for HiBOMP-P to select a correct atom in the current iteration for the $t$ th hierarchical mode is 
	\begin{align}
		&\delta_{\sigma_{\min}}\|\mathbf{x}^l_{[\mathbf{\Xi}^*\backslash\mathbf{\Theta},\mathbf{\Theta}^{*\Delta}]}\|_{(d^*+d^{*\Delta})2,\infty}\nonumber\\
		&-(\delta_{d^{\circ},d^*+d^{*\Delta}}+\delta_{d^{\circ},d^{\Delta}})^{\frac{1}{2}}\|\mathbf{x}^l_{\mathbf{\Xi}^{\circ}\backslash\mathbf{\Theta}^{\circ}}\|_{(d^*+d^{*\Delta}+d^{\Delta})2,\infty}\nonumber\\
		&\qquad\qquad\qquad\qquad\qquad\qquad>\frac{2\sqrt{\overline{d}}\epsilon}{1-(\overline{G}_{*}+\overline{G}_{\circ})},\label{theo4mainmain}
	\end{align}
where $\delta_{\sigma_{\min}}$, $\delta_{d^{\circ},d^*+d^{*\Delta}}$, $\delta_{d^{\circ},d^{\Delta}}$, $\overline{G}_{*}$, and $\overline{G}_{\circ}$ are given in \textbf{Theorem \ref{theo6}}.
\end{theoremmaindescription11}

Similarly, the following corollary offers a general version of \textbf{Theorem \ref{theo11}} that applies to arbitrary hierarchical modes, achieved through the combination of results from \textbf{Theorem \ref{theo11}} for each hierarchical mode and original hierarchical block.

\begin{Corollary8}\label{Corollary814}
	Let $\mathbf{x}\in\mathbb{C}^{N_1N_2\cdots N_{n} d}$ be an $n$-mode hierarchically block-sparse vector, and let $\mathbf{y}=\mathbf{D}\mathbf{x}+\mathbf{n}$ for a given matrix $\mathbf{D}\in\mathbb{C}^{M\times N_1N_2\cdots N_{n} d}$ with $\|\mathbf{n}\|_2\leq\epsilon$. Suppose that $(d^*+d^{*\Delta}-1)\nu_{d^*+d^{*\Delta}}+(\lceil\frac{rd}{d^*+d^{*\Delta}}\rceil-1)(d^*+d^{*\Delta})\mu_{d^*+d^{*\Delta}}<1$, $(d^*+d^{*\Delta}-1)\nu_{d^*+d^{*\Delta}}  +(k_t-\overline{\alpha}-1)(d^*+d^{*\Delta})\mu_{d^*+d^{*\Delta}}<1$, and $(d^{\circ}-1)\nu_{d^{\circ}}+(\lceil\frac{rd}{d^{\circ}}\rceil-1)d^{\circ}\mu_{d^{\circ}}<1$. A sufficient condition for HiBOMP-P to select a correct atom in the current iteration for the $t$ th hierarchical mode is 
	\begin{align}
		&\frac{\delta_{\sigma_{\min}}\|\mathbf{x}^l_{[\mathbf{\Xi}^*\backslash\mathbf{\Theta},\mathbf{\Theta}^{*\Delta}]}\|_{2}}{\sqrt{k_t}}-(\delta_{d^{\circ},d^*+d^{*\Delta}}+\delta_{d^{\circ},d^{\Delta}})^{\frac{1}{2}}\|\mathbf{x}^l_{\mathbf{\Xi}^{\circ}\backslash\mathbf{\Theta}^{\circ}}\|_{2}\nonumber\\
		&\qquad\qquad\qquad\qquad\qquad\qquad\qquad>\frac{2\sqrt{\overline{d}}\epsilon}{1-(\overline{G}_{*}+\overline{G}_{\circ})},\forall i, t,\nonumber
	\end{align}
	where $\delta_{\sigma_{\min}}$, $\delta_{d^{\circ},d^*+d^{*\Delta}}$, $\delta_{d^{\circ},d^{\Delta}}$, $\overline{G}_{*}$, and $\overline{G}_{\circ}$ are given in \textbf{Theorem \ref{theo6}}, and $k_t$ denotes the sparsity level in the $t$th hierarchical mode.
\end{Corollary8}

\textbf{Corollary \ref{Corollary814}} follows from that
\begin{align}
	\|\mathbf{x}^l_{[\mathbf{\Xi}^*\backslash\mathbf{\Theta},\mathbf{\Theta}^{*\Delta}]}\|_{(d^*+d^{*\Delta})2,\infty}\geq\frac{\|\mathbf{x}^l_{[\mathbf{\Xi}^*\backslash\mathbf{\Theta},\mathbf{\Theta}^{*\Delta}]}\|_{2}}{\sqrt{k_t}},\nonumber
\end{align}
 and
 \begin{align}
 	\|\mathbf{x}^l_{\mathbf{\Xi}^{\circ}\backslash\mathbf{\Theta}^{\circ}}\|_{(d^*+d^{*\Delta}+d^{\Delta})2,\infty}\leq\|\mathbf{x}^l_{\mathbf{\Xi}^{\circ}\backslash\mathbf{\Theta}^{\circ}}\|_{2}.\nonumber
 \end{align} 
It can be observed that when the norm of the outside support blocks is small, i.e., $\|\mathbf{x}^l_{\mathbf{\Xi}^{\circ}\backslash\mathbf{\Theta}^{\circ}}\|_{2}$ is small, the condition in \textbf{Corollary \ref{Corollary814}} becomes easier to satisfy. This implies that in hierarchically block-sparse recovery, the outside support blocks are regarded as interference, leading to a similar conclusion as the noiseless recovery condition. Interestingly, we notice that
\begin{align}
	&\lim_{\mu_{d^{\circ}}\to0}\delta_{d^{\circ},d^*+d^{*\Delta}}=0,\nonumber\\
	&\lim_{\mu_{d^{\circ}}\to0}\delta_{d^{\circ},d^{\Delta}}=0.\nonumber
\end{align}
These findings suggest that interference caused by the outside support blocks can be eliminated when the hierarchical block coherence of the measurement matrix is sufficiently low, as $\lim_{\mu_{d^{\circ}}\to0}(\delta_{d^{\circ},d^*+d^{*\Delta}}+\delta_{d^{\circ},d^{\Delta}})^{\frac{1}{2}}\|\mathbf{x}^l_{\mathbf{\Xi}^{\circ}\backslash\mathbf{\Theta}^{\circ}}\|_{2}=0$. Following these asymptotic analyses, we further have
\begin{align}
	\lim_{\mu_{d^{*}+d^{\Delta}},\nu_{d^{*}+d^{\Delta}},\mu_{d^{\circ}}\to0}\delta_{\sigma_{\min}}=1,\nonumber\\
	\lim_{\mu_{d^{*}+d^{\Delta}},\nu_{d^{*}+d^{\Delta}},\mu_{d^{\circ}}\to0}(\overline{G}_{*}+\overline{G}_{\circ})=0.\nonumber
\end{align}
Then, the condition in \textbf{Corollary \ref{Corollary814}} becomes that
\begin{align}
\|\mathbf{x}^l_{[\mathbf{\Xi}^*\backslash\mathbf{\Theta},\mathbf{\Theta}^{*\Delta}]}\|_{2}>2\sqrt{k_t\overline{d}}\epsilon,\forall i, t.\nonumber
\end{align}
This implies that if the norm of the support blocks is larger than a bound related to the hierarchical block sparsity, the block length of non support blocks, and the noise power, HiBOMP-P would reliably perform recovery in asymptotic cases. A special case is when 
$\epsilon$ is set to 0. In this case, $\|\mathbf{x}^l_{[\mathbf{\Xi}^*\backslash\mathbf{\Theta},\mathbf{\Theta}^{*\Delta}]}\|_{2}>0$ is sufficient for the exact recovery of HiBOMP-P.

\subsection{Optimal Hierarchical Structure}\label{secoptimalhier}

In hierarchically block-sparse recovery formulation, support blocks are partitioned into different individual blocks. In each iteration, the HiBOMP-P's reliably searching for support indices should not only consider the influence of non support blocks within the current hierarchical block, but should also consider the influence of non support blocks outside the current hierarchical block. This is because the residual vector contains the support components in both the current and the outside hierarchical blocks, which will cause interference to support selection in the current hierarchical block. Thus, in revealing the optimal hierarchical structure, the number of the outside hierarchical blocks should be zero.  Considering these factors, we present the optimal hierarchical structure in the following proposition.

\begin{proposition3}\label{prop3}
	Let $\mathbf{x}\in\mathbb{C}^{N_1N_2\cdots N_{n} d}$ be an $n$-mode hierarchically block-sparse vector with minimum unit block length $d$, and let $\mathbf{y}=\mathbf{D}\mathbf{x}+\mathbf{n}$ for a given matrix $\mathbf{D}\in\mathbb{C}^{M\times N_1N_2\cdots N_{n} d}$. If $k_t=1$ $(t\in\{\mathring{(n-1)}\})$, then the associated hierarchical structure is optimal for reliable recovery.
\end{proposition3}

\begin{figure*}[hb]
		\rule[0pt]{18.2cm}{0.05em}
	\begin{align}
		k_t(d^*+d^{*\Delta})<\frac{\mu_{d^*+d^{*\Delta}}^{-1}-(d^*+d^{*\Delta}-1)\frac{\nu_{d^*+d^{*\Delta}}}{\mu_{d^*+d^{*\Delta}}}+d^*+d^{*\Delta}}{(\lceil\frac{\overline{d}}{d^*+d^{*\Delta}}\rceil+1)(\frac{1-(d^*+d^{*\Delta}-1)\nu_{d^*+d^{*\Delta}}+(d^*+d^{*\Delta})\mu_{d^*+d^{*\Delta}}}{1-(d^*+d^{*\Delta}-1)\nu_{d^*+d^{*\Delta}}-(\lceil\frac{rd}{d^*+d^{*\Delta}}\rceil-1)(d^*+d^{*\Delta})\mu_{d^*+d^{*\Delta}}})}+\overline{\alpha}(d^*+d^{*\Delta}),\label{theo7mainmain}
	\end{align}
\end{figure*}

As we can see, the optimal hierarchical structure indicates that the hierarchical block sparsity is equal to 1 except for the last hierarchical mode. This is consistent with the intuition, also with the existing results \cite{greed2004}, that a lower sparsity level would induce a better recovery performance.
Note that \textbf{Proposition~\ref{prop3}} provides the optimal hierarchical structure based on the minimum unit block length as $d$. It is worth mentioning that a larger block length induces a stronger block structure characteristic, leading to improved recovery performance of block-sparse recovery. Then, it is natural to have that if the $c$ support blocks are continuous within hierarchical blocks, i.e., the minimum unit block length becomes $cd$ and $k_n=1$, hierarchically block-sparse recovery would offers more desirable performance. In what follows, we present recovery conditions related to the optimal hierarchical structure, especially including the ERC and a sufficient condition for its establishment.

\begin{theoremmaindescription13}\label{theo13}
	Let $\mathbf{x}\in\mathbb{C}^{N_1N_2\cdots N_{n} d}$ be an $n$-mode hierarchically block-sparse vector, and let $\mathbf{y}=\mathbf{D}\mathbf{x}$ for a given matrix $\mathbf{D}\in\mathbb{C}^{M\times N_1N_2\cdots N_{n} d}$. Suppose that $\mathbf{x}$ exhibits the optimal hierarchical structure in \textbf{Proposition \ref{prop3}}. For the $t$th $(t\in\{\mathring{n}\})$ hierarchical mode, suppose that $\mathbf{D}_{\mathbf{\Xi}_{<i>}\cup\mathbf{\Theta}}$ is full rank.  
	Then, the ERC for HiBOMP-P is given by 
	\begin{align}
		\rho_{c(d^*+d^{*\Delta},\overline{d})}(\ddot{\mathbf{A}}^{\dagger}_{[\mathbf{\Xi}^*\backslash\mathbf{\Theta},\mathbf{\Theta}^{*\Delta}]}\ddot{\mathbf{A}}_{\overline{\mathbf{\Xi}}_{<i>}\backslash\mathbf{\Theta}})<1,\forall i, t.\nonumber
	\end{align}
\end{theoremmaindescription13}

It can be observed that \textbf{Theorem \ref{theo13}} drops the dependence on the terms related to the support blocks outside the current block based on the optimal hierarchical structure in \textbf{Proposition \ref{prop3}}. In this case, the ERC of HiBOMP-P mainly depends on the resistance between the term related to the support block $\ddot{\mathbf{A}}^{\dagger}_{[\mathbf{\Xi}^*\backslash\mathbf{\Theta},\mathbf{\Theta}^{*\Delta}]}$ and the term related to the non support block $\ddot{\mathbf{A}}_{\overline{\mathbf{\Xi}}_{<i>}\backslash\mathbf{\Theta}}$ in the current hierarchical block. Similarly, the condition in \textbf{Theorem \ref{theo13}} is not useful as it relies on the specific positions of the support blocks. To address this issue, the following theorem provides the reconstructible sparsity, which serves as a sufficient condition of the ERC in \textbf{Theorem~\ref{theo13}}.

\begin{theoremmaindescription7}\label{theo7}
	Let $\mathbf{x}\in\mathbb{C}^{N_1N_2\cdots N_{n} d}$ be an $n$-mode hierarchically block-sparse vector, and let $\mathbf{y}=\mathbf{D}\mathbf{x}$ for a given matrix $\mathbf{D}\in\mathbb{C}^{M\times N_1N_2\cdots N_{n} d}$. Suppose that $\mathbf{x}$ exhibits the optimal hierarchical structure in \textbf{Proposition \ref{prop3}}, and $(d^*+d^{*\Delta}-1)\nu_{d^*+d^{*\Delta}}+(\lceil\frac{rd}{d^*+d^{*\Delta}}\rceil-1)(d^*+d^{*\Delta})\mu_{d^*+d^{*\Delta}}<1$. A sufficient condition for the establishment of the ERC in \textbf{Theorem \ref{theo13}} is given in (\ref{theo7mainmain}),
where $r=|\mathbf{\Theta}|=\alpha^*+\alpha^{\Delta}+\overline{\alpha}+\beta$.
\end{theoremmaindescription7}

The condition in \textbf{Theorem \ref{theo7}} is derived from \textbf{Theorem~\ref{theo6}}.
In deriving the aforementioned theorems, we make the assumption that the hierarchical structures in terms of various block lengths, such as $d^*$, $d^{\Delta}$ and $\overline{d}$, are known in advance. This is an extended assumption compared with that in the existing studies on block-sparse recovery, as they typically assume knowledge of the regular block length of the block-sparse signal \cite{Eldar2010,liyang2022}. Moreover, given an PSI index set, the block lengths are determined according to the definitions of these block lengths.

Based on \textbf{Theorem \ref{theo7}}, the following corollary holds by setting $\mathbf{\Theta}=\mathbf{\emptyset}$, which leads to that $\alpha^*=0$, $\alpha^{\Delta}=0$, $\overline{\alpha}=0$, and $\beta=0$, for HiBOMP, i.e., the algorithm that does not rely on the PSI. 

\begin{Corollary2}
	If the PSI $\mathbf{\Theta}=\mathbf{\emptyset}$, i.e., $r=0$, a sufficient condition for the establishment of (\ref{theo6main}) for the $t$th hierarchical mode is
	\begin{align}
		k_t(d^*+d^{\Delta})<\frac{\mu_{d^*+d^{\Delta}}^{-1}-((d^*+d^{\Delta})-1)\frac{\nu_{d^*+d^{\Delta}}}{\mu_{d^*+d^{\Delta}}}+d^*+d^{\Delta}}{\lceil\frac{\overline{d}}{d^*+d^{\Delta}}\rceil+1}.\label{coro3main}
	\end{align}
\end{Corollary2}

\begin{rmk}\label{rmkk6}
	\emph{Actually, (\ref{coro3main}) represents an upper bound on the reconstructible block sparsity in one support block of the $t$th $(t\in\{\mathring{n}\})$ hierarchical mode.  
		It can be transformed into the reconstructible block sparsity related to the true block sparsity of the $t$th hierarchical mode by multiplying $k_0k_1\cdots k_{t-1}$ on both sides, which is given by 
		\begin{align}
			kd^*<&k_0k_1\cdots k_{t-1}\nonumber\\
			&\times\frac{\mu_{d^*+d^{\Delta}}^{-1}-((d^*+d^{\Delta})-1)\frac{\nu_{d^*+d^{\Delta}}}{\mu_{d^*+d^{\Delta}}}+d^*+d^{\Delta}}{\lceil\frac{\overline{d}}{d^*+d^{\Delta}}\rceil+1},\label{truereconstructibleblocksparsity}
		\end{align}
		where $k=k_0k_1\cdots k_t$ denotes the true block sparsity of the $t$th hierarchical mode.
	As for the true block sparsity, the existing result with the sub-coherence being equal to 0 as derived in \cite{Eldar2010} is given by
		\begin{align}
			kd<\frac{1}{2}\bigg(\frac{1}{\mu_B}+d\bigg)\triangleq \overline{K},\nonumber
		\end{align}
		which applies to conventional block-sparse recovery. Meanwhile, considering that the measurement matrix satisfies hierarchical block orthogonality, the true reconstructible block sparsity in (\ref{truereconstructibleblocksparsity}) with $\nu_{d^*+d^{\Delta}}=0$ is given by
		\begin{align}
			k(d^*+d^{\Delta})<k_0k_1\cdots k_{t-1}
			\times\frac{\mu_{d^*+d^{\Delta}}^{-1}+d^*+d^{\Delta}}{\lceil\frac{\overline{d}}{d^*+d^{\Delta}}\rceil+1} \triangleq \overline{K}^*.\nonumber
		\end{align}
	As illustrated in Remark \ref{rmk11}, $k_0k_1\cdots k_{t-1}\geq1$ and $\mu_{d^*+d^{\Delta}}\leq\mu_B$, which indicates that $\overline{K}^*\geq\overline{K}$ when $\lceil\frac{\overline{d}}{d^*+d^{\Delta}}\rceil\leq1$. This demonstrates that the optimal hierarchical structure provides superior recovery performance for hierarchically block-sparse recovery compared to the conventional approach, owing to its hierarchical structure. This result aligns with the intuition that as the non support component becomes stronger (i.e., when $\overline{d}$ increases), the recovery performance will decrease. Conversely, the performance improves when the support component becomes stronger (i.e., when $d^{*}+d^{*\Delta}$ increases). 
	}
\end{rmk}

\begin{rmk6}\label{rmk66}
	\emph{This remark develops some relations between the true reconstructible block sparsity bound and the compression rate $\frac{M}{N}$. As reported in \cite{Strohmer2003}, the conventional matrix coherence defined in \textbf{Definition \ref{defconventionalmip}} can be lower bounded by $\mu\geq\sqrt{\frac{N-M}{M(N-1)}}$. Since $\nu_{d^*+d^{\Delta}}\leq\mu$, we have
	\begin{align}
		\nu_{d^*+d^{\Delta}}\leq\sqrt{\frac{N-M}{M(N-1)}},\nonumber
	\end{align}
when the conventional matrix coherence $\mu$ achieves this bound. In this case, the true reconstructible block sparsity given in (\ref{truereconstructibleblocksparsity}) changes into
\begin{align}
	kd^*&<k_0k_1\cdots k_{t-1}\nonumber\\
	&\quad\!\!\times\!\!\frac{\mu_{d^*+d^{\Delta}}^{-1}\Big(1-((d^*+d^{\Delta})-1)\sqrt{\frac{1-\omega}{M(1-\frac{1}{N})}}\Big)+d^*+d^{\Delta}}{\lceil\frac{\overline{d}}{d^*+d^{\Delta}}\rceil+1}\nonumber\\
	&\triangleq \overline{K}^*_{\circ},\nonumber
\end{align}
where $\omega=\frac{M}{N}$ denotes the compression rate. The following limit holds:
\begin{align}
	&\lim_{\frac{M}{N}=\omega,N\to\infty}\overline{K}^*_{\circ}=k_0k_1\cdots k_{t-1}\nonumber\\
	&\qquad\qquad\frac{\mu_{d^*+d^{\Delta}}^{-1}\Big(1-((d^*+d^{\Delta})-1)\sqrt{\frac{1-\omega}{M}}\Big)+d^*+d^{\Delta}}{\lceil\frac{\overline{d}}{d^*+d^{\Delta}}\rceil+1}.\nonumber
\end{align}
In this scenario, it can be observed that the true reconstructible block sparsity $\overline{K}^*_{\circ}$ improves with the increase in the compression ratio $\omega$. Additionally, as the number of measurements increases, the true reconstructible block sparsity improves correspondingly. These results indicate that HiBOMP-P, with a higher compression rate or a greater number of measurements, performs well in scenarios characterized by higher sparsity.}
\end{rmk6}

	Based on the optimal hierarchical structure derived in \textbf{Proposition \ref{prop3}}, we present an intuitive result for hierarchically block-sparse recovery with $d=1$ in the following remark. 
	\begin{rmk7} 
		\emph{Given that the PSI index set satisfies $|\mathbf{\Theta}|=\overline{\alpha}+\beta$, there are two types of PSI components corresponding to the overlap parameters $\overline{\alpha}$ and $\beta$, respectively. Therefore, we have $r=\overline{\alpha}+\beta$. Under this configuration, the condition in (5) becomes $\hat{G}_{*}<1$, where
	\begin{align}
		\hat{G}_{*}=&\bigg(\bigg\lceil\frac{\overline{d}}{d^*}\bigg\rceil(k_t-\overline{\alpha})d^*\mu_{d^*}\nonumber\\
		&\quad+\frac{(k_t-\overline{\alpha})(d^*)^2\mu^2_{d^*}\lceil\frac{\overline{d}}{d^*}\rceil \lceil\frac{rd}{d^*}\rceil}{1-(d^*-1)\nu_{d^*}-(\lceil\frac{rd}{d^*}\rceil-1)(d^*)\mu_{d^*}}\bigg)\nonumber\\
		&
		\times
		\bigg((1-(d^*-1)\nu_{d^*}  -(k_t-\overline{\alpha}-1)d^*\mu_{d^*})\nonumber\\
		&\qquad-\frac{(d^*)^2\mu_{d^*}^2\lceil\frac{rd}{d^*} \rceil (k_t-\overline{\alpha})}{1-(d^*-1)\nu_{d^*}-(\lceil\frac{rd}{d^*}\rceil-1)(d^*)\mu_{d^*}}\bigg)^{-1}.\nonumber
	\end{align}
	For the $n$th hierarchical mode, by setting $d=1$ for hierarchically sparse recovery, we obtain $\overline{d}=d^{*}=1$. Consequently, the aforementioned condition could be expressed as
	\begin{align}
		\frac{(k_n-\overline{\alpha})\Big(\mu_n
			+\frac{\mu^2_{n} (\overline{\alpha}+\beta)}{1-(\overline{\alpha}+\beta-1)\mu_{n}}\Big)}{(1  -(k_n-\overline{\alpha}-1)\mu_{n})
			-\frac{\mu_{n}^2(\overline{\alpha}+\beta) (k_n-\overline{\alpha})}{1-(\overline{\alpha}+\beta-1)\mu_{n}}}<1,\nonumber
	\end{align}
	where $\mu_{n}$ denotes the matrix coherence of the column-block submatrix of $\mathbf{D}\in\mathbb{C}^{M\times N_1N_2\cdots N_{n} }$ in the $n$th hierarchical mode (as defined in \textbf{Definition \ref{defconventionalmip}}), which can be further simplified as
	\begin{align}
		\frac{(k_n-\overline{\alpha})\mu_n}{1-(k_n+\beta-1)\mu_n}<1.\nonumber
	\end{align}
	This implies that if the matrix coherence for the $n$th hierarchical mode satisfies
	\begin{align}
		\mu_n < \frac{1}{2k_n - \overline{\alpha} + \beta - 1}, \label{conven}
	\end{align}
	 HiBOMP-P is capable of selecting the correct support set. It can be observed that the condition in (\ref{conven}) converges to the condition given in \cite[Eqn. (1)]{Herzet2013}, which has been proven to be tight. In other words, for the measurement matrix $\mathbf{D}\in\mathbb{C}^{M\times N_1N_2\cdots N_{n} }$, if the matrix coherence of its column-block submatrix $\mathbf{D}_{[i]}\in\mathbb{C}^{M \times N_{n} }$ satisfies
	\begin{align}
		\mu_n = \frac{1}{2k_n - \overline{\alpha} + \beta - 1}, \nonumber
	\end{align}
	 HiBOMP-P may select an incorrect atom in the first iteration for the $n$th hierarchical mode, resulting in a failure to exactly recover the underlying signal. This demonstrates that the condition in (\ref{conven}) is tight for HiBOMP-P in the $n$th hierarchical mode when $d = 1$.}
	\end{rmk7}

\begin{figure*}[hb]
	\rule[0pt]{18.2cm}{0.05em}
	\begin{align}
		&G_{*}(\mathbf{r}^{l})\triangleq\frac{\|\ddot{\mathbf{A}}^{\rm H}_{\overline{\mathbf{\Xi}}_{<i>}\backslash\mathbf{\Theta}}\overline{\mathbf{r}}^l\|_{(\overline{d})2,\infty}}{\|\ddot{\mathbf{A}}^{\rm H}_{\mathbf{\Xi}_{<i>}\backslash\mathbf{\Theta}}\overline{\mathbf{r}}^l\|_{(d^*+d^{*\Delta}+d^{\Delta})2,\infty}-\|\ddot{\mathbf{A}}^{\rm H}_{\mathbf{\Xi}_{<i>}\backslash\mathbf{\Theta}}\mathbf{r}^{\circ^l}\|_{(d^*+d^{*\Delta}+d^{\Delta})2,\infty}},\nonumber\\
		&G_{\circ}(\mathbf{r}^{l})\triangleq\frac{\|\ddot{\mathbf{A}}^{\rm H}_{\overline{\mathbf{\Xi}}_{<i>}\backslash\mathbf{\Theta}}\mathbf{r}^{\circ^l}\|_{(\overline{d})2,\infty}}{\|\ddot{\mathbf{A}}^{\rm H}_{\mathbf{\Xi}_{<i>}\backslash\mathbf{\Theta}}\overline{\mathbf{r}}^l\|_{(d^*+d^{*\Delta}+d^{\Delta})2,\infty}-\|\ddot{\mathbf{A}}^{\rm H}_{\mathbf{\Xi}_{<i>}\backslash\mathbf{\Theta}}\mathbf{r}^{\circ^l}\|_{(d^*+d^{*\Delta}+d^{\Delta})2,\infty}}.\label{GXINGXING1}
	\end{align}
\end{figure*}	

\begin{figure*}[hb]
		\rule[0pt]{18.2cm}{0.05em}
	\begin{align}
		G_{*}(\mathbf{r}^{l})
		&=\frac{\|\ddot{\mathbf{A}}^{\rm H}_{\overline{\mathbf{\Xi}}_{<i>}\backslash\mathbf{\Theta}}(\ddot{\mathbf{A}}^{\dagger}_{\mathbf{\Xi}^*\backslash\mathbf{\Theta}\cup\mathbf{\Theta}^{*\Delta}})^{\rm H}\ddot{\mathbf{A}}^{\rm H}_{\mathbf{\Xi}^*\backslash\mathbf{\Theta}\cup\mathbf{\Theta}^{*\Delta}}\overline{\mathbf{r}}^l\|_{(\overline{d})2,\infty}}{\|\ddot{\mathbf{A}}^{\rm H}_{\mathbf{\Xi}_{<i>}\backslash\mathbf{\Theta}}\overline{\mathbf{r}}^l\|_{(d^*+d^{*\Delta}+d^{\Delta})2,\infty}-\|\ddot{\mathbf{A}}^{\rm H}_{\mathbf{\Xi}_{<i>}\backslash\mathbf{\Theta}}\mathbf{r}^{\circ^l}\|_{(d^*+d^{*\Delta}+d^{\Delta})2,\infty}}\nonumber\\
		&=\frac{\|\ddot{\mathbf{A}}^{\rm H}_{\overline{\mathbf{\Xi}}_{<i>}\backslash\mathbf{\Theta}}(\ddot{\mathbf{A}}^{\dagger}_{[\mathbf{\Xi}^*\backslash\mathbf{\Theta},\mathbf{\Theta}^{*\Delta}]})^{\rm H}\ddot{\mathbf{A}}^{\rm H}_{[\mathbf{\Xi}^*\backslash\mathbf{\Theta},\mathbf{\Theta}^{*\Delta}]}\overline{\mathbf{r}}^l\|_{(\overline{d})2,\infty}}{\|\ddot{\mathbf{A}}^{\rm H}_{[\mathbf{\Xi}^*\backslash\mathbf{\Theta},\mathbf{\Theta}^{*\Delta},\mathbf{\Xi}^{\Delta}\backslash\mathbf{\Theta}]}\overline{\mathbf{r}}^l\|_{(d^*+d^{*\Delta}+d^{\Delta})2,\infty}-\|\ddot{\mathbf{A}}^{\rm H}_{[\mathbf{\Xi}^*\backslash\mathbf{\Theta},\mathbf{\Theta}^{*\Delta},\mathbf{\Xi}^{\Delta}\backslash\mathbf{\Theta}]}\mathbf{r}^{\circ^l}\|_{(d^*+d^{*\Delta}+d^{\Delta})2,\infty}}\nonumber\\
		&\leq\frac{\|\ddot{\mathbf{A}}^{\rm H}_{\overline{\mathbf{\Xi}}_{<i>}\backslash\mathbf{\Theta}}(\ddot{\mathbf{A}}^{\dagger}_{[\mathbf{\Xi}^*\backslash\mathbf{\Theta},\mathbf{\Theta}^{*\Delta}]})^{\rm H}\ddot{\mathbf{A}}^{\rm H}_{[\mathbf{\Xi}^*\backslash\mathbf{\Theta},\mathbf{\Theta}^{*\Delta}]}\overline{\mathbf{r}}^l\|_{(\overline{d})2,\infty}}{\|\ddot{\mathbf{A}}^{\rm H}_{[\mathbf{\Xi}^*\backslash\mathbf{\Theta},\mathbf{\Theta}^{*\Delta}]}\overline{\mathbf{r}}^l\|_{(d^*+d^{*\Delta})2,\infty}-\|\ddot{\mathbf{A}}^{\rm H}_{[\mathbf{\Xi}^*\backslash\mathbf{\Theta},\mathbf{\Theta}^{*\Delta},\mathbf{\Xi}^{\Delta}\backslash\mathbf{\Theta}]}\mathbf{r}^{\circ^l}\|_{(d^*+d^{*\Delta}+d^{\Delta})2,\infty}}\nonumber\\
		&\leq\frac{\rho_{r(\overline{d},d^*+d^{*\Delta})}(\ddot{\mathbf{A}}^{\rm H}_{\overline{\mathbf{\Xi}}_{<i>}\backslash\mathbf{\Theta}}(\ddot{\mathbf{A}}^{\dagger}_{[\mathbf{\Xi}^*\backslash\mathbf{\Theta},\mathbf{\Theta}^{*\Delta}]})^{\rm H})\|\ddot{\mathbf{A}}^{\rm H}_{[\mathbf{\Xi}^*\backslash\mathbf{\Theta},\mathbf{\Theta}^{*\Delta}]}\overline{\mathbf{r}}^l\|_{(d^*+d^{*\Delta})2,\infty}}{\|\ddot{\mathbf{A}}^{\rm H}_{[\mathbf{\Xi}^*\backslash\mathbf{\Theta},\mathbf{\Theta}^{*\Delta}]}\overline{\mathbf{r}}^l\|_{(d^*+d^{*\Delta})2,\infty}-\|\ddot{\mathbf{A}}^{\rm H}_{[\mathbf{\Xi}^*\backslash\mathbf{\Theta},\mathbf{\Theta}^{*\Delta},\mathbf{\Xi}^{\Delta}\backslash\mathbf{\Theta}]}\mathbf{r}^{\circ^l}\|_{(d^*+d^{*\Delta}+d^{\Delta})2,\infty}}\nonumber\\
		&=\frac{\rho_{c(d^*+d^{*\Delta},\overline{d})}(\ddot{\mathbf{A}}^{\dagger}_{[\mathbf{\Xi}^*\backslash\mathbf{\Theta},\mathbf{\Theta}^{*\Delta}]}\ddot{\mathbf{A}}_{\overline{\mathbf{\Xi}}_{<i>}\backslash\mathbf{\Theta}})}{1-\frac{\|\ddot{\mathbf{A}}^{\rm H}_{[\mathbf{\Xi}^*\backslash\mathbf{\Theta},\mathbf{\Theta}^{*\Delta},\mathbf{\Xi}^{\Delta}\backslash\mathbf{\Theta}]}\mathbf{r}^{\circ^l}\|_{(d^*+d^{*\Delta}+d^{\Delta})2,\infty}}{\|\ddot{\mathbf{A}}^{\rm H}_{[\mathbf{\Xi}^*\backslash\mathbf{\Theta},\mathbf{\Theta}^{*\Delta}]}\overline{\mathbf{r}}^l\|_{(d^{*\Delta})2,\infty}}}.\label{gg1}
	\end{align}
\end{figure*}

\section{Proofs of Theorems}\label{secproofstheorems}
\subsection{Proof of \textbf{Theorem 1}}\label{profoftheorem1}
	\begin{IEEEproof}
	As we can see from Algorithm \ref{alg:HiBOXX}, after the significant support block has been chosen in the first hierarchical mode, the HiBOMP-P algorithm will recursively search for this block until all the nonzero supports inside are selected with the help of the PSI. The residual is only updated once during this process, specifically at the time of support selection in the first hierarchical mode. For the $l$th iteration of the $t$th hierarchical mode, choosing a correct support block by HiBOMP-P is equivalent to requiring that
	\begin{align}
		G(\mathbf{r}^{l})&=\frac{\|\ddot{\mathbf{A}}^{\rm H}_{\overline{\mathbf{\Xi}}_{<i>}\backslash\mathbf{\Theta}}\mathbf{r}^{l}\|_{(\overline{d})2,\infty}}{\|\ddot{\mathbf{A}}^{\rm H}_{\mathbf{\Xi}_{<i>}\backslash\mathbf{\Theta}}\mathbf{r}^{l}\|_{(d^*+d^{*\Delta}+d^{\Delta})2,\infty}}\nonumber\\
		&=\frac{\|\ddot{\mathbf{A}}^{\rm H}_{\overline{\mathbf{\Xi}}_{<i>}\backslash\mathbf{\Theta}}(\mathbf{r}^{*^l}+\mathbf{D}_{\mathbf{\Theta}^{*\Delta}}\mathbf{x}^l_{*\Delta}+\mathbf{r}^{\circ^l})\|_{(\overline{d})2,\infty}}{\|\ddot{\mathbf{A}}^{\rm H}_{\mathbf{\Xi}_{<i>}\backslash\mathbf{\Theta}}(\mathbf{r}^{*^l}+\mathbf{D}_{\mathbf{\Theta}^{*\Delta}}\mathbf{x}^l_{*\Delta}+\mathbf{r}^{\circ^l})\|_{(d^*+d^{*\Delta}+d^{\Delta})2,\infty}}\nonumber\\
		&=\frac{\|\ddot{\mathbf{A}}^{\rm H}_{\overline{\mathbf{\Xi}}_{<i>}\backslash\mathbf{\Theta}}(\overline{\mathbf{r}}^l+\mathbf{r}^{\circ^l})\|_{(\overline{d})2,\infty}}{\|\ddot{\mathbf{A}}^{\rm H}_{\mathbf{\Xi}_{<i>}\backslash\mathbf{\Theta}}(\overline{\mathbf{r}}^l+\mathbf{r}^{\circ^l})\|_{(d^*+d^{*\Delta}+d^{\Delta})2,\infty}}\nonumber\\
		&\leq
		\Big(\|\ddot{\mathbf{A}}^{\rm H}_{\overline{\mathbf{\Xi}}_{<i>}\backslash\mathbf{\Theta}}\overline{\mathbf{r}}^l\|_{(\overline{d})2,\infty}+\|\ddot{\mathbf{A}}^{\rm H}_{\overline{\mathbf{\Xi}}_{<i>}\backslash\mathbf{\Theta}}\mathbf{r}^{\circ^l}\|_{(\overline{d})2,\infty}\Big)\nonumber\\
		&\quad\times\Big(\|\ddot{\mathbf{A}}^{\rm H}_{\mathbf{\Xi}_{<i>}\backslash\mathbf{\Theta}}\overline{\mathbf{r}}^l\|_{(d^*+d^{*\Delta}+d^{\Delta})2,\infty}\nonumber\\
		&\qquad\enspace-\|\ddot{\mathbf{A}}^{\rm H}_{\mathbf{\Xi}_{<i>}\backslash\mathbf{\Theta}}\mathbf{r}^{\circ^l}\|_{(d^*+d^{*\Delta}+d^{\Delta})2,\infty}\Big)^{-1}\nonumber\\
		&\triangleq G_{*}(\mathbf{r}^{l})+G_{\circ}(\mathbf{r}^{l})<1,\label{mainderiv}
	\end{align}
	where $G_{*}(\mathbf{r}^{l})$ and $G_{\circ}(\mathbf{r}^{l})$ are given in (\ref{GXINGXING1}),
	$\overline{\mathbf{r}}^l$ represents $\mathbf{r}^{*^l}+\mathbf{D}_{\mathbf{\Theta}^{*\Delta}}\mathbf{x}^l_{\mathbf{\Theta}^{*\Delta}}$, $\mathbf{r}^{\circ^l}$ denotes the residual vector composed of the supports that lie outside the current hierarchical block, $\overline{\mathbf{\Xi}}_{<i>}$ and $\mathbf{\Xi}_{<i>}$ represent the non support and support index sets of the $t$th hierarchical mode from the $i$th significant block in the last hierarchical mode, $\mathbf{\Xi}_{<i>}$ consists of $\mathbf{\Xi}^*$ and $\mathbf{\Xi}^{\Delta}$, $\mathbf{\Xi}^*$ denotes the true support index set, $\mathbf{\Xi}^{\Delta} = \mathbf{\Xi}_{<i>}\backslash\mathbf{\Xi}^*$, $\mathbf{\Theta}$ is the PSI set with $\mathbf{\Theta}=\mathbf{\Theta}^*\cup\mathbf{\Theta}^{\Delta}\cup\mathbf{\Theta}^{-}$, $\mathbf{\Theta}^{*\Delta}$ is the PSI set corresponding to the additional support block with $\mathbf{\Theta}^{*\Delta}\cap\mathbf{\Theta}=\mathbf{\emptyset}$, and $\mathbf{x}^l_{*\Delta}$ is the corresponding PSI vector. Since $|\overline{\mathbf{\Xi}}_{<i>}\cap\mathbf{\Theta}|=\beta +\overline{\beta}$, the column-wise block length of each block in $\ddot{\mathbf{A}}_{\overline{\mathbf{\Xi}}_{<i>}\cap\mathbf{\Theta}}=\frac{\beta (N_t-k_t)}{N_t-k_t}=\beta$. Note that the column-wise block length of each block in $\ddot{\mathbf{A}}_{\overline{\mathbf{\Xi}}_{<i>}}$ of the $t$th hierarchical mode is equal to $N_{t+1}N_{t+2}\cdots N_{n}d$. Thus, we obtain that the column-wise block length of $\ddot{\mathbf{A}}^{\rm H}_{\overline{\mathbf{\Xi}}_{<i>}\backslash\mathbf{\Theta}}$ is $N_{t+1}N_{t+2}\cdots N_{n}d-\beta d$, which leads to the same block length value of $\ddot{\mathbf{A}}^{\rm H}_{\overline{\mathbf{\Xi}}_{<i>}\backslash\mathbf{\Theta}}\mathbf{r}^l$. The block length value of $\ddot{\mathbf{A}}^{\rm H}_{\mathbf{\Xi}_{<i>}\backslash\mathbf{\Theta}}\mathbf{r}^l$ can be demonstrated similarly.
	In the following, we derive the upper bounds of $G_{*}(\mathbf{r}^{l})$ and $G_{\circ}(\mathbf{r}^{l})$, respectively.
	
	Since $\overline{\mathbf{r}}^l=\mathbf{r}^{*^l}+\mathbf{D}_{\mathbf{\Theta}^{*\Delta}}\mathbf{x}^l_{*\Delta}=\mathbf{P}^{\bot}_{\mathbf{D}_{\mathbf{\Theta}}}\mathbf{y}^*+\mathbf{D}_{\mathbf{\Theta}^{*\Delta}}\mathbf{x}^l_{\mathbf{\Theta}^{*\Delta}}$ and $\mathbf{y}^*\in\text{span}(\mathbf{D}_{\mathbf{\Xi}^*})$, $\mathbf{r}^{*^l}\in\text{span}(\mathbf{D}_{\mathbf{\Xi}^*\backslash\mathbf{\Theta}})=\text{span}(\ddot{\mathbf{A}}_{\mathbf{\Xi}^*\backslash\mathbf{\Theta}})$, where $\mathbf{\Xi}^*$ denotes the true support index set. Meanwhile, $\mathbf{D}_{\mathbf{\Theta}^{*\Delta}}\mathbf{x}^l_{\mathbf{\Theta}^{*\Delta}}\in\text{span}(\mathbf{D}_{\mathbf{\Theta}^{*\Delta}})$. Thus, we have  $\ddot{\mathbf{A}}_{\mathbf{\Xi}^*\backslash\mathbf{\Theta}\cup\mathbf{\Theta}^{*\Delta}}\ddot{\mathbf{A}}^{\dagger}_{\mathbf{\Xi}^*\backslash\mathbf{\Theta}\cup\mathbf{\Theta}^{*\Delta}}\overline{\mathbf{r}}^l=\overline{\mathbf{r}}^l$, where $\ddot{\mathbf{A}}_{\mathbf{\Xi}^*\backslash\mathbf{\Theta}\cup\mathbf{\Theta}^{*\Delta}}\ddot{\mathbf{A}}^{\dagger}_{\mathbf{\Xi}^*\backslash\mathbf{\Theta}\cup\mathbf{\Theta}^{*\Delta}}$ denotes the orthogonal projector onto $\text{span}(\ddot{\mathbf{A}}_{\mathbf{\Xi}^*\backslash\mathbf{\Theta}\cup\mathbf{\Theta}^{*\Delta}})$. As $\ddot{\mathbf{A}}_{\mathbf{\Xi}^*\backslash\mathbf{\Theta}\cup\mathbf{\Theta}^{*\Delta}}\ddot{\mathbf{A}}^{\dagger}_{\mathbf{\Xi}^*\backslash\mathbf{\Theta}\cup\mathbf{\Theta}^{*\Delta}}$ is Hermitian matrix, it indicates that
	\begin{align}
		(\ddot{\mathbf{A}}^{\dagger}_{\mathbf{\Xi}^*\backslash\mathbf{\Theta}\cup\mathbf{\Theta}^{*\Delta}})^{\rm H}\ddot{\mathbf{A}}^{\rm H}_{\mathbf{\Xi}^*\backslash\mathbf{\Theta}\cup\mathbf{\Theta}^{*\Delta}}\overline{\mathbf{r}}^l=\overline{\mathbf{r}}^l.\nonumber
	\end{align} 
 Hence, the relation in (\ref{gg1}) holds,
	where we denote $[\mathbf{\Xi}^*\backslash\mathbf{\Theta},\mathbf{\Theta}^{*\Delta},\mathbf{\Xi}^{\Delta}\backslash\mathbf{\Theta}]$ by
	\begin{align}
		&[(\mathbf{\Xi}^*\backslash\mathbf{\Theta})_{(1)},\mathbf{\Theta}^{*\Delta}_{(1)},(\mathbf{\Xi}^{\Delta}\backslash\mathbf{\Theta})_{(1)},(\mathbf{\Xi}^*\backslash\mathbf{\Theta})_{(2)},\mathbf{\Theta}^{*\Delta}_{(2)},(\mathbf{\Xi}^{\Delta}\backslash\mathbf{\Theta})_{(2)},\nonumber\\
		&\enspace\cdots,(\mathbf{\Xi}^*\backslash\mathbf{\Theta})_{(k_t)},\mathbf{\Theta}^{*\Delta}_{(k_t)},(\mathbf{\Xi}^{\Delta}\backslash\mathbf{\Theta})_{(k_t)}],\nonumber
	\end{align} 
 $\mathbf{\Xi}^*\backslash\mathbf{\Theta}$ by $[(\mathbf{\Xi}^*\backslash\mathbf{\Theta})_{(1)},(\mathbf{\Xi}^*\backslash\mathbf{\Theta})_{(2)},\cdots,(\mathbf{\Xi}^*\backslash\mathbf{\Theta})_{(k_t)}]$, $\mathbf{\Xi}^{\Delta}\backslash\mathbf{\Theta}$ by $[(\mathbf{\Xi}^{\Delta}\backslash\mathbf{\Theta})_{(1)},(\mathbf{\Xi}^{\Delta}\backslash\mathbf{\Theta})_{(2)},\cdots,(\mathbf{\Xi}^{\Delta}\backslash\mathbf{\Theta})_{(k_t)}]$, and the last inequality follows from \textbf{Corollary \ref{corollary1}}. 
	
	Note that, based on \textbf{Corollary {\ref{corollary1}}}, 
	\begin{align}
		&\|\ddot{\mathbf{A}}^{\rm H}_{[\mathbf{\Xi}^*\backslash\mathbf{\Theta},\mathbf{\Theta}^{*\Delta},\mathbf{\Xi}^{\Delta}\backslash\mathbf{\Theta}]}\mathbf{r}^{\circ^l}\|_{(d^*+d^{*\Delta}+d^{\Delta})2,\infty}\nonumber\\
		&\leq\big(\|\ddot{\mathbf{A}}^{\rm H}_{[\mathbf{\Xi}^*\backslash\mathbf{\Theta},\mathbf{\Theta}^{*\Delta}]}\mathbf{r}^{\circ^l}\|^2_{(d^*+d^{*\Delta})2,\infty}+\|\ddot{\mathbf{A}}^{\rm H}_{\mathbf{\Xi}^{\Delta}\backslash\mathbf{\Theta}}\mathbf{r}^{\circ^l}\|^2_{(d^{\Delta})2,\infty}\big)^{\frac{1}{2}}\nonumber\\
		&=\big(\|\ddot{\mathbf{A}}^{\rm H}_{[\mathbf{\Xi}^*\backslash\mathbf{\Theta},\mathbf{\Theta}^{*\Delta}]}\mathbf{P}^{\bot}_{\mathbf{D}_{\mathbf{\Theta}}}\mathbf{y}^{\circ}\|^2_{(d^*+d^{*\Delta})2,\infty}\nonumber\\
		&\quad\enspace+\|\ddot{\mathbf{A}}^{\rm H}_{\mathbf{\Xi}^{\Delta}\backslash\mathbf{\Theta}}\mathbf{P}^{\bot}_{\mathbf{D}_{\mathbf{\Theta}}}\mathbf{y}^{\circ}\|^2_{(d^{\Delta})2,\infty}\big)^{\frac{1}{2}}\nonumber\\
		&=\big(\|\ddot{\mathbf{A}}^{\rm H}_{[\mathbf{\Xi}^*\backslash\mathbf{\Theta},\mathbf{\Theta}^{*\Delta}]}\mathbf{P}^{\bot}_{\mathbf{D}_{\mathbf{\Theta}}}\mathbf{D}_{\mathbf{\Xi}^{\circ}\backslash\mathbf{\Theta}^{\circ}}\mathbf{x}^l_{\mathbf{\Xi}^{\circ}\backslash\mathbf{\Theta}^{\circ}}\|^2_{(d^*+d^{*\Delta})2,\infty}\nonumber\\
		&\quad\enspace+\|\ddot{\mathbf{A}}^{\rm H}_{\mathbf{\Xi}_{\Delta}\backslash\mathbf{\Theta}}\mathbf{P}^{\bot}_{\mathbf{D}_{\mathbf{\Theta}}}\mathbf{D}_{\mathbf{\Xi}^{\circ}\backslash\mathbf{\Theta}^{\circ}}\mathbf{x}^l_{\mathbf{\Xi}^{\circ}\backslash\mathbf{\Theta}^{\circ}}\|^2_{(d^{\Delta})2,\infty}\big)^{\frac{1}{2}}\nonumber\\
		&\leq\big(\rho_{r(d^*+d^{*\Delta},d^{\circ})}(\ddot{\mathbf{A}}^{\rm H}_{[\mathbf{\Xi}^*\backslash\mathbf{\Theta},\mathbf{\Theta}^{*\Delta}]}\ddot{\mathbf{A}}_{\mathbf{\Xi}^{\circ}\backslash\mathbf{\Theta}^{\circ}})\|\mathbf{x}^l_{\mathbf{\Xi}^{\circ}\backslash\mathbf{\Theta}^{\circ}}\|^2_{(d^{\circ})2,\infty}\nonumber\\
		&\quad\enspace+\rho_{r(d^{\Delta},d^{\circ})}(\ddot{\mathbf{A}}^{\rm H}_{\mathbf{\Xi}_{\Delta}\backslash\mathbf{\Theta}}\ddot{\mathbf{A}}_{\mathbf{\Xi}^{\circ}\backslash\mathbf{\Theta}^{\circ}})\|\mathbf{x}^l_{\mathbf{\Xi}^{\circ}\backslash\mathbf{\Theta}^{\circ}}\|^2_{(d^{\circ})2,\infty}\big)^{\frac{1}{2}}\nonumber\\
		&=\big(\rho_{c(d^{\circ},d^*+d^{*\Delta})}(\ddot{\mathbf{A}}^{\rm H}_{\mathbf{\Xi}^{\circ}\backslash\mathbf{\Theta}^{\circ}}\ddot{\mathbf{A}}_{[\mathbf{\Xi}^*\backslash\mathbf{\Theta},\mathbf{\Theta}^{*\Delta}]})\|\mathbf{x}^l_{\mathbf{\Xi}^{\circ}\backslash\mathbf{\Theta}^{\circ}}\|^2_{(d^{\circ})2,\infty}\nonumber\\
		&\quad\enspace+\rho_{c(d^{\circ},d^{\Delta})}(\ddot{\mathbf{A}}^{\rm H}_{\mathbf{\Xi}^{\circ}\backslash\mathbf{\Theta}^{\circ}}\ddot{\mathbf{A}}_{\mathbf{\Xi}_{\Delta}\backslash\mathbf{\Theta}})\|\mathbf{x}^l_{\mathbf{\Xi}^{\circ}\backslash\mathbf{\Theta}^{\circ}}\|^2_{(d^{\circ})2,\infty}\big)^{\frac{1}{2}},\label{upperbound1}
	\end{align}
	where the last inequality follows from $\ddot{\mathbf{A}}_{\mathbf{\Xi}^{\circ}\backslash\mathbf{\Theta}^{\circ}}=\mathbf{P}^{\bot}_{\mathbf{D}_{\mathbf{\Theta}}}\mathbf{D}_{\mathbf{\Xi}^{\circ}\backslash\mathbf{\Theta}^{\circ}}$. Meanwhile,
	\begin{align}
		&\|\ddot{\mathbf{A}}^{\rm H}_{[\mathbf{\Xi}^*\backslash\mathbf{\Theta},\mathbf{\Theta}^{*\Delta}]}\overline{\mathbf{r}}^l\|_{(d^*+d^{*\Delta})2,\infty}\nonumber\\
		&=\|\ddot{\mathbf{A}}^{\rm H}_{[\mathbf{\Xi}^*\backslash\mathbf{\Theta},\mathbf{\Theta}^{*\Delta}]}(\mathbf{r}^{*^l}+\mathbf{D}_{\mathbf{\Theta}^{*\Delta}}\mathbf{x}^l_{*\Delta})\|_{(d^*+d^{*\Delta})2,\infty}\nonumber\\
		&=\|\ddot{\mathbf{A}}^{\rm H}_{[\mathbf{\Xi}^*\backslash\mathbf{\Theta},\mathbf{\Theta}^{*\Delta}]}(\mathbf{P}^{\bot}_{\mathbf{D}_{\mathbf{\Theta}}}\mathbf{y}^*+\mathbf{D}_{\mathbf{\Theta}^{*\Delta}}\mathbf{x}^l_{\mathbf{\Theta}^{*\Delta}})\|_{(d^*+d^{*\Delta})2,\infty}\nonumber\\
		&=\|\ddot{\mathbf{A}}^{\rm H}_{[\mathbf{\Xi}^*\backslash\mathbf{\Theta},\mathbf{\Theta}^{*\Delta}]}(\mathbf{P}^{\bot}_{\mathbf{D}_{\mathbf{\Theta}}}\mathbf{y}^*+\mathbf{P}^{\bot}_{\mathbf{D}_{\mathbf{\Theta}}}\mathbf{D}_{\mathbf{\Theta}^{*\Delta}}\mathbf{x}^l_{\mathbf{\Theta}^{*\Delta}})\|_{(d^*+d^{*\Delta})2,\infty}\nonumber\\
		&=\|\ddot{\mathbf{A}}^{\rm H}_{[\mathbf{\Xi}^*\backslash\mathbf{\Theta},\mathbf{\Theta}^{*\Delta}]}\mathbf{P}^{\bot}_{\mathbf{D}_{\mathbf{\Theta}}}\mathbf{D}_{[\mathbf{\Xi}^*\backslash\mathbf{\Theta},\mathbf{\Theta}^{*\Delta}]}\mathbf{x}^l_{[\mathbf{\Xi}^*\backslash\mathbf{\Theta},\mathbf{\Theta}^{*\Delta}]}\|_{(d^*+d^{*\Delta})2,\infty}\nonumber\\
		&\geq\sigma_{\min}(\ddot{\mathbf{A}}^{\rm H}_{[\mathbf{\Xi}^*\backslash\mathbf{\Theta},\mathbf{\Theta}^{*\Delta}]}\ddot{\mathbf{A}}_{[\mathbf{\Xi}^*\backslash\mathbf{\Theta},\mathbf{\Theta}^{*\Delta}]})\nonumber\\
		&\quad\times\|\mathbf{x}^l_{[\mathbf{\Xi}^*\backslash\mathbf{\Theta},\mathbf{\Theta}^{*\Delta}]}\|_{(d^*+d^{*\Delta})2,\infty}.\label{upperbound2}
	\end{align}
Combining (\ref{upperbound1}) and (\ref{upperbound2}) yields 
\begin{align}
	G_{*}(\mathbf{r}^{l})\leq&\rho_{c(d^*+d^{*\Delta},\overline{d})}(\ddot{\mathbf{A}}^{\dagger}_{[\mathbf{\Xi}^*\backslash\mathbf{\Theta},\mathbf{\Theta}^{*\Delta}]}\ddot{\mathbf{A}}_{\overline{\mathbf{\Xi}}_{<i>}\backslash\mathbf{\Theta}})\nonumber\\
	&\times\bigg(1-\big(\rho_{c(d^{\circ},d^*+d^{*\Delta})}(\ddot{\mathbf{A}}^{\rm H}_{\mathbf{\Xi}^{\circ}\backslash\mathbf{\Theta}^{\circ}}\ddot{\mathbf{A}}_{[\mathbf{\Xi}^*\backslash\mathbf{\Theta},\mathbf{\Theta}^{*\Delta}]})\nonumber\\
	&\qquad\qquad+\rho_{c(d^{\circ},d^{\Delta})}(\ddot{\mathbf{A}}^{\rm H}_{\mathbf{\Xi}^{\circ}\backslash\mathbf{\Theta}^{\circ}}\ddot{\mathbf{A}}_{\mathbf{\Xi}^{\Delta}\backslash\mathbf{\Theta}})\big)^{\frac{1}{2}}\nonumber\\
	&\qquad\quad\enspace\times\big(\sigma_{\min}(\ddot{\mathbf{A}}^{\rm H}_{[\mathbf{\Xi}^*\backslash\mathbf{\Theta},\mathbf{\Theta}^{*\Delta}]}\ddot{\mathbf{A}}_{[\mathbf{\Xi}^*\backslash\mathbf{\Theta},\mathbf{\Theta}^{*\Delta}]})\big)^{-1}\nonumber\\
	&\qquad\quad\enspace\times\frac{\|\mathbf{x}^l_{\mathbf{\Xi}^{\circ}\backslash\mathbf{\Theta}^{\circ}}\|_{(d^{\circ})2,\infty}}{\|\mathbf{x}^l_{[\mathbf{\Xi}^*\backslash\mathbf{\Theta},\mathbf{\Theta}^{*\Delta}]}\|_{(d^*+d^{*\Delta})2,\infty}}\bigg)^{-1}\triangleq{	G}_{*}.\label{gxing1}
\end{align}

Now, we come to derive the upper bound of $G_{\circ}(\mathbf{r}^{l})$. The following inequality hods:
\begin{align}
	&\|\ddot{\mathbf{A}}^{\rm H}_{\overline{\mathbf{\Xi}}_{<i>}\backslash\mathbf{\Theta}}\mathbf{r}^{\circ^l}\|_{(\overline{d})2,\infty}\nonumber\\
	&=\|\ddot{\mathbf{A}}^{\rm H}_{\overline{\mathbf{\Xi}}_{<i>}\backslash\mathbf{\Theta}}\mathbf{P}^{\bot}_{\mathbf{D}_{\mathbf{\Theta}}}\mathbf{y}^{\circ}\|_{(\overline{d})2,\infty}\nonumber\\
	&=\|\ddot{\mathbf{A}}^{\rm H}_{\overline{\mathbf{\Xi}}_{<i>}\backslash\mathbf{\Theta}}\mathbf{P}^{\bot}_{\mathbf{D}_{\mathbf{\Theta}}}\mathbf{D}_{\mathbf{\Xi}^{\circ}\backslash\mathbf{\Theta}^{\circ}}\mathbf{x}^l_{\mathbf{\Xi}^{\circ}\backslash\mathbf{\Theta}^{\circ}}\|_{(\overline{d})2,\infty}\nonumber\\
	&\leq\rho_{r(\overline{d},d^{\circ})}(\ddot{\mathbf{A}}^{\rm H}_{\overline{\mathbf{\Xi}}_{<i>}\backslash\mathbf{\Theta}}\ddot{\mathbf{A}}_{\mathbf{\Xi}^{\circ}\backslash\mathbf{\Theta}^{\circ}})\|\mathbf{x}^l_{\mathbf{\Xi}^{\circ}\backslash\mathbf{\Theta}^{\circ}}\|_{(\overline{d})2,\infty}\nonumber\\
	&=\rho_{c(d^{\circ},\overline{d})}(\ddot{\mathbf{A}}^{\rm H}_{\mathbf{\Xi}^{\circ}\backslash\mathbf{\Theta}^{\circ}}\ddot{\mathbf{A}}_{\overline{\mathbf{\Xi}}_{<i>}\backslash\mathbf{\Theta}})\|\mathbf{x}^l_{\mathbf{\Xi}^{\circ}\backslash\mathbf{\Theta}^{\circ}}\|_{(\overline{d})2,\infty}.\nonumber
\end{align} 
Moreover, based on \textbf{Corollary \ref{corollary1}}, we have
\begin{align}
	&\|\ddot{\mathbf{A}}^{\rm H}_{\mathbf{\Xi}_{<i>}\backslash\mathbf{\Theta}}\overline{\mathbf{r}}^l\|_{(d^*+d^{*\Delta}+d^{\Delta})2,\infty}\nonumber\\
	&=\|\ddot{\mathbf{A}}^{\rm H}_{[\mathbf{\Xi}^*\backslash\mathbf{\Theta},\mathbf{\Theta}^{*\Delta},\mathbf{\Xi}^{\Delta}\backslash\mathbf{\Theta}]}\overline{\mathbf{r}}^l\|_{(d^*+d^{*\Delta}+d^{\Delta})2,\infty}\nonumber\\
	&\geq\|\ddot{\mathbf{A}}^{\rm H}_{[\mathbf{\Xi}^*\backslash\mathbf{\Theta},\mathbf{\Theta}^{*\Delta}]}\overline{\mathbf{r}}^l\|_{(d^*+d^{*\Delta})2,\infty}\nonumber\\
	&\geq\sigma_{\min}(\ddot{\mathbf{A}}^{\rm H}_{[\mathbf{\Xi}^*\backslash\mathbf{\Theta},\mathbf{\Theta}^{*\Delta}]}\ddot{\mathbf{A}}_{[\mathbf{\Xi}^*\backslash\mathbf{\Theta},\mathbf{\Theta}^{*\Delta}]})\nonumber\\
	&\quad\times\|\mathbf{x}^l_{[\mathbf{\Xi}^*\backslash\mathbf{\Theta},\mathbf{\Theta}^{*\Delta}]}\|_{(d^*+d^{*\Delta})2,\infty},\nonumber
\end{align}
where the first inequality is because the minimum value of $\|\ddot{\mathbf{A}}^{\rm H}_{\mathbf{\Xi}^{\Delta}\backslash\mathbf{\Theta}}\overline{\mathbf{r}}^l\|_{(d^*+d^{*\Delta})2,\infty}$ is equal to 0, and the last inequality follows from (\ref{upperbound2}).

Based on these bounds, we have
\begin{align}
	G_{\circ}(\mathbf{r}^{l})\leq&\rho_{c(d^{\circ},\overline{d})}(\ddot{\mathbf{A}}^{\rm H}_{\mathbf{\Xi}^{\circ}\backslash\mathbf{\Theta}^{\circ}}\ddot{\mathbf{A}}_{\overline{\mathbf{\Xi}}_{<i>}\backslash\mathbf{\Theta}})\nonumber\\
	&\times\bigg(\sigma_{\min}(\ddot{\mathbf{A}}^{\rm H}_{[\mathbf{\Xi}^*\backslash\mathbf{\Theta},\mathbf{\Theta}^{*\Delta}]}\ddot{\mathbf{A}}_{[\mathbf{\Xi}^*\backslash\mathbf{\Theta},\mathbf{\Theta}^{*\Delta}]})\nonumber\\
	&\qquad\times\frac{\|\mathbf{x}^l_{[\mathbf{\Xi}^*\backslash\mathbf{\Theta},\mathbf{\Theta}^{*\Delta}]}\|_{(d^*+d^{*\Delta})2,\infty}}{\|\mathbf{x}^l_{\mathbf{\Xi}^{\circ}\backslash\mathbf{\Theta}^{\circ}}\|_{(d^{\circ})2,\infty}}\nonumber\\
	&\qquad-\big(\rho_{c(d^{\circ},d^*+d^{*\Delta})}(\ddot{\mathbf{A}}^{\rm H}_{\mathbf{\Xi}^{\circ}\backslash\mathbf{\Theta}^{\circ}}\ddot{\mathbf{A}}_{[\mathbf{\Xi}^*\backslash\mathbf{\Theta},\mathbf{\Theta}^{*\Delta}]})\nonumber\\
	&\qquad+\rho_{c(d^{\circ},d^{\Delta})}(\ddot{\mathbf{A}}^{\rm H}_{\mathbf{\Xi}^{\circ}\backslash\mathbf{\Theta}^{\circ}}\ddot{\mathbf{A}}_{\mathbf{\Xi}_{\Delta}\backslash\mathbf{\Theta}})\big)^{\frac{1}{2}}\bigg)^{-1}\triangleq	{G}_{\circ}.\label{sigmalastlast}
\end{align}

Finally, by combining (\ref{mainderiv}), (\ref{gxing1}) and (\ref{sigmalastlast}), the proof is completed.
\end{IEEEproof}

\subsection{Proof of \textbf{Theorem \ref{theo6}}}
\begin{IEEEproof}
The proof proceeds by deriving the bounds of the unknown terms in $\overline{G}_{*}$ and $\overline{G}_{\circ}$, which are provided in the following content.

\textbf{\emph{Lower bound of $\sigma_{\min}(\ddot{\mathbf{A}}^{\rm H}_{[\mathbf{\Xi}^*\backslash\mathbf{\Theta},\mathbf{\Theta}^{*\Delta}]}\ddot{\mathbf{A}}_{[\mathbf{\Xi}^*\backslash\mathbf{\Theta},\mathbf{\Theta}^{*\Delta}]})$:}}

Based on \textbf{Lemma \ref{lemma1}}, we have the inequality in (\ref{deltadelta22}).
\begin{figure*}[hb]
	\rule[0pt]{18.2cm}{0.05em}
\begin{align}
	\sigma_{\min}(\ddot{\mathbf{A}}^{\rm H}_{[\mathbf{\Xi}^*\backslash\mathbf{\Theta},\mathbf{\Theta}^{*\Delta}]}\ddot{\mathbf{A}}_{[\mathbf{\Xi}^*\backslash\mathbf{\Theta},\mathbf{\Theta}^{*\Delta}]})
	\geq&1-(d^*+d^{*\Delta}-1)\nu_{d^*+d^{*\Delta}}  -(k_t-\overline{\alpha}-1)(d^*+d^{*\Delta})\mu_{d^*+d^{*\Delta}}\nonumber\\
	&-\frac{(d^*+d^{*\Delta})^2\mu_{d^*+d^{*\Delta}}^2\lceil\frac{rd}{d^*+d^{*\Delta}} \rceil (k_t-\overline{\alpha})}{1-(d^*+d^{*\Delta}-1)\nu_{d^*+d^{*\Delta}}-(\lceil\frac{rd}{d^*+d^{*\Delta}}\rceil-1)(d^*+d^{*\Delta})\mu_{d^*+d^{*\Delta}}}.\label{deltadelta22}
\end{align}
\end{figure*}

\emph{\textbf{Upper bound of $\rho_{c(d^*+d^{*\Delta},\overline{d})}(\ddot{\mathbf{A}}^{\dagger}_{[\mathbf{\Xi}^*\backslash\mathbf{\Theta},\mathbf{\Theta}^{*\Delta}]}\ddot{\mathbf{A}}_{\overline{\mathbf{\Xi}}_{<i>}\backslash\mathbf{\Theta}})$:}}

	Observe that	
	\begin{align}
		&\rho_{c(d^*+d^{*\Delta},\overline{d})}(\ddot{\mathbf{A}}^{\dagger}_{[\mathbf{\Xi}^*\backslash\mathbf{\Theta},\mathbf{\Theta}^{*\Delta}]}\ddot{\mathbf{A}}_{\overline{\mathbf{\Xi}}_{<i>}\backslash\mathbf{\Theta}})\nonumber\\
		&=\rho_{c(d^*+d^{*\Delta},\overline{d})}((\ddot{\mathbf{A}}_{[\mathbf{\Xi}^*\backslash\mathbf{\Theta},\mathbf{\Theta}^{*\Delta}]}^{\rm H}\ddot{\mathbf{A}}_{[\mathbf{\Xi}^*\backslash\mathbf{\Theta},\mathbf{\Theta}^{*\Delta}]})^{-1}\nonumber\\
		&\qquad\qquad\qquad\enspace\times\ddot{\mathbf{A}}^{\rm H}_{[\mathbf{\Xi}^*\backslash\mathbf{\Theta},\mathbf{\Theta}^{*\Delta}]}\ddot{\mathbf{A}}_{\overline{\mathbf{\Xi}}_{<i>}\backslash\mathbf{\Theta}})\nonumber\\
		&\leq\rho_{c(d^*+d^{*\Delta},d^*+d^{*\Delta})}((\ddot{\mathbf{A}}_{[\mathbf{\Xi}^*\backslash\mathbf{\Theta},\mathbf{\Theta}^{*\Delta}]}^{\rm H}\ddot{\mathbf{A}}_{[\mathbf{\Xi}^*\backslash\mathbf{\Theta},\mathbf{\Theta}^{*\Delta}]})^{-1})\nonumber\\
		&\quad\times\rho_{c(d^*+d^{*\Delta},\overline{d})}(\ddot{\mathbf{A}}^{\rm H}_{[\mathbf{\Xi}^*\backslash\mathbf{\Theta},\mathbf{\Theta}^{*\Delta}]}\ddot{\mathbf{A}}_{\overline{\mathbf{\Xi}}_{<i>}\backslash\mathbf{\Theta}})\nonumber\\
		&\leq\frac{1}{\sigma_{\min}(\ddot{\mathbf{A}}_{[\mathbf{\Xi}^*\backslash\mathbf{\Theta},\mathbf{\Theta}^{*\Delta}]}^{\rm H}\ddot{\mathbf{A}}_{[\mathbf{\Xi}^*\backslash\mathbf{\Theta},\mathbf{\Theta}^{*\Delta}]})}\nonumber\\
		&\quad\times\rho_{c(d^*+d^{*\Delta},\overline{d})}(\ddot{\mathbf{A}}^{\rm H}_{[\mathbf{\Xi}^*\backslash\mathbf{\Theta},\mathbf{\Theta}^{*\Delta}]}\ddot{\mathbf{A}}_{\overline{\mathbf{\Xi}}_{<i>}\backslash\mathbf{\Theta}})\nonumber\\
		&\leq
		\rho_{c(d^*+d^{*\Delta},\overline{d})}(\ddot{\mathbf{A}}^{\rm H}_{[\mathbf{\Xi}^*\backslash\mathbf{\Theta},\mathbf{\Theta}^{*\Delta}]}\ddot{\mathbf{A}}_{\overline{\mathbf{\Xi}}_{<i>}\backslash\mathbf{\Theta}})\nonumber\\
		&
		\quad\times
		\Bigg(1-(d^*+d^{*\Delta}-1)\nu_{d^*+d^{*\Delta}} \nonumber\\ &\qquad\enspace-(k_t-\overline{\alpha}-1)(d^*+d^{*\Delta})\mu_{d^*+d^{*\Delta}}\nonumber\\
		&-\bigg((d^*+d^{*\Delta})^2\mu_{d^*+d^{*\Delta}}^2\lceil\frac{rd}{d^*+d^{*\Delta}} \rceil (k_t-\overline{\alpha})\bigg)\nonumber\\
		&\times\Big(1-(d^*+d^{*\Delta}-1)\nu_{d^*+d^{*\Delta}}\nonumber\\
		&\qquad-(\lceil\frac{rd}{d^*+d^{*\Delta}}\rceil-1)(d^*+d^{*\Delta})\mu_{d^*+d^{*\Delta}}\Big)^{-1}\Bigg)^{-1},\nonumber
	\end{align}
	where the last inequality follows from \textbf{Lemma \ref{lemma1}}.
	
	It remains to derive the upper bound of $\rho_{c(d^*+d^{*\Delta},\overline{d})}(\ddot{\mathbf{A}}^{\rm H}_{[\mathbf{\Xi}^*\backslash\mathbf{\Theta},\mathbf{\Theta}^{*\Delta}]}\ddot{\mathbf{A}}_{\overline{\mathbf{\Xi}}_{<i>}\backslash\mathbf{\Theta}})$. Note that 
	\begin{align}
		&\rho_{c(d^*+d^{*\Delta},\overline{d})}(\ddot{\mathbf{A}}^{\rm H}_{[\mathbf{\Xi}^*\backslash\mathbf{\Theta},\mathbf{\Theta}^{*\Delta}]}\ddot{\mathbf{A}}_{\overline{\mathbf{\Xi}}_{<i>}\backslash\mathbf{\Theta}})\nonumber\\
		&=\rho_{c(d^*+d^{*\Delta},\overline{d})}(\mathbf{D}^{\rm H}_{[\mathbf{\Xi}^*\backslash\mathbf{\Theta},\mathbf{\Theta}^{*\Delta}]}\mathbf{P}^{\bot}_{\mathbf{D}_{\mathbf{\Theta}}}\mathbf{D}_{\overline{\mathbf{\Xi}}_{<i>}\backslash\mathbf{\Theta}})\nonumber\\
		&=\rho_{c(d^*+d^{*\Delta},\overline{d})}(\mathbf{D}^{\rm H}_{[\mathbf{\Xi}^*\backslash\mathbf{\Theta},\mathbf{\Theta}^{*\Delta}]}(\mathbf{I}-\mathbf{P}_{\mathbf{D}_{\mathbf{\Theta}}})\mathbf{D}_{\overline{\mathbf{\Xi}}_{<i>}\backslash\mathbf{\Theta}})\nonumber\\
		&\leq\rho_{c(d^*+d^{*\Delta},\overline{d})}(\mathbf{D}^{\rm H}_{[\mathbf{\Xi}^*\backslash\mathbf{\Theta},\mathbf{\Theta}^{*\Delta}]}\mathbf{D}_{\overline{\mathbf{\Xi}}_{<i>}\backslash\mathbf{\Theta}})\nonumber\\
		&\quad+\rho_{c(d^*+d^{*\Delta},\overline{d})}(\mathbf{D}^{\rm H}_{[\mathbf{\Xi}^*\backslash\mathbf{\Theta},\mathbf{\Theta}^{*\Delta}]}\mathbf{P}_{\mathbf{D}_{\mathbf{\Theta}}}\mathbf{D}_{\overline{\mathbf{\Xi}}_{<i>}\backslash\mathbf{\Theta}}).\nonumber
	\end{align}
	In what follows, we respectively provide the upper bounds of the terms $\rho_{c(d^*+d^{*\Delta},\overline{d})}(\mathbf{D}^{\rm H}_{[\mathbf{\Xi}^*\backslash\mathbf{\Theta},\mathbf{\Theta}^{*\Delta}]}\mathbf{D}_{\overline{\mathbf{\Xi}}_{<i>}\backslash\mathbf{\Theta}})$ and $\rho_{c(d^*+d^{*\Delta},\overline{d})}(\mathbf{D}^{\rm H}_{[\mathbf{\Xi}^*\backslash\mathbf{\Theta},\mathbf{\Theta}^{*\Delta}]}\mathbf{P}_{\mathbf{D}_{\mathbf{\Theta}}}\mathbf{D}_{\overline{\mathbf{\Xi}}_{<i>}\backslash\mathbf{\Theta}})$.
	
	Firstly, the following inequality holds:
	\begin{align}
		&\rho_{c(d^*+d^{*\Delta},\overline{d})}(\mathbf{D}^{\rm H}_{[\mathbf{\Xi}^*\backslash\mathbf{\Theta},\mathbf{\Theta}^{*\Delta}]}\mathbf{D}_{\overline{\mathbf{\Xi}}_{<i>}\backslash\mathbf{\Theta}})\nonumber\\
		&\leq\bigg\lceil\frac{\overline{d}}{d^*+d^{*\Delta}}\bigg\rceil\rho_{c(d^*+d^{*\Delta},\overline{d})}(\mathbf{D}^{\rm H}_{[\mathbf{\Xi}^*\backslash\mathbf{\Theta},\mathbf{\Theta}^{*\Delta}]}\mathbf{D}_{\overline{\mathbf{\Xi}}_{<i>}\backslash\mathbf{\Theta}})\nonumber\\
		&\leq\bigg\lceil\frac{\overline{d}}{d^*+d^{*\Delta}}\bigg\rceil(k_t-\overline{\alpha})(d^*+d^{*\Delta})\mu_{d^*+d^{*\Delta}}\nonumber.
	\end{align}
	
	Secondly, we have
	\begin{align}
		&\rho_{c(d^*+d^{*\Delta},\overline{d})}(\mathbf{D}^{\rm H}_{[\mathbf{\Xi}^*\backslash\mathbf{\Theta},\mathbf{\Theta}^{*\Delta}]}\mathbf{P}_{\mathbf{D}_{\mathbf{\Theta}}}\mathbf{D}_{\overline{\mathbf{\Xi}}_{<i>}\backslash\mathbf{\Theta}})\nonumber\\
		&=\rho_{c(d^*+d^{*\Delta},\overline{d})}(\mathbf{D}^{\rm H}_{[\mathbf{\Xi}^*\backslash\mathbf{\Theta},\mathbf{\Theta}^{*\Delta}]}\mathbf{D}_{\mathbf{\Theta}}\mathbf{D}^{\dagger}_{\mathbf{\Theta}}\mathbf{D}_{\overline{\mathbf{\Xi}}_{<i>}\backslash\mathbf{\Theta}})\nonumber\\
		&=\rho_{c(d^*+d^{*\Delta},\overline{d})}(\mathbf{D}^{\rm H}_{[\mathbf{\Xi}^*\backslash\mathbf{\Theta},\mathbf{\Theta}^{*\Delta}]}\mathbf{D}_{\mathbf{\Theta}}(\mathbf{D}^{\rm H}_{\mathbf{\Theta}}\mathbf{D}_{\mathbf{\Theta}})^{-1}\mathbf{D}^{\rm H}_{\mathbf{\Theta}}\mathbf{D}_{\overline{\mathbf{\Xi}}_{<i>}\backslash\mathbf{\Theta}})\nonumber\\
		&\leq\rho_{c(d^*+d^{*\Delta},d^*+d^{*\Delta})}(\mathbf{D}^{\rm H}_{[\mathbf{\Xi}^*\backslash\mathbf{\Theta},\mathbf{\Theta}^{*\Delta}]}\mathbf{D}_{\mathbf{\Theta}})\nonumber\\
		&\quad\times\rho_{c(d^*+d^{*\Delta},d^*+d^{*\Delta})}((\mathbf{D}^{\rm H}_{\mathbf{\Theta}}\mathbf{D}_{\mathbf{\Theta}})^{-1})\nonumber\\
		&\quad\times\rho_{c(d^*+d^{*\Delta},\overline{d})}(\mathbf{D}_{\mathbf{\Theta}}^{\rm H}\mathbf{D}_{\overline{\mathbf{\Xi}}_{<i>}\backslash\mathbf{\Theta}}).\label{threeterms}
	\end{align}
	Note that the three terms in (\ref{threeterms}) satisfy that
	\begin{align}
		&\rho_{c(d^*+d^{*\Delta},d^*+d^{*\Delta})}(\mathbf{D}^{\rm H}_{[\mathbf{\Xi}^*\backslash\mathbf{\Theta},\mathbf{\Theta}^{*\Delta}]}\mathbf{D}_{\mathbf{\Theta}})\nonumber\\
		&\leq (k_t-\overline{\alpha})(d^*+d^{*\Delta})\mu_{d^*+d^{*\Delta}},\nonumber\\
		&\rho_{c(d^*+d^{*\Delta},d^*+d^{*\Delta})}((\mathbf{D}^{\rm H}_{\mathbf{\Theta}}\mathbf{D}_{\mathbf{\Theta}})^{-1})\nonumber\\
		&\leq\frac{1}{\sigma_{\min}(\mathbf{D}^{\rm H}_{\mathbf{\Theta}}\mathbf{D}_{\mathbf{\Theta}})}\nonumber\\
		&\leq\bigg(1-(d^*+d^{*\Delta}-1)\nu_{d^*+d^{*\Delta}}\nonumber\\
		&\quad\quad-(\lceil\frac{rd}{d^*+d^{*\Delta}}\rceil-1)(d^*+d^{*\Delta})\mu_{d^*+d^{*\Delta}}\bigg)^{-1},\nonumber\\
		&\rho_{c(d^*+d^{*\Delta},\overline{d})}(\mathbf{D}_{\mathbf{\Theta}}^{\rm H}\mathbf{D}_{\overline{\mathbf{\Xi}}_{<i>}\backslash\mathbf{\Theta}})\nonumber\\
		&
		\leq\bigg\lceil\frac{\overline{d}}{d^*+d^{*\Delta}}\bigg\rceil\rho_{c(d^*+d^{*\Delta},d^*+d^{*\Delta})}(\mathbf{D}_{\mathbf{\Theta}}^{\rm H}\mathbf{D}_{\overline{\mathbf{\Xi}}_{<i>}\backslash\mathbf{\Theta}})\nonumber\\
		&\leq \bigg\lceil\frac{\overline{d}}{d^*+d^{*\Delta}}\bigg\rceil \bigg\lceil\frac{rd}{d^*+d^{*\Delta}}\bigg\rceil (d^*+d^{*\Delta})\mu_{d^*+d^{*\Delta}}.\nonumber
	\end{align}
	Thus, (\ref{threeterms}) becomes the inequality in (\ref{rhorho22}).
	\begin{figure*}[hb]
		\rule[0pt]{18.2cm}{0.05em}
	\begin{align}
		\rho_{c(d^*+d^{*\Delta},\overline{d})}(\mathbf{D}^{\rm H}_{[\mathbf{\Xi}^*\backslash\mathbf{\Theta},\mathbf{\Theta}^{*\Delta}]}\mathbf{P}_{\mathbf{D}_{\mathbf{\Theta}}}\mathbf{D}_{\overline{\mathbf{\Xi}}_{<i>}\backslash\mathbf{\Theta}})
		\leq\frac{(k_t-\overline{\alpha})(d^*+d^{*\Delta})^2\mu^2_{d^*+d^{*\Delta}}\lceil\frac{\overline{d}}{d^*+d^{*\Delta}}\rceil \lceil\frac{rd}{d^*+d^{*\Delta}}\rceil}{1-(d^*+d^{*\Delta}-1)\nu_{d^*+d^{*\Delta}}-(\lceil\frac{rd}{d^*+d^{*\Delta}}\rceil-1)(d^*+d^{*\Delta})\mu_{d^*+d^{*\Delta}}}.\label{rhorho22}
	\end{align}
	\end{figure*}
	The aforementioned derivation indicates  that the relation in (\ref{longlongequ}) holds.
\begin{figure*}[hb]
	\rule[0pt]{18.2cm}{0.05em}
	\begin{align}
		&\rho_{c(d^*+d^{*\Delta},\overline{d})}(\ddot{\mathbf{A}}^{\dagger}_{[\mathbf{\Xi}^*\backslash\mathbf{\Theta},\mathbf{\Theta}^{*\Delta}]}\ddot{\mathbf{A}}_{\overline{\mathbf{\Xi}}_{<i>}\backslash\mathbf{\Theta}})\nonumber\\
		&\leq\bigg(\bigg\lceil\frac{\overline{d}}{d^*+d^{*\Delta}}\bigg\rceil(k_t-\overline{\alpha})(d^*+d^{*\Delta})\mu_{d^*+d^{*\Delta}}
		+\frac{(k_t-\overline{\alpha})(d^*+d^{*\Delta})^2\mu^2_{d^*+d^{*\Delta}}\lceil\frac{\overline{d}}{d^*+d^{*\Delta}}\rceil \lceil\frac{rd}{d^*+d^{*\Delta}}\rceil}{1-(d^*+d^{*\Delta}-1)\nu_{d^*+d^{*\Delta}}-(\lceil\frac{rd}{d^*+d^{*\Delta}}\rceil-1)(d^*+d^{*\Delta})\mu_{d^*+d^{*\Delta}}}\bigg)\nonumber\\
		&\quad
		\times
		\bigg((1-(d^*+d^{*\Delta}-1)\nu_{d^*+d^{*\Delta}}  -(k_t-\overline{\alpha}-1)(d^*+d^{*\Delta})\mu_{d^*+d^{*\Delta}})\nonumber\\
		&\qquad\quad-\frac{(d^*+d^{*\Delta})^2\mu_{d^*+d^{*\Delta}}^2\lceil\frac{rd}{d^*+d^{*\Delta}} \rceil (k_t-\overline{\alpha})}{1-(d^*+d^{*\Delta}-1)\nu_{d^*+d^{*\Delta}}-(\lceil\frac{rd}{d^*+d^{*\Delta}}\rceil-1)(d^*+d^{*\Delta})\mu_{d^*+d^{*\Delta}}}\bigg)^{-1}.\label{longlongequ}
	\end{align}
	\end{figure*}
	
	\emph{\textbf{Upper bound of $\rho_{c(d^{\circ},d^*+d^{*\Delta})}(\ddot{\mathbf{A}}^{\rm H}_{\mathbf{\Xi}^{\circ}\backslash\mathbf{\Theta}^{\circ}}\ddot{\mathbf{A}}_{[\mathbf{\Xi}^*\backslash\mathbf{\Theta},\mathbf{\Theta}^{*\Delta}]})$:}}
	
	The following inequality holds:
	\begin{align}
		&\rho_{c(d^{\circ},d^*+d^{*\Delta})}(\ddot{\mathbf{A}}^{\rm H}_{\mathbf{\Xi}^{\circ}\backslash\mathbf{\Theta}^{\circ}}\ddot{\mathbf{A}}_{[\mathbf{\Xi}^*\backslash\mathbf{\Theta},\mathbf{\Theta}^{*\Delta}]})\nonumber\\
		&=\rho_{c(d^{\circ},d^*+d^{*\Delta})}(\mathbf{D}^{\rm H}_{\mathbf{\Xi}^{\circ}\backslash\mathbf{\Theta}^{\circ}}\mathbf{P}^{\bot}_{\mathbf{D}_{\mathbf{\Theta}}}\mathbf{D}_{[\mathbf{\Xi}^*\backslash\mathbf{\Theta},\mathbf{\Theta}^{*\Delta}]})\nonumber\\
		&=\rho_{c(d^{\circ},d^*+d^{*\Delta})}(\mathbf{D}^{\rm H}_{\mathbf{\Xi}^{\circ}\backslash\mathbf{\Theta}^{\circ}}(\mathbf{I}-\mathbf{P}_{\mathbf{D}_{\mathbf{\Theta}}})\mathbf{D}_{[\mathbf{\Xi}^*\backslash\mathbf{\Theta},\mathbf{\Theta}^{*\Delta}]})\nonumber\\
		&\leq\rho_{c(d^{\circ},d^*+d^{*\Delta})}(\mathbf{D}^{\rm H}_{\mathbf{\Xi}^{\circ}\backslash\mathbf{\Theta}^{\circ}}\mathbf{D}_{[\mathbf{\Xi}^*\backslash\mathbf{\Theta},\mathbf{\Theta}^{*\Delta}]})\nonumber\\
		&\quad+\rho_{c(d^{\circ},d^*+d^{*\Delta})}(\mathbf{D}^{\rm H}_{\mathbf{\Xi}^{\circ}\backslash\mathbf{\Theta}^{\circ}}\mathbf{P}_{\mathbf{D}_{\mathbf{\Theta}}}\mathbf{D}_{[\mathbf{\Xi}^*\backslash\mathbf{\Theta},\mathbf{\Theta}^{*\Delta}]})\nonumber\\
		&\leq\bigg\lceil\frac{d^{*}+d^{*\Delta}}{d^{\circ}}\bigg\rceil\rho_{c(d^{\circ},d^{\circ})}(\mathbf{D}^{\rm H}_{\mathbf{\Xi}^{\circ}\backslash\mathbf{\Theta}^{\circ}}\mathbf{D}_{[\mathbf{\Xi}^*\backslash\mathbf{\Theta},\mathbf{\Theta}^{*\Delta}]})\nonumber\\
		&\quad+\bigg\lceil\frac{d^{*}+d^{*\Delta}}{d^{\circ}}\bigg\rceil\rho_{c(d^{\circ},d^{\circ})}(\mathbf{D}^{\rm H}_{\mathbf{\Xi}^{\circ}\backslash\mathbf{\Theta}^{\circ}}\mathbf{P}_{\mathbf{D}_{\mathbf{\Theta}}}\mathbf{D}_{[\mathbf{\Xi}^*\backslash\mathbf{\Theta},\mathbf{\Theta}^{*\Delta}]}).\nonumber
	\end{align}
Note that
\begin{align}
	&\rho_{c(d^{\circ},d^{\circ})}(\mathbf{D}^{\rm H}_{\mathbf{\Xi}^{\circ}\backslash\mathbf{\Theta}^{\circ}}\mathbf{D}_{[\mathbf{\Xi}^*\backslash\mathbf{\Theta},\mathbf{\Theta}^{*\Delta}]})\leq k^{\circ}_{t}d^{\circ}\mu_{d^{\circ}},\nonumber\\
	&\rho_{c(d^{\circ},d^{\circ})}(\mathbf{D}^{\rm H}_{\mathbf{\Xi}^{\circ}\backslash\mathbf{\Theta}^{\circ}}\mathbf{P}_{\mathbf{D}_{\mathbf{\Theta}}}\mathbf{D}_{[\mathbf{\Xi}^*\backslash\mathbf{\Theta},\mathbf{\Theta}^{*\Delta}]})\nonumber\\
	&=\rho_{c(d^{\circ},d^{\circ})}(\mathbf{D}^{\rm H}_{\mathbf{\Xi}^{\circ}\backslash\mathbf{\Theta}^{\circ}}\mathbf{D}_{\mathbf{\Theta}}(\mathbf{D}^{\rm H}_{\mathbf{\Theta}}\mathbf{D}_{\mathbf{\Theta}})^{-1}\mathbf{D}^{\rm H}_{\mathbf{\Theta}}\mathbf{D}_{[\mathbf{\Xi}^*\backslash\mathbf{\Theta},\mathbf{\Theta}^{*\Delta}]})\nonumber\\
	&\leq\rho_{c(d^{\circ},d^{\circ})}(\mathbf{D}^{\rm H}_{\mathbf{\Xi}^{\circ}\backslash\mathbf{\Theta}^{\circ}}\mathbf{D}_{\mathbf{\Theta}})\rho_{c(d^{\circ},d^{\circ})}((\mathbf{D}^{\rm H}_{\mathbf{\Theta}}\mathbf{D}_{\mathbf{\Theta}})^{-1})\nonumber\\
	&\quad\times\rho_{c(d^{\circ},d^{\circ})}(\mathbf{D}^{\rm H}_{\mathbf{\Theta}}\mathbf{D}_{[\mathbf{\Xi}^*\backslash\mathbf{\Theta},\mathbf{\Theta}^{*\Delta}]})\nonumber\\
	&\leq\frac{(k_t^{\circ}-\gamma)(d^{\circ})^2\mu^2_{d^{\circ}} \lceil\frac{rd}{d^{\circ}}\rceil}{1-(d^{\circ}-1)\nu_{d^{\circ}}-(\lceil\frac{rd}{d^{\circ}}\rceil-1)d^{\circ}\mu_{d^{\circ}}},\nonumber
\end{align}
	where the upper bounds of $\rho_{c(d^{\circ},d^{\circ})}(\mathbf{D}^{\rm H}_{\mathbf{\Xi}^{\circ}\backslash\mathbf{\Theta}^{\circ}}\mathbf{D}_{\mathbf{\Theta}})$, $\rho_{c(d^{\circ},d^{\circ})}((\mathbf{D}^{\rm H}_{\mathbf{\Theta}}\mathbf{D}_{\mathbf{\Theta}})^{-1})$, and $\rho_{c(d^{\circ},d^{\circ})}(\mathbf{D}^{\rm H}_{\mathbf{\Theta}}\mathbf{D}_{[\mathbf{\Xi}^*\backslash\mathbf{\Theta},\mathbf{\Theta}^{*\Delta}]})$ can be derived similarly based on the aforementioned analysis. Thus, we obtain 
	\begin{align}
		&\rho_{c(d^{\circ},d^*+d^{*\Delta})}(\ddot{\mathbf{A}}^{\rm H}_{\mathbf{\Xi}^{\circ}\backslash\mathbf{\Theta}^{\circ}}\ddot{\mathbf{A}}_{[\mathbf{\Xi}^*\backslash\mathbf{\Theta},\mathbf{\Theta}^{*\Delta}]})\nonumber\\
		&\leq\bigg\lceil\frac{d^{*}+d^{*\Delta}}{d^{\circ}}\bigg\rceil\nonumber\\
		&\quad\times\bigg(k^{\circ}_{t}d^{\circ}\mu_{d^{\circ}}+\frac{(k_t^{\circ}-\gamma)(d^{\circ})^2\mu^2_{d^{\circ}} \lceil\frac{rd}{d^{\circ}}\rceil}{1-(d^{\circ}-1)\nu_{d^{\circ}}-(\lceil\frac{rd}{d^{\circ}}\rceil-1)d^{\circ}\mu_{d^{\circ}}}\bigg).\nonumber
	\end{align}
	
	\textbf{\emph{Upper bound of $\rho_{c(d^{\circ},d^{\Delta})}(\ddot{\mathbf{A}}^{\rm H}_{\mathbf{\Xi}^{\circ}\backslash\mathbf{\Theta}^{\circ}}\ddot{\mathbf{A}}_{\mathbf{\Xi}^{\Delta}\backslash\mathbf{\Theta}})$:}}
	
	The proof is similar to that of the upper bound of $\rho_{c(d^{\circ},d^*+d^{*\Delta})}(\ddot{\mathbf{A}}^{\rm H}_{\mathbf{\Xi}^{\circ}\backslash\mathbf{\Theta}^{\circ}}\ddot{\mathbf{A}}_{[\mathbf{\Xi}^*\backslash\mathbf{\Theta},\mathbf{\Theta}^{*\Delta}]})$.
	Observe that
	\begin{align}
		&\rho_{c(d^{\circ},d^{\Delta})}(\ddot{\mathbf{A}}^{\rm H}_{\mathbf{\Xi}^{\circ}\backslash\mathbf{\Theta}^{\circ}}\ddot{\mathbf{A}}_{\mathbf{\Xi}^{\Delta}\backslash\mathbf{\Theta}})\nonumber\\
		&=\rho_{c(d^{\circ},d^{\Delta})}(\mathbf{D}^{\rm H}_{\mathbf{\Xi}^{\circ}\backslash\mathbf{\Theta}^{\circ}}\mathbf{P}^{\bot}_{\mathbf{D}_{\mathbf{\Theta}}}\mathbf{D}_{\mathbf{\Xi}^{\Delta}\backslash\mathbf{\Theta}})\nonumber\\
		&=\rho_{c(d^{\circ},d^{\Delta})}(\mathbf{D}^{\rm H}_{\mathbf{\Xi}^{\circ}\backslash\mathbf{\Theta}^{\circ}}(\mathbf{I}-\mathbf{P}_{\mathbf{D}_{\mathbf{\Theta}}})\mathbf{D}_{\mathbf{\Xi}^{\Delta}\backslash\mathbf{\Theta}})\nonumber\\
		&\leq\rho_{c(d^{\circ},d^{\Delta})}(\mathbf{D}^{\rm H}_{\mathbf{\Xi}^{\circ}\backslash\mathbf{\Theta}^{\circ}}\mathbf{D}_{\mathbf{\Xi}^{\Delta}\backslash\mathbf{\Theta}})\nonumber\\
		&\quad+\rho_{c(d^{\circ},d^{\Delta})}(\mathbf{D}^{\rm H}_{\mathbf{\Xi}^{\circ}\backslash\mathbf{\Theta}^{\circ}}\mathbf{P}_{\mathbf{D}_{\mathbf{\Theta}}}\mathbf{D}_{\mathbf{\Xi}^{\Delta}\backslash\mathbf{\Theta}})\nonumber\\
		&\leq\bigg\lceil\frac{d^{\Delta}}{d^{\circ}}\bigg\rceil\rho_{c(d^{\circ},d^{\circ})}(\mathbf{D}^{\rm H}_{\mathbf{\Xi}^{\circ}\backslash\mathbf{\Theta}^{\circ}}\mathbf{D}_{\mathbf{\Xi}^{\Delta}\backslash\mathbf{\Theta}})\nonumber\\
		&\quad+\bigg\lceil\frac{d^{\Delta}}{d^{\circ}}\bigg\rceil\rho_{c(d^{\circ},d^{\circ})}(\mathbf{D}^{\rm H}_{\mathbf{\Xi}^{\circ}\backslash\mathbf{\Theta}^{\circ}}\mathbf{P}_{\mathbf{D}_{\mathbf{\Theta}}}\mathbf{D}_{\mathbf{\Xi}^{\Delta}\backslash\mathbf{\Theta}}).\nonumber
	\end{align}

Since
\begin{align}
	&\rho_{c(d^{\circ},d^{\circ})}(\mathbf{D}^{\rm H}_{\mathbf{\Xi}^{\circ}\backslash\mathbf{\Theta}^{\circ}}\mathbf{D}_{\mathbf{\Xi}^{\Delta}\backslash\mathbf{\Theta}})\leq k_t^{\circ}d^{\circ}\mu_{d^{\circ}},\nonumber\\
	&\rho_{c(d^{\circ},d^{\circ})}(\mathbf{D}^{\rm H}_{\mathbf{\Xi}^{\circ}\backslash\mathbf{\Theta}^{\circ}}\mathbf{P}_{\mathbf{D}_{\mathbf{\Theta}}}\mathbf{D}_{\mathbf{\Xi}^{\Delta}\backslash\mathbf{\Theta}})\nonumber\\
	&=\rho_{c(d^{\circ},d^{\circ})}(\mathbf{D}^{\rm H}_{\mathbf{\Xi}^{\circ}\backslash\mathbf{\Theta}^{\circ}}\mathbf{D}_{\mathbf{\Theta}}(\mathbf{D}^{\rm H}_{\mathbf{\Theta}}\mathbf{D}_{\mathbf{\Theta}})^{-1}\mathbf{D}^{\rm H}_{\mathbf{\Theta}}\mathbf{D}_{\mathbf{\Xi}^{\Delta}\backslash\mathbf{\Theta}})\nonumber\\
	&\leq\frac{(k_t^{\circ}-\gamma)(d^{\circ})^2\mu^2_{d^{\circ}} \lceil\frac{rd}{d^{\circ}}\rceil}{1-(d^{\circ}-1)\nu_{d^{\circ}}-(\lceil\frac{rd}{d^{\circ}}\rceil-1)d^{\circ}\mu_{d^{\circ}}},\nonumber
\end{align}
we have
\begin{align}
	&\rho_{c(d^{\circ},d^{\Delta})}(\ddot{\mathbf{A}}^{\rm H}_{\mathbf{\Xi}^{\circ}\backslash\mathbf{\Theta}^{\circ}}\ddot{\mathbf{A}}_{\mathbf{\Xi}^{\Delta}\backslash\mathbf{\Theta}})\nonumber\\
	&\leq\bigg\lceil\frac{d^{\Delta}}{d^{\circ}}\bigg\rceil\bigg(k_t^{\circ}d^{\circ}\mu_{d^{\circ}}+\frac{(k_t^{\circ}-\gamma)(d^{\circ})^2\mu^2_{d^{\circ}} \lceil\frac{rd}{d^{\circ}}\rceil}{1-(d^{\circ}-1)\nu_{d^{\circ}}-(\lceil\frac{rd}{d^{\circ}}\rceil-1)d^{\circ}\mu_{d^{\circ}}}\bigg).\nonumber
\end{align}

\textbf{\emph{Upper bound of $\rho_{c(d^{\circ},\overline{d})}(\ddot{\mathbf{A}}^{\rm H}_{\mathbf{\Xi}^{\circ}\backslash\mathbf{\Theta}^{\circ}}\ddot{\mathbf{A}}_{\overline{\mathbf{\Xi}}_{<i>}\backslash\mathbf{\Theta}})$:}}

The following inequalities hold:
\begin{align}
	&\rho_{c(d^{\circ},\overline{d})}(\ddot{\mathbf{A}}^{\rm H}_{\mathbf{\Xi}^{\circ}\backslash\mathbf{\Theta}^{\circ}}\ddot{\mathbf{A}}_{\overline{\mathbf{\Xi}}_{<i>}\backslash\mathbf{\Theta}})\nonumber\\
	&=\rho_{c(d^{\circ},\overline{d})}(\mathbf{D}^{\rm H}_{\mathbf{\Xi}^{\circ}\backslash\mathbf{\Theta}^{\circ}}\mathbf{P}^{\bot}_{\mathbf{D}_{\mathbf{\Theta}}}\mathbf{D}_{\overline{\mathbf{\Xi}}_{<i>}\backslash\mathbf{\Theta}})\nonumber\\
	&=\rho_{c(d^{\circ},d^{\Delta})}(\mathbf{D}^{\rm H}_{\mathbf{\Xi}^{\circ}\backslash\mathbf{\Theta}^{\circ}}(\mathbf{I}-\mathbf{P}_{\mathbf{D}_{\mathbf{\Theta}}})\mathbf{D}_{\overline{\mathbf{\Xi}}_{<i>}\backslash\mathbf{\Theta}})\nonumber\\
	&\leq\rho_{c(d^{\circ},\overline{d})}(\mathbf{D}^{\rm H}_{\mathbf{\Xi}^{\circ}\backslash\mathbf{\Theta}^{\circ}}\mathbf{D}_{\overline{\mathbf{\Xi}}_{<i>}\backslash\mathbf{\Theta}})\nonumber\\
	&\quad+\rho_{c(d^{\circ},d^{\Delta})}(\mathbf{D}^{\rm H}_{\mathbf{\Xi}^{\circ}\backslash\mathbf{\Theta}^{\circ}}\mathbf{P}_{\mathbf{D}_{\mathbf{\Theta}}}\mathbf{D}_{\overline{\mathbf{\Xi}}_{<i>}\backslash\mathbf{\Theta}})\nonumber\\
	&\leq\bigg\lceil\frac{\overline{d}}{d^{\circ}}\bigg\rceil\rho_{c(d^{\circ},d^{\circ})}(\mathbf{D}^{\rm H}_{\mathbf{\Xi}^{\circ}\backslash\mathbf{\Theta}^{\circ}}\mathbf{D}_{\overline{\mathbf{\Xi}}_{<i>}\backslash\mathbf{\Theta}})\nonumber\\
	&\quad+\bigg\lceil\frac{\overline{d}}{d^{\circ}}\bigg\rceil\rho_{c(d^{\circ},d^{\circ})}(\mathbf{D}^{\rm H}_{\mathbf{\Xi}^{\circ}\backslash\mathbf{\Theta}^{\circ}}\mathbf{P}_{\mathbf{D}_{\mathbf{\Theta}}}\mathbf{D}_{\overline{\mathbf{\Xi}}_{<i>}\backslash\mathbf{\Theta}}),\nonumber\\
	&\rho_{c(d^{\circ},d^{\circ})}(\mathbf{D}^{\rm H}_{\mathbf{\Xi}^{\circ}\backslash\mathbf{\Theta}^{\circ}}\mathbf{D}_{\overline{\mathbf{\Xi}}_{<i>}\backslash\mathbf{\Theta}})\leq k_t^{\circ}d^{\circ}\mu_{d^{\circ}},\nonumber\\
	&\rho_{c(d^{\circ},d^{\circ})}(\mathbf{D}^{\rm H}_{\mathbf{\Xi}^{\circ}\backslash\mathbf{\Theta}^{\circ}}\mathbf{P}_{\mathbf{D}_{\mathbf{\Theta}}}\mathbf{D}_{\overline{\mathbf{\Xi}}_{<i>}\backslash\mathbf{\Theta}})\nonumber\\
	&=\rho_{c(d^{\circ},d^{\circ})}(\mathbf{D}^{\rm H}_{\mathbf{\Xi}^{\circ}\backslash\mathbf{\Theta}^{\circ}}\mathbf{D}_{\mathbf{\Theta}}(\mathbf{D}^{\rm H}_{\mathbf{\Theta}}\mathbf{D}_{\mathbf{\Theta}})^{-1}\mathbf{D}^{\rm H}_{\mathbf{\Theta}}\mathbf{D}_{\overline{\mathbf{\Xi}}_{<i>}\backslash\mathbf{\Theta}})\nonumber\\
	&\leq\frac{(k_t^{\circ}-\gamma)(d^{\circ})^2\mu^2_{d^{\circ}} \lceil\frac{rd}{d^{\circ}}\rceil}{1-(d^{\circ}-1)\nu_{d^{\circ}}-(\lceil\frac{rd}{d^{\circ}}\rceil-1)d^{\circ}\mu_{d^{\circ}}}.\nonumber
\end{align}
Therefore, we have
\begin{align}
	&\rho_{c(d^{\circ},\overline{d})}(\ddot{\mathbf{A}}^{\rm H}_{\mathbf{\Xi}^{\circ}\backslash\mathbf{\Theta}^{\circ}}\ddot{\mathbf{A}}_{\overline{\mathbf{\Xi}}_{<i>}\backslash\mathbf{\Theta}})\nonumber\\
	&\leq\bigg\lceil\frac{\overline{d}}{d^{\circ}}\bigg\rceil\bigg(k_t^{\circ}d^{\circ}\mu_{d^{\circ}}+\frac{(k_t^{\circ}-\gamma)(d^{\circ})^2\mu^2_{d^{\circ}} \lceil\frac{rd}{d^{\circ}}\rceil}{1-(d^{\circ}-1)\nu_{d^{\circ}}-(\lceil\frac{rd}{d^{\circ}}\rceil-1)d^{\circ}\mu_{d^{\circ}}}\bigg).\nonumber
\end{align}

Finally, by combining the aforementioned upper bounds with the definitions of ${G}_{*}$ and ${G}_{\circ}$ given in \textbf{Theorem \ref{theorem1}}, we obtain the upper bounds of ${G}_{*}$ and ${G}_{\circ}$, which are denoted by $\overline{G}_{*}$ and $\overline{G}_{\circ}$, respectively. This completes the whole proof.
\end{IEEEproof}

\subsection{Proof of \textbf{Theorem \ref{theo10}}}
\begin{IEEEproof}
Using the symbols from \textbf{Section \ref{ERCsresults}}, to prove that HiBOMP-P selects a correct atom in the current iteration under noisy scenario, it is necessary to have
\begin{align}
	\|\ddot{\mathbf{A}}^{\rm H}_{\mathbf{\Xi}_{<i>}\backslash\mathbf{\Theta}}\mathbf{r}^{l}\|_{(d^*+d^{*\Delta}+d^{\Delta})2,\infty}>\|\ddot{\mathbf{A}}^{\rm H}_{\overline{\mathbf{\Xi}}_{<i>}\backslash\mathbf{\Theta}}\mathbf{r}^{l}\|_{(\overline{d})2,\infty},\nonumber
\end{align}
or more specifically, 
\begin{align}
	&\|\ddot{\mathbf{A}}^{\rm H}_{\mathbf{\Xi}_{<i>}\backslash\mathbf{\Theta}}(\overline{\mathbf{r}}^l+\mathbf{r}^{\circ^l})\|_{(d^*+d^{*\Delta}+d^{\Delta})2,\infty}\nonumber\\
	&\qquad\qquad\qquad\qquad>\|\ddot{\mathbf{A}}^{\rm H}_{\overline{\mathbf{\Xi}}_{<i>}\backslash\mathbf{\Theta}}(\overline{\mathbf{r}}^l+\mathbf{r}^{\circ^l})\|_{(\overline{d})2,\infty}.\label{nositymaincondition}
\end{align}
Note that, in noisy settings,
\begin{align}
	\overline{\mathbf{r}}^l+\mathbf{r}^{\circ^l} &= \mathbf{P}^{\bot}_{\mathbf{D}_{\mathbf{\Theta}}}(\mathbf{D}_{\mathbf{\Xi}^*\backslash\mathbf{\Theta}}\mathbf{x}_{\mathbf{\Xi}^*\backslash\mathbf{\Theta}}+\mathbf{D}_{\mathbf{\Theta}^{*\Delta}}\mathbf{x}^l_{\mathbf{\Theta}^{*\Delta}}\nonumber\\
	&\qquad\qquad+\mathbf{D}_{\mathbf{\Xi}^{\circ}\backslash\mathbf{\Theta}^{\circ}}\mathbf{x}^l_{\mathbf{\Xi}^{\circ}\backslash\mathbf{\Theta}^{\circ}}+\mathbf{n}).\nonumber
\end{align}
Thus, a sufficient condition of the establishment of (\ref{nositymaincondition}) is that
\begin{align}
	&\|\ddot{\mathbf{A}}^{\rm H}_{\mathbf{\Xi}_{<i>}\backslash\mathbf{\Theta}}\mathbf{P}^{\bot}_{\mathbf{D}_{\mathbf{\Theta}}}\nonumber\\
	&\times(\mathbf{D}_{\mathbf{\Xi}^*\backslash\mathbf{\Theta}}\mathbf{x}_{\mathbf{\Xi}^*\backslash\mathbf{\Theta}}+\mathbf{D}_{\mathbf{\Theta}^{*\Delta}}\mathbf{x}^l_{\mathbf{\Theta}^{*\Delta}})\|_{(d^*+d^{*\Delta}+d^{\Delta})2,\infty}\nonumber\\
	&\quad-	\|\ddot{\mathbf{A}}^{\rm H}_{\mathbf{\Xi}_{<i>}\backslash\mathbf{\Theta}}\mathbf{P}^{\bot}_{\mathbf{D}_{\mathbf{\Theta}}}\mathbf{D}_{\mathbf{\Xi}^{\circ}\backslash\mathbf{\Theta}^{\circ}}\mathbf{x}^l_{\mathbf{\Xi}^{\circ}\backslash\mathbf{\Theta}^{\circ}}\|_{(d^*+d^{*\Delta}+d^{\Delta})2,\infty}\nonumber\\
	&\quad-\|\ddot{\mathbf{A}}^{\rm H}_{\mathbf{\Xi}_{<i>}\backslash\mathbf{\Theta}}\mathbf{P}^{\bot}_{\mathbf{D}_{\mathbf{\Theta}}}\mathbf{n}\|_{(d^*+d^{*\Delta}+d^{\Delta})2,\infty}\nonumber\\
	&>\|\ddot{\mathbf{A}}^{\rm H}_{\overline{\mathbf{\Xi}}_{<i>}\backslash\mathbf{\Theta}}(\mathbf{D}_{\mathbf{\Xi}^*\backslash\mathbf{\Theta}}\mathbf{x}_{\mathbf{\Xi}^*\backslash\mathbf{\Theta}}+\mathbf{D}_{\mathbf{\Theta}^{*\Delta}}\mathbf{x}^l_{\mathbf{\Theta}^{*\Delta}})\|_{(\overline{d})2,\infty}\nonumber\\
	&+\|\ddot{\mathbf{A}}^{\rm H}_{\overline{\mathbf{\Xi}}_{<i>}\backslash\mathbf{\Theta}}\mathbf{D}_{\mathbf{\Xi}^{\circ}\backslash\mathbf{\Theta}^{\circ}}\mathbf{x}^l_{\mathbf{\Xi}^{\circ}\backslash\mathbf{\Theta}^{\circ}}\|_{(\overline{d})2,\infty}
	+\|\ddot{\mathbf{A}}^{\rm H}_{\overline{\mathbf{\Xi}}_{<i>}\backslash\mathbf{\Theta}}\mathbf{n}\|_{(\overline{d})2,\infty},\nonumber
\end{align}
i.e.,
\begin{align}
	&\|\ddot{\mathbf{A}}^{\rm H}_{\mathbf{\Xi}_{<i>}\backslash\mathbf{\Theta}}\mathbf{P}^{\bot}_{\mathbf{D}_{\mathbf{\Theta}}}\nonumber\\
	&\times(\mathbf{D}_{\mathbf{\Xi}^*\backslash\mathbf{\Theta}}\mathbf{x}_{\mathbf{\Xi}^*\backslash\mathbf{\Theta}}
	+\mathbf{D}_{\mathbf{\Theta}^{*\Delta}}\mathbf{x}^l_{\mathbf{\Theta}^{*\Delta}})\|_{(d^*+d^{*\Delta}+d^{\Delta})2,\infty}\nonumber\\
	&-	\|\ddot{\mathbf{A}}^{\rm H}_{\mathbf{\Xi}_{<i>}\backslash\mathbf{\Theta}}\mathbf{P}^{\bot}_{\mathbf{D}_{\mathbf{\Theta}}}\mathbf{D}_{\mathbf{\Xi}^{\circ}\backslash\mathbf{\Theta}^{\circ}}\mathbf{x}^l_{\mathbf{\Xi}^{\circ}\backslash\mathbf{\Theta}^{\circ}}\|_{(d^*+d^{*\Delta}+d^{\Delta})2,\infty}\nonumber\\
	&-(\|\ddot{\mathbf{A}}^{\rm H}_{\overline{\mathbf{\Xi}}_{<i>}\backslash\mathbf{\Theta}}(\mathbf{D}_{\mathbf{\Xi}^*\backslash\mathbf{\Theta}}\mathbf{x}_{\mathbf{\Xi}^*\backslash\mathbf{\Theta}}+\mathbf{D}_{\mathbf{\Theta}^{*\Delta}}\mathbf{x}^l_{\mathbf{\Theta}^{*\Delta}})\|_{(\overline{d})2,\infty}\nonumber\\
	&+\|\ddot{\mathbf{A}}^{\rm H}_{\overline{\mathbf{\Xi}}_{<i>}\backslash\mathbf{\Theta}}\mathbf{D}_{\mathbf{\Xi}^{\circ}\backslash\mathbf{\Theta}^{\circ}}\mathbf{x}^l_{\mathbf{\Xi}^{\circ}\backslash\mathbf{\Theta}^{\circ}}\|_{(\overline{d})2,\infty})\nonumber\\
	&\qquad\qquad\qquad\qquad\qquad\qquad>2\|\ddot{\mathbf{A}}^{\rm H}_{\overline{\mathbf{\Xi}}_{<i>}\backslash\mathbf{\Theta}}\mathbf{n}\|_{(\overline{d})2,\infty}.\label{noisyderivation}
\end{align}
From \textbf{Theorem \ref{theo6}}, we have
\begin{align}
	&(\overline{G}_{*}+\overline{G}_{\circ})(\|\ddot{\mathbf{A}}^{\rm H}_{\mathbf{\Xi}_{<i>}\backslash\mathbf{\Theta}}\mathbf{P}^{\bot}_{\mathbf{D}_{\mathbf{\Theta}}}\nonumber\\
	&\quad\enspace\enspace\enspace\enspace\times(\mathbf{D}_{\mathbf{\Xi}^*\backslash\mathbf{\Theta}}\mathbf{x}_{\mathbf{\Xi}^*\backslash\mathbf{\Theta}}
	+\mathbf{D}_{\mathbf{\Theta}^{*\Delta}}\mathbf{x}^l_{\mathbf{\Theta}^{*\Delta}})\|_{(d^*+d^{*\Delta}+d^{\Delta})2,\infty}\nonumber\\
	&-	\|\ddot{\mathbf{A}}^{\rm H}_{\mathbf{\Xi}_{<i>}\backslash\mathbf{\Theta}}\mathbf{P}^{\bot}_{\mathbf{D}_{\mathbf{\Theta}}}\mathbf{D}_{\mathbf{\Xi}^{\circ}\backslash\mathbf{\Theta}^{\circ}}\mathbf{x}^l_{\mathbf{\Xi}^{\circ}\backslash\mathbf{\Theta}^{\circ}}\|_{(d^*+d^{*\Delta}+d^{\Delta})2,\infty})\nonumber\\
	&>\|\ddot{\mathbf{A}}^{\rm H}_{\overline{\mathbf{\Xi}}_{<i>}\backslash\mathbf{\Theta}}(\mathbf{D}_{\mathbf{\Xi}^*\backslash\mathbf{\Theta}}\mathbf{x}_{\mathbf{\Xi}^*\backslash\mathbf{\Theta}}+\mathbf{D}_{\mathbf{\Theta}^{*\Delta}}\mathbf{x}^l_{\mathbf{\Theta}^{*\Delta}})\|_{(\overline{d})2,\infty}\nonumber\\
	&\quad+\|\ddot{\mathbf{A}}^{\rm H}_{\overline{\mathbf{\Xi}}_{<i>}\backslash\mathbf{\Theta}}\mathbf{D}_{\mathbf{\Xi}^{\circ}\backslash\mathbf{\Theta}^{\circ}}\mathbf{x}^l_{\mathbf{\Xi}^{\circ}\backslash\mathbf{\Theta}^{\circ}}\|_{(\overline{d})2,\infty}.\nonumber
\end{align}
It is worth mentioning that $\overline{G}_{*}+\overline{G}_{\circ}<1$ is a sufficient condition of (\ref{mainmain}). Then, by direct calculation, we obtain that
\begin{align}
	&\|\ddot{\mathbf{A}}^{\rm H}_{\mathbf{\Xi}_{<i>}\backslash\mathbf{\Theta}}\mathbf{P}^{\bot}_{\mathbf{D}_{\mathbf{\Theta}}}\nonumber\\
	&\times(\mathbf{D}_{\mathbf{\Xi}^*\backslash\mathbf{\Theta}}\mathbf{x}_{\mathbf{\Xi}^*\backslash\mathbf{\Theta}}+\mathbf{D}_{\mathbf{\Theta}^{*\Delta}}\mathbf{x}^l_{\mathbf{\Theta}^{*\Delta}})\|_{(d^*+d^{*\Delta}+d^{\Delta})2,\infty}\nonumber\\
	&-	\|\ddot{\mathbf{A}}^{\rm H}_{\mathbf{\Xi}_{<i>}\backslash\mathbf{\Theta}}\mathbf{P}^{\bot}_{\mathbf{D}_{\mathbf{\Theta}}}\mathbf{D}_{\mathbf{\Xi}^{\circ}\backslash\mathbf{\Theta}^{\circ}}\mathbf{x}^l_{\mathbf{\Xi}^{\circ}\backslash\mathbf{\Theta}^{\circ}}\|_{(d^*+d^{*\Delta}+d^{\Delta})2,\infty}\nonumber\\
	&-(\|\ddot{\mathbf{A}}^{\rm H}_{\overline{\mathbf{\Xi}}_{<i>}\backslash\mathbf{\Theta}}(\mathbf{D}_{\mathbf{\Xi}^*\backslash\mathbf{\Theta}}\mathbf{x}_{\mathbf{\Xi}^*\backslash\mathbf{\Theta}}
	+\mathbf{D}_{\mathbf{\Theta}^{*\Delta}}\mathbf{x}^l_{\mathbf{\Theta}^{*\Delta}})\|_{(\overline{d})2,\infty}\nonumber\\
	&\quad\enspace+\|\ddot{\mathbf{A}}^{\rm H}_{\overline{\mathbf{\Xi}}_{<i>}\backslash\mathbf{\Theta}}\mathbf{D}_{\mathbf{\Xi}^{\circ}\backslash\mathbf{\Theta}^{\circ}}\mathbf{x}^l_{\mathbf{\Xi}^{\circ}\backslash\mathbf{\Theta}^{\circ}}\|_{(\overline{d})2,\infty})\nonumber\\
	&>(1-(\overline{G}_{*}+\overline{G}_{\circ}))(\|\ddot{\mathbf{A}}^{\rm H}_{\mathbf{\Xi}_{<i>}\backslash\mathbf{\Theta}}\mathbf{P}^{\bot}_{\mathbf{D}_{\mathbf{\Theta}}}\nonumber\\
	&\qquad\quad\times(\mathbf{D}_{\mathbf{\Xi}^*\backslash\mathbf{\Theta}}\mathbf{x}_{\mathbf{\Xi}^*\backslash\mathbf{\Theta}}+\mathbf{D}_{\mathbf{\Theta}^{*\Delta}}\mathbf{x}^l_{\mathbf{\Theta}^{*\Delta}})\|_{(d^*+d^{*\Delta}+d^{\Delta})2,\infty}\nonumber\\
	&-	\|\ddot{\mathbf{A}}^{\rm H}_{\mathbf{\Xi}_{<i>}\backslash\mathbf{\Theta}}\mathbf{P}^{\bot}_{\mathbf{D}_{\mathbf{\Theta}}}\mathbf{D}_{\mathbf{\Xi}^{\circ}\backslash\mathbf{\Theta}^{\circ}}\mathbf{x}^l_{\mathbf{\Xi}^{\circ}\backslash\mathbf{\Theta}^{\circ}}\|_{(d^*+d^{*\Delta}+d^{\Delta})2,\infty}).\label{mainproofM}
\end{align}
By letting the right-hand side of (\ref{noisyderivation}) being smaller than the right-hand side of (\ref{mainproofM}), we obtain that 
\begin{align}
	&\|\ddot{\mathbf{A}}^{\rm H}_{\mathbf{\Xi}_{<i>}\backslash\mathbf{\Theta}}\mathbf{P}^{\bot}_{\mathbf{D}_{\mathbf{\Theta}}}\nonumber\\
	&\enspace\times(\mathbf{D}_{\mathbf{\Xi}^*\backslash\mathbf{\Theta}}\mathbf{x}_{\mathbf{\Xi}^*\backslash\mathbf{\Theta}}+\mathbf{D}_{\mathbf{\Theta}^{*\Delta}}\mathbf{x}^l_{\mathbf{\Theta}^{*\Delta}})\|_{(d^*+d^{*\Delta}+d^{\Delta})2,\infty}\nonumber\\
	&-	\|\ddot{\mathbf{A}}^{\rm H}_{\mathbf{\Xi}_{<i>}\backslash\mathbf{\Theta}}\mathbf{P}^{\bot}_{\mathbf{D}_{\mathbf{\Theta}}}\mathbf{D}_{\mathbf{\Xi}^{\circ}\backslash\mathbf{\Theta}^{\circ}}\mathbf{x}^l_{\mathbf{\Xi}^{\circ}\backslash\mathbf{\Theta}^{\circ}}\|_{(d^*+d^{*\Delta}+d^{\Delta})2,\infty}\nonumber\\
	&\qquad\qquad\qquad\qquad\qquad\qquad>\frac{2\|\ddot{\mathbf{A}}^{\rm H}_{\overline{\mathbf{\Xi}}_{<i>}\backslash\mathbf{\Theta}}\mathbf{n}\|_{(\overline{d})2,\infty}}{1-(\overline{G}_{*}+\overline{G}_{\circ})} \label{asuffcondition}
\end{align}
is a sufficient condition of the establishment of the correct block selection of HiBOMP-P.

As illustrated in \textbf{Section \ref{profoftheorem1}},
\begin{align}
	&\|\ddot{\mathbf{A}}^{\rm H}_{\mathbf{\Xi}_{<i>}\backslash\mathbf{\Theta}}\mathbf{P}^{\bot}_{\mathbf{D}_{\mathbf{\Theta}}}(\mathbf{D}_{\mathbf{\Xi}^*\backslash\mathbf{\Theta}}\mathbf{x}_{\mathbf{\Xi}^*\backslash\mathbf{\Theta}}\nonumber\\
	&\enspace+\mathbf{D}_{\mathbf{\Theta}^{*\Delta}}\mathbf{x}^l_{\mathbf{\Theta}^{*\Delta}})\|_{(d^*+d^{*\Delta}+d^{\Delta})2,\infty}\nonumber\\
	&\geq\delta_{\sigma_{\min}}\|\mathbf{x}^l_{[\mathbf{\Xi}^*\backslash\mathbf{\Theta},\mathbf{\Theta}^{*\Delta}]}\|_{(d^*+d^{*\Delta})2,\infty},\nonumber\\
	&\|\ddot{\mathbf{A}}^{\rm H}_{\mathbf{\Xi}_{<i>}\backslash\mathbf{\Theta}}\mathbf{P}^{\bot}_{\mathbf{D}_{\mathbf{\Theta}}}\mathbf{D}_{\mathbf{\Xi}^{\circ}\backslash\mathbf{\Theta}^{\circ}}\mathbf{x}^l_{\mathbf{\Xi}^{\circ}\backslash\mathbf{\Theta}^{\circ}}\|_{(d^*+d^{*\Delta}+d^{\Delta})2,\infty}\nonumber\\
	&\leq(\delta_{d^{\circ},d^*+d^{*\Delta}}+\delta_{d^{\circ},d^{\Delta}})^{\frac{1}{2}}\|\mathbf{x}^l_{\mathbf{\Xi}^{\circ}\backslash\mathbf{\Theta}^{\circ}}\|_{(d^*+d^{*\Delta}+d^{\Delta})2,\infty}.\label{midbounds}
\end{align}
Therefore, combining (\ref{asuffcondition}) and (\ref{midbounds}) yields that
(\ref{theo4main})
is a sufficient condition for choosing a correct block in the current iteration.
\end{IEEEproof}

\subsection{Proof of \textbf{Theorem \ref{theo11}}}
\begin{IEEEproof}
Suppose that $j$ is the block index contributing to the largest value in $\Big\{\|\ddot{\mathbf{A}}^{\rm H}_{(\overline{\mathbf{\Xi}}_{<i>}\backslash\mathbf{\Theta})_{[j]}}\mathbf{n}\|_2\Big\}$. Denote $j_1,j_2,\cdots,j_{\overline{d}}$ as the indices within the $j$th block. Observe that
\begin{align}
	&\|\ddot{\mathbf{A}}^{\rm H}_{\overline{\mathbf{\Xi}}_{<i>}\backslash\mathbf{\Theta}}\mathbf{n}\|_{(\overline{d})2,\infty}\nonumber\\
	&=\|\ddot{\mathbf{A}}^{\rm H}_{(\overline{\mathbf{\Xi}}_{<i>}\backslash\mathbf{\Theta})_{[j]}}\mathbf{n}\|_{(\overline{d})2,\infty}\nonumber\\
	&=\sqrt{\sum_{l=j_1}^{j_{\overline{d}}}|\ddot{\mathbf{A}}^{\rm H}_{(\overline{\mathbf{\Xi}}_{<i>}\backslash\mathbf{\Theta})_{(l)}}\mathbf{n}|^2}\nonumber\\
	&=\sqrt{\sum_{l=j_1}^{j_{\overline{d}}}|\mathbf{D}^{\rm H}_{(\overline{\mathbf{\Xi}}_{<i>}\backslash\mathbf{\Theta})_{(l)}}\mathbf{P}^{\bot}_{\mathbf{\Theta}}\mathbf{n}|^2}\nonumber\\
	&\leq\sqrt{\sum_{l=j_1}^{j_{\overline{d}}}\|\mathbf{D}^{\rm H}_{(\overline{\mathbf{\Xi}}_{<i>}\backslash\mathbf{\Theta})_{(l)}}\|^2_2\|\mathbf{P}^{\bot}_{\mathbf{\Theta}}\mathbf{n}\|_2^2}\nonumber\\
	&\leq\sqrt{\sum_{l=j_1}^{j_{\overline{d}}}\|\mathbf{n}\|_2^2}\leq\sqrt{\overline{d}\epsilon^2}=\sqrt{\overline{d}}\epsilon,\nonumber
\end{align}
where the second inequality is because $\mathbf{D}$ is normalized to have unit column norm, i.e., $\|\mathbf{D}^{\rm H}_{(\overline{\mathbf{\Xi}}_{<i>}\backslash\mathbf{\Theta})_{(l)}}\|_2=1$ and $\|\mathbf{P}^{\bot}_{\mathbf{\Theta}}\mathbf{n}\|_2\leq\|\mathbf{n}\|_2$, and the last inequality follows from the assumption that $\|\mathbf{n}\|_2\leq\epsilon$. Thus, combining the result of (\ref{theo4main}) in \textbf{Theorem \ref{theo10}}, we obtain that if (\ref{theo4mainmain})
holds, then HiBOMP-P selects a correct variable in the current iteration.

Now we come to the stopping rule. Suppose that all the support blocks in the support index set $\mathbf{\Xi}$ have been selected in the $q$th iteration. Then, the residual vector satisfies that 
\begin{align}
	\|\mathbf{r}^q\|_2&=\|\mathbf{P}_{\mathbf{\Xi}\cup\mathbf{\Theta}}^{\bot}\mathbf{y}\|_2\nonumber\\
	&=\|\mathbf{P}_{\mathbf{\Xi}\cup\mathbf{\Theta}}^{\bot}(\mathbf{D}_{\mathbf{\Xi}}\mathbf{x}^q_{\mathbf{\Xi}}+\mathbf{n})\|_2\nonumber\\
	&=\|\mathbf{P}_{\mathbf{\Xi}\cup\mathbf{\Theta}}^{\bot}\mathbf{n}\|_2\leq\|\mathbf{n}\|_2\leq\epsilon.\nonumber
\end{align}  
Hence, when all the correct support indices have been chosen, $\|\mathbf{r}\|_2$ will be less than $\epsilon$, and the HiBOMP-P stops iteration. Then, suppose that HiBOMP-P has performed for $c$ iterations. Then, we have
\begin{align}
	\|\mathbf{r}^c\|_2&=\|\mathbf{P}_{\mathbf{\Xi}^c\cup\mathbf{\Theta}}^{\bot}\mathbf{y}\|_2\nonumber\\
	&=\|\mathbf{P}_{\mathbf{\Xi}^c\cup\mathbf{\Theta}}^{\bot}(\mathbf{D}_{\mathbf{\Xi}}\mathbf{x}^c_{\mathbf{\Xi}}+\mathbf{n})\|_2\nonumber\\
	&\geq\|\mathbf{P}_{\mathbf{\Xi}^c\cup\mathbf{\Theta}}^{\bot}\mathbf{D}_{\mathbf{\Xi}}\mathbf{x}^c_{\mathbf{\Xi}}\|_2-\|\mathbf{P}_{\mathbf{\Xi}^c\cup\mathbf{\Theta}}^{\bot}\mathbf{n}\|_2\nonumber\\
	&\geq\|\mathbf{P}_{\mathbf{\Xi}^c\cup\mathbf{\Theta}}^{\bot}\mathbf{D}_{\mathbf{\Xi}}\mathbf{x}^c_{\mathbf{\Xi}}\|_2-\epsilon.\nonumber
\end{align}
Note that
\begin{align}
	&\|\mathbf{P}_{\mathbf{\Xi}^c\cup\mathbf{\Theta}}^{\bot}\mathbf{D}_{\mathbf{\Xi}}\mathbf{x}^c_{\mathbf{\Xi}}\|_2\nonumber\\
	&\geq\sigma_{\min}(\mathbf{D}^{\rm H}_{\mathbf{\Xi}}\mathbf{D}_{\mathbf{\Xi}})\|\mathbf{x}^c_{\mathbf{\Xi}}\|_2\nonumber\\
	&\geq(1-(d*+d^{\Delta}+d^{*\Delta}-1)\nu_{d*+d^{\Delta}+d^{*\Delta}}\nonumber\\
	&\quad-(k-1)(d*+d^{\Delta}+d^{*\Delta})\mu_{d*+d^{\Delta}+d^{*\Delta}})\frac{2\sqrt{\overline{d}}\epsilon}{1-(\overline{G}_{*}+\overline{G}_{\circ})}\nonumber\\
	&>2\epsilon.\nonumber
\end{align}
Therefore, we have 
\begin{align}
	\|\mathbf{r}^c\|_2\geq\|\mathbf{P}_{\mathbf{\Xi}^c\cup\mathbf{\Theta}}^{\bot}\mathbf{D}_{\mathbf{\Xi}}\mathbf{x}^c_{\mathbf{\Xi}}\|_2-\epsilon>\epsilon.\nonumber
\end{align}
This indicates that HiBOMP-P does not stop early, which competes the proof.
\end{IEEEproof}

\section{Simulation Examples}

\begin{figure}[!tp]
	\centering
	\subfigure[$M=80$, $N=400$, $d_{out}=16$, $d=4$, $k_{in}=2$, and $k_{out}=1$.]{\label{noiselessa}
		\includegraphics[width=1\linewidth]{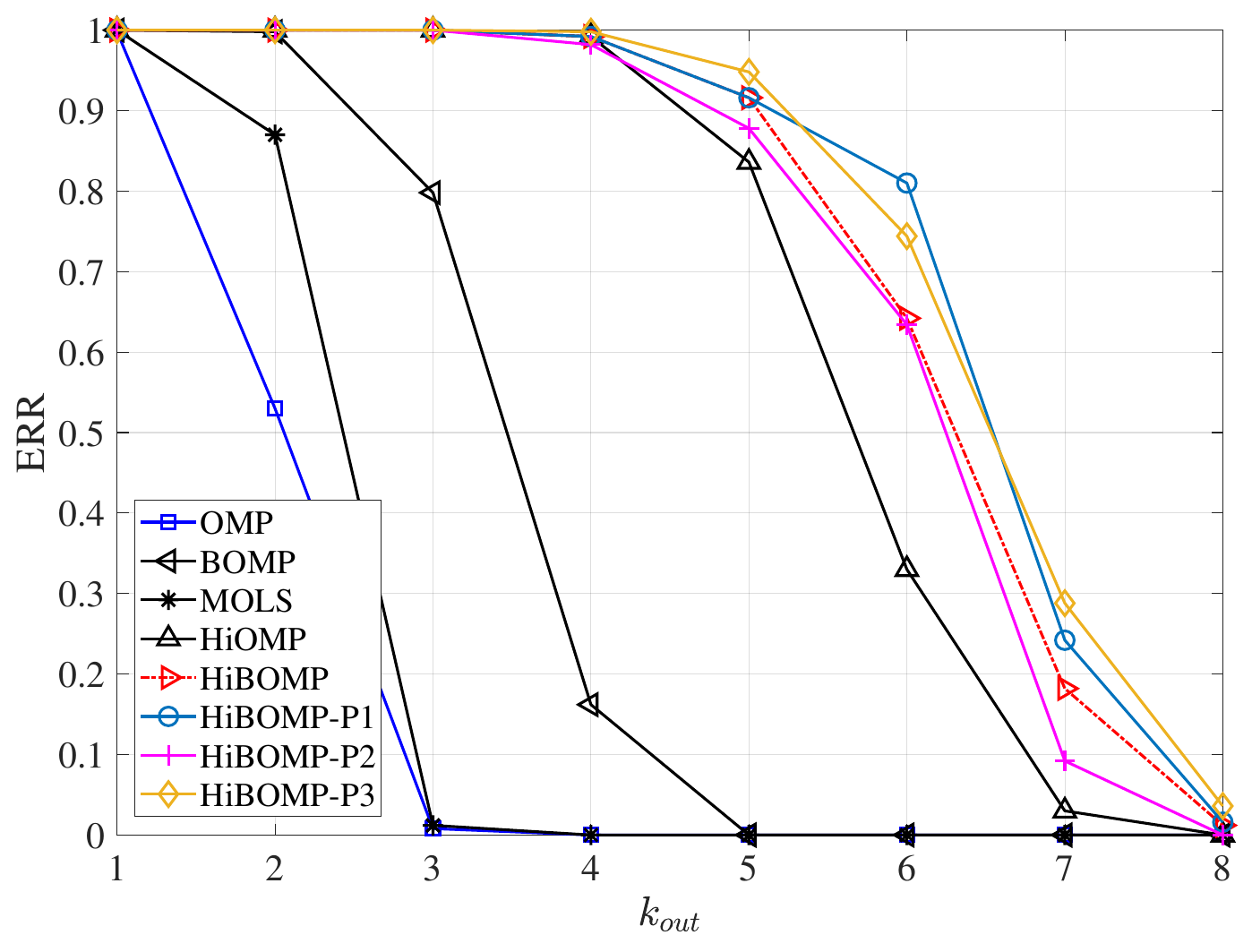}}
	\hspace{0.01\linewidth}
	\subfigure[$M=40$, $N=400$, $d_{out}=16$, $d=4$, $k_{in}=2$, and $k_{out}=2$.]{\label{noiselessb}
		\includegraphics[width=1\linewidth]{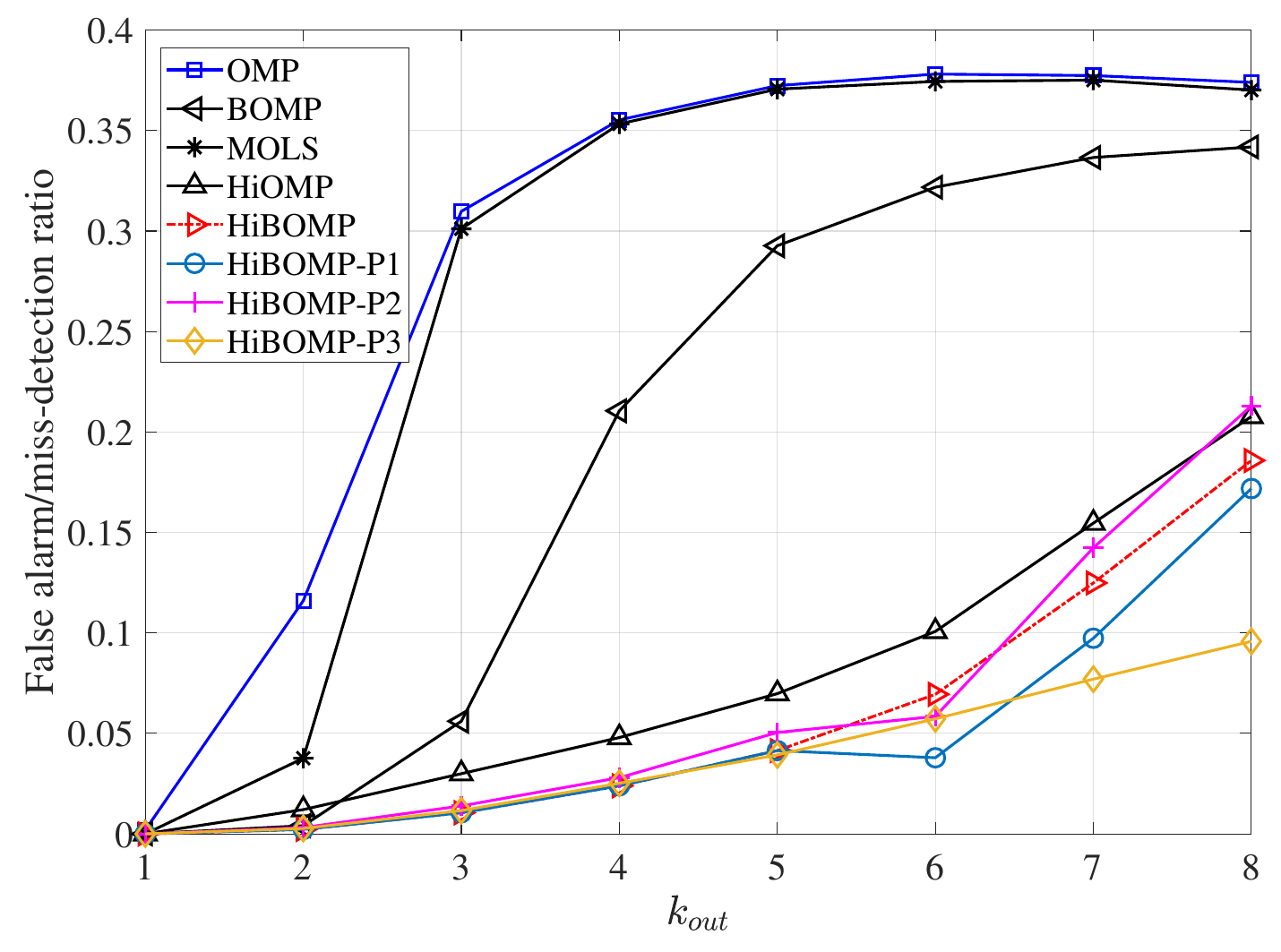}}
	\vspace*{-1mm}
	\caption{ERR and false alarm/miss-detection ratio for recovering hierarchically block-sparse 2-PAM signals as a function of $k_{out}$ with $M=128$, $N=512$, $d_{out}=16$, $d=2$, and $k_{in}=6$.}
	\label{noiselessfig} % Fig.5
	\vspace*{2mm}
\end{figure}

\begin{figure}[!tp]
	\centering
	\subfigure[$M=80$, $N=400$, $d_{out}=16$, $d=4$, $k_{in}=2$, and $k_{out}=1$.]{\label{noisya}
		\includegraphics[width=1\linewidth]{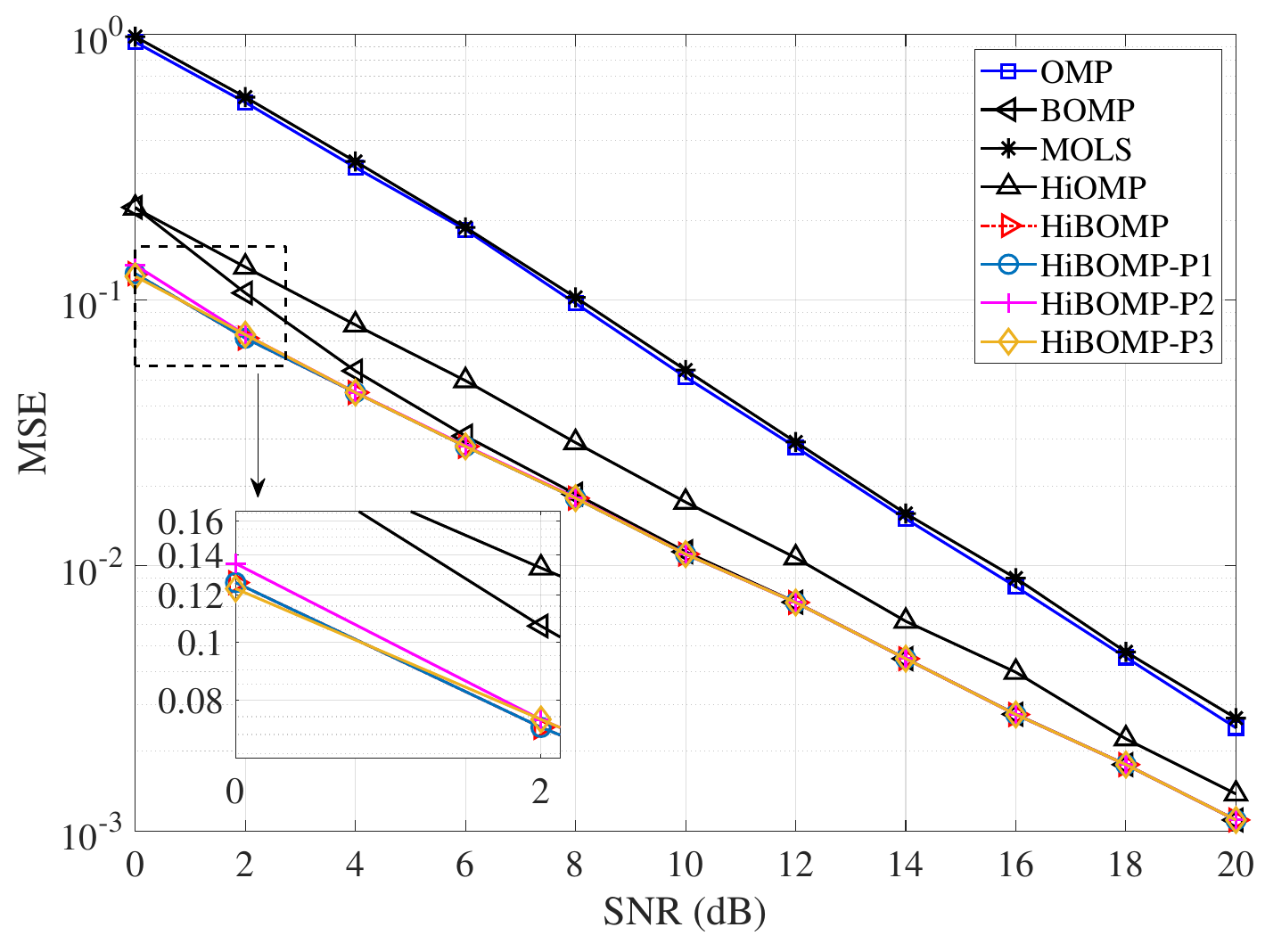}}
	\hspace{0.01\linewidth}
	\subfigure[$M=40$, $N=400$, $d_{out}=16$, $d=4$, $k_{in}=2$, and $k_{out}=2$.]{\label{noisyb}
		\includegraphics[width=1\linewidth]{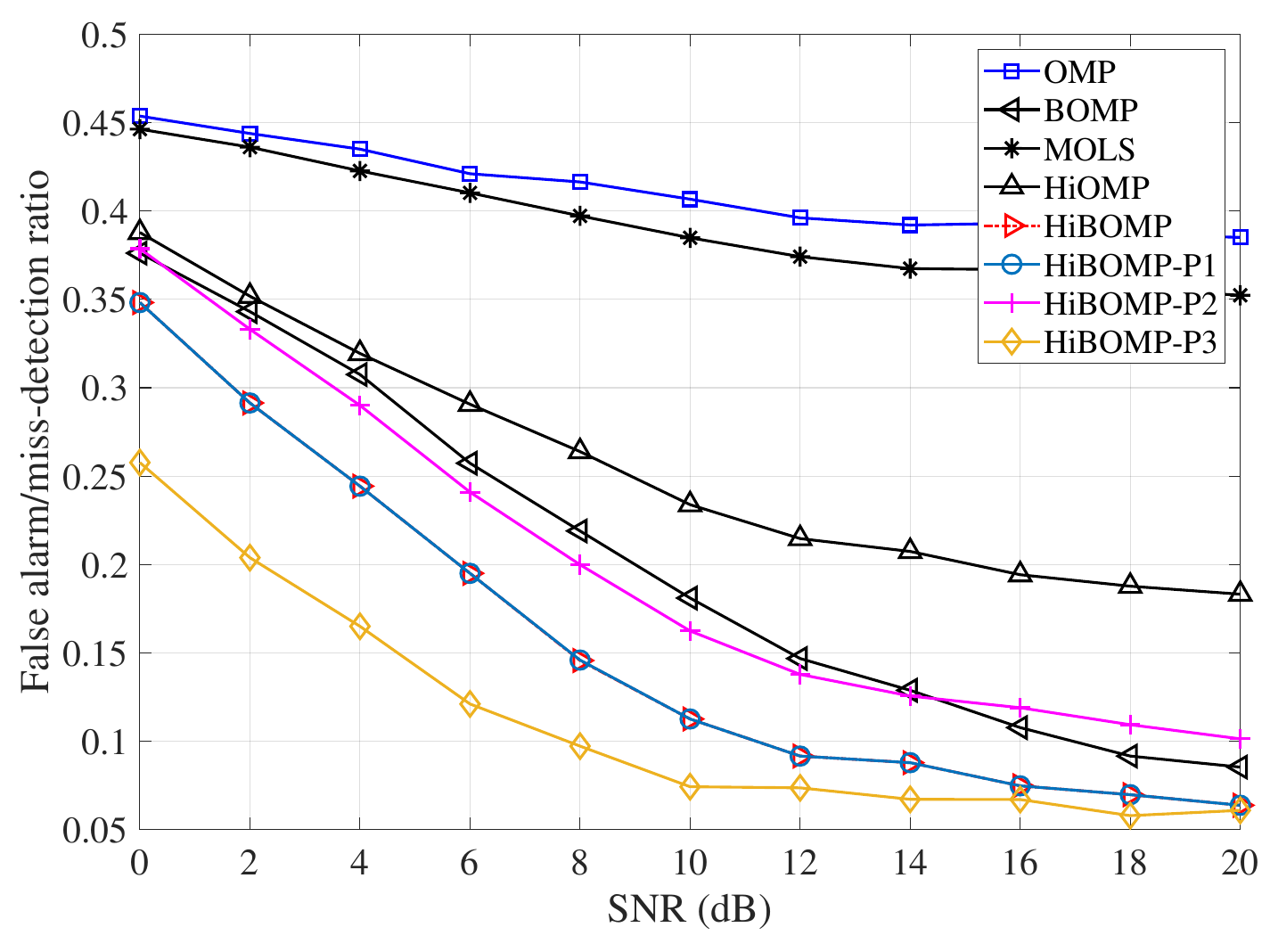}}
	\vspace*{-1mm}
	\caption{MSE and false alarm/miss-detection ratio for recovering hierarchically block-sparse Gaussian signals as a function of SNR.}
	\label{noisyfig} % Fig.5
	\vspace*{2mm}
\end{figure}

We empirically investigate the recovery performance of HiBOMP-P in reconstructing hierarchically block-sparse signals under both noiseless and noisy conditions. In each trial, we generate an $M\times N$ measurement matrix with each element an independent and identically distributed (i.i.d.) draw of a Gaussian distribution, with zero mean and $\frac{1}{M}$ variance. The measurement matrices are normalized to have unit column norm. For each realization of a hierarchically block-sparse signal, its block support is chosen uniformly at random, and nonzero elements are 1) drawn independently from a standard Gaussian distribution, or 2) drawn independently from the set $\{\pm1\}$. These two types of tensors are referred to as the block-sparse Gaussian signal and the block-sparse 2-ary pulse amplitude modulation (2-PAM) signal, respectively \cite{jwang2017}. 
	The hierarchical mode in each trial is fixed at 2, as 2-mode hierarchically block-sparse recovery offers greater clarity for simulation and is more commonly used in practical applications. For clarity of symbols, we denote the block sparsity of the first hierarchical mode as $k_{out}$, and denote the block sparsity of the second hierarchical mode as $k_{in}$, i.e., the block sparsity that is within one support block from the first hierarchical mode. The block lengths of the first and the second hierarchical modes are represented by $d_{out}$ and $d$, respectively. The exact recovery ratio (ERR) and the false alarm/miss-detection ratio of hierarchically block-sparse reconstruction are examined as functions of block sparsity in the noise-free scenario. In the noisy scenario, the normalized mean square error (NMSE) and the false alarm/miss-detection ratio are employed to assess the recovery performance of various approaches. More specifically, the ERR is defined by \cite{Kim2020}
\begin{align} 
	\text{ERR} = \frac{\text{number of exact reconstructions}}{\text{number of total trials}},\nonumber
\end{align}
the false alarm/miss-detection ratio is expressed as
\begin{align} 
	\text{false alarm/miss-detection ratio}=\frac{|\hat{\mathbf{\Xi}}\setminus\mathbf{\Xi}|}{k_{out}k_{in}d},\nonumber
\end{align}
and the NMSE is given by
\begin{align} 
	\text{NMSE} = \frac{\|\hat{\mathbf{x}}-\mathbf{x}\|^2_2}{\|\mathbf{x}\|^2_2},\nonumber
\end{align}
  where $\hat{\mathbf{\Xi}}$ is the index set consisting of the indices corresponding to nonzero elements of the estimated signal, $\mathbf{\Xi}$ denotes the correct support index set, $\hat{\mathbf{x}}$ represents the estimated hierarchically block-sparse signal, and $\mathbf{x}$ denotes the true hierarchically block-sparse signal. Note that the false alarm ratio is equivalent to the miss-detection ratio since $|\hat{\mathbf{\Xi}}|=|\mathbf{\Xi}|=k_{out}k_{in}d$.
  In our experiments, we perform 1,000 independent trials for each point of the approaches, and consider the following recovery algorithms: 1)~OMP \cite{greed2004}; 2)~BOMP \cite{Eldar2010}; 3)~MOLS \cite{jwang2017}; 4)~HiOMP; 5)~HiBOMP; 6)~HiBOMP-P with $\overline{\alpha}= \lceil 0.2k_{out}\rceil$, which is named as HiBOMP-P1; 7)~HiBOMP-P with $\overline{\alpha}= \lceil 0.2k_{out}\rceil$ and $\beta=\lceil0.2 k_{in}\rceil$, which is named as HiBOMP-P2; 8)~HiBOMP-P with $\alpha^{*\Delta}=\frac{d_{out}}{d}-k_{in}$, which is named as HiBOMP-P3.  
  	To control variables, all other unspecified PSI parameters are set to zero.
 
 In the noiseless case, the recovery performance of all the compared approaches in terms of ERR and false alarm/miss-detection ratio decreases as the $k_{out}$ increases. This is consistent with the theoretical analysis that a lower sparsity level would contribute a better recovery performance, which highlights the importance of reconstructible sparsity analysis in revealing the recoverability of various algorithms. In general, the hierarchical recovery algorithms perform better than the other algorithms compared. By leveraging beneficial PSI, HiBOMP-P1 and HiBOMP-P3 achieve higher ERRs and lower false alarm/miss-detection ratios than those of HiBOMP within all the $k_{out}$ levels. Meanwhile, due to the PSI input related to non support indices, HiBOMP-P2 exhibits a worse performance. It is worth mentioning that in the implementation of HiBOMP-P3, the PSI used does not indicate the true support index. This confirms the statement that HCS could enjoy improved recovery performance even when the PSI and true support do not overlap. Then, we come to verify the theoretical results in terms of reconstructible sparsity presented in this paper. Firstly, we verify the correctness of the sufficiency of the reconstructible block sparsity in terms of $k_{out}$. For $M=128$ and $N=512$, we have $\mu_{d^*+d^{\Delta}}=\mu_{d_{out}}=\mu_{16}\approx0.05$. Meanwhile, consider that the conventional matrix coherence $\mu$ achieve the lower bound of $\sqrt{\frac{N-M}{M(N-1)}}\approx0.077$. Then, the hierarchical sub-coherence satisfies that $\nu_{d^*+d^{\Delta}}\leq0.077$. In this case, the reconstructible block sparsity bound in \emph{Remark \ref{rmk66}} indicates that $k_{out}(d^*+d^{\Delta})<6.45\Rightarrow k_{out}=0$ is sufficient for exact recovery of HiBOMP-P. When setting $\nu_{d^*+d^{\Delta}}=0$, the bound in \emph{Remark \ref{rmkk6}} reveals that $k_{out}(d^*+d^{\Delta})<18\Rightarrow k_{out}\leq1$ is sufficient for exact recovery of HiBOMP-P. It can be observed that these results are consistent with the empirical results as presented in Fig. \ref{noiselessa}, wherein the empirical results indicate that $k_{out}\leq3$ is sufficient for the exact recovery of HiBOMP-P3. It is worth mentioning that the results with $\nu_{d^*+d^{\Delta}}\leq0.077$ and $\nu_{d^*+d^{\Delta}}=0$ are less restricted than the main results in \textbf{Theorems \ref{theorem1}} to \textbf{\ref{theo12}}, thus the correctness of the results with $\nu_{d^*+d^{\Delta}}\leq0.077$ and $\nu_{d^*+d^{\Delta}}=0$ also indicates the correctness of the results in \textbf{Theorems \ref{theorem1}} to \textbf{\ref{theo12}}. Secondly, we compare the derived results with those presented in \cite{Eldar2010}. For fair comparison, we illustrate the results corresponding to the true reconstructible block sparsity bounds. For the existing results with $\nu=0$ for conventional block-sparse recovery in \cite{Eldar2010}, they indicate that $\overline{k}d<4.57\Rightarrow \overline{k}\leq2$ is sufficient for exact recovery of BOMP, as the conventional block coherence satisfies that $\mu_B\approx0.14$. For the results derived in \emph{Remark \ref{rmkk6}}, the true reconstructible block sparsity in a specific hierarchical block satisfies that $k_{in}d<4.57\Rightarrow k_{in}\leq2$. This indicates that if the true reconstructible block sparsity satisfies that $k_{in}k_{out}<2 k_{out}$, then HiBOMP-P performs exact recovery, where $k_{in}k_{out}$ is the true reconstructible block sparsity of HiBOMP-P. It can be observed that $2 k_{out}\geq2$, which reveals a better recovery guarantee in terms of the true reconstructible block sparsity. However, the analytical results presented are more pessimistic than the numerical results, since our analyses based on the MIP framework, providing significant improvements over the existing results anyway, essentially consider worst-case scenarios \cite{Eldar2010,jwang2017}. 

In the noisy case, we compare MSE and false alarm/miss-detection ratio as functions of SNR. Overall, the algorithms using block structure perform better than those do not exploit the block structure. Note that in noisy scenarios, the performance of BOMP is better than that of HiOMP due to the use of block structure. Meanwhile, it can be observed from Fig. \ref{noisyb} that HiBOMP-P1 exhibits similar performance to HiBOMP, which indicates that the PSI level set with $\overline{\alpha}=\lceil0.2k_{out}\rceil$ could not provide significant gain. On the contrary, HiBOMP-P3 offers significant performance improvements over that of HiBOMP even if there is no overlap between the input PSI and and the true support set. 
	Furthermore, HiBOMP-P3 exhibits superior performance compared to HiBOMP-P1, indicating that the additional PSI index set plays a more effective role in support selection under the given simulation settings. These results align with the theoretical outcomes, demonstrating that the hierarchical structure effectively leverages the beneficial information from the PSI associated with the additional support index set. The practical utility of HiBOMP-P is further highlighted, given that PSI may not always coincide with the true support set in real-world applications.

\section{Conclusions}

This paper has studied recovery conditions related to the MIP of HCS with the assistance of PSI. Specifically, a general hierarchically block-sparse recovery model in terms of an $n$-mode hierarchical structure was formulated, and hierarchical MIP concepts were introduced that offer improvements in revealing uncertain relations within the measurement matrix with a hierarchical structure. We have elaborated on a HiBOMP-P algorithm, which can recursively search all correct support blocks. Theoretical analyses in terms of ERCs have been presented, wherein reconstructible sparsity levels are derived as sufficient guarantees, revealing desirable recoverability of HiBOMP-P in noiseless scenarios. Furthermore, we have provided several insights based on these MIP-related results, involving noisy recovery conditions and optimal hierarchical structure analysis. In particular, HCS can provide improved recovery performance even if the PSI and true support set are different. Simulation results have verified the analysis, and all outcomes, including theoretical and experimental results, confirm the advantages of HCS in recovering hierarchically block-sparse signals.

\appendix

\subsection{Useful Lemmas and Corollaries}

\begin{lemma13}\label{lemma13}
	Given a matrix $\mathbf{D}$ and its column-block submatrix $\underline{\mathbf{D}}$, the following inequality holds:
	\begin{align}
		\|\underline{\mathbf{D}}\|_2\leq\|\mathbf{D}\|_2\nonumber.
	\end{align}
\end{lemma13}

\begin{IEEEproof}
	Without loss of generality, assume that $\mathbf{D}=[\underline{\mathbf{D}},\overline{\mathbf{D}}]$. Let $\overline{\mathbf{D}}_i$ be the $i$th column of $\overline{\mathbf{D}}$. Denote $\underline{\mathbf{D}}^i=[\underline{\mathbf{D}},\overline{\mathbf{D}}_i]$. Then, we have
	\begin{align}
		\|\underline{\mathbf{D}}^i\|_2 &= \sqrt{\lambda_{\max}((\underline{\mathbf{D}}^i)^{\rm H}\underline{\mathbf{D}}^i)},\nonumber\\
		\|\underline{\mathbf{D}}\|_2 &= \sqrt{\lambda_{\max}(\underline{\mathbf{D}}^{\rm H}\underline{\mathbf{D}})}.\nonumber
	\end{align}
Note that
\begin{align}
	(\underline{\mathbf{D}}^i)^{\rm H}\underline{\mathbf{D}}^i=\left[\begin{array}{cc}
		\underline{\mathbf{D}}^{\rm H}\underline{\mathbf{D}} & \underline{\mathbf{D}}^{\rm H}\overline{\mathbf{D}}_i \\
		\overline{\mathbf{D}}_i^{\rm H}\underline{\mathbf{D}} & \overline{\mathbf{D}}_i^{\rm H}\overline{\mathbf{D}}_i
	\end{array}\right],\nonumber
\end{align}
where $\underline{\mathbf{D}}^{\rm H}\underline{\mathbf{D}}$ and $(\underline{\mathbf{D}}^i)^{\rm H}\underline{\mathbf{D}}^i$ are Hermitian matrices. Suppose that the eigenvalues of $\underline{\mathbf{D}}^{\rm H}\underline{\mathbf{D}}$ are $(\mu_1,\mu_2,\cdots,\mu_n)$, and the eigenvalues of $(\underline{\mathbf{D}}^i)^{\rm H}\underline{\mathbf{D}}^i$ are $(\lambda_1,\lambda_2,\cdots,\lambda_{n+1})$. Based on Cauchy interlace theorem \cite{matrixanalysis}, we have 
\begin{align}
	\lambda_1\leq\mu_1\leq\lambda_2\leq\mu_2\leq\cdots\leq\mu_n\leq\lambda_{n+1}.\nonumber
\end{align}
Thus, the following inequality holds:
\begin{align}
	\|\underline{\mathbf{D}}^i\|_2=\sqrt{\lambda_{n+1}}\geq\|\underline{\mathbf{D}}\|_2=\sqrt{\mu_n}.\nonumber
\end{align}

By analogy, we finally obtain that $\|\underline{\mathbf{D}}\|_2\leq\|\mathbf{D}\|_2$.
\end{IEEEproof}

\begin{lemma2}\label{lemma2}
	(\cite[Lemma 1]{liyang2023}) Given a matrix $\mathbf{D}\in\mathbb{R}^{M\times k d}$ that consists of $k$ column-block submatrices of size ${M\times d}$, define $\lambda_{\min}$ and $\lambda_{\max}$ as the minimum and maximum eigenvalues of the matrix $\mathbf{D}^{\rm H}\mathbf{D}\in\mathbb{R}^{kd\times kd}$. When $(d-1)\nu+(k-1)d\mu_B<1$, we have 
	\begin{align}
		&1-(d-1)\nu  -(k-1)d\mu_B\leq\lambda_{\min}\nonumber\\ &\qquad\qquad\qquad\leq\lambda_{\max}\leq1+(d-1)\nu+(k-1)d\mu_B,\nonumber
	\end{align}
	where $\nu$ and $\mu_B$ are the sub-coherence and block coherence of $\mathbf{D}$, respectively.
\end{lemma2}

\begin{lemma1}\label{lemma1}
	Given a matrix $\mathbf{D}$ with a block length of $d$ for its column-block submatrices, suppose that $|\mathbf{\Theta}|=r$, $|\mathbf{\Xi}|=s$, $\mathbf{\Theta}\cap\mathbf{\Xi}=\mathbf{\emptyset}$. When $(d-1)\nu+(s-1)d\mu_B<1$ and $(d-1)\nu+(r-1)d\mu_B<1$, $\forall\mathbf{x}_{\mathbf{\Xi}}$ we have
	\begin{align}
		&\bigg((1-(d-1)\nu  -(s-1)d\mu_B)\nonumber\\
		&-\frac{d^2\mu_B^2rs}{1-(d-1)\nu-(r-1)d\mu_B}\bigg)\|\mathbf{x}_{\mathbf{\Xi}}\|_2^2\nonumber\\
		&\leq\|\ddot{\mathbf{A}}^{\mathbf{\Theta}}_{\mathbf{\Xi}}\mathbf{x}_{\mathbf{\Xi}}\|_2^2\leq\bigg(1+(d-1)\nu+(s-1)d\mu_B\bigg)\|\mathbf{x}_{\mathbf{\Xi}}\|_2^2,\label{lemma6in1}\\
		&\sigma_{\min}((\ddot{\mathbf{A}}^{\mathbf{\Theta}}_{\mathbf{\Xi}})^{\rm H}\ddot{\mathbf{A}}^{\mathbf{\Theta}}_{\mathbf{\Xi}})\geq\bigg((1-(d-1)\nu  -(s-1)d\mu_B)\nonumber\\
		&\qquad\qquad\qquad\qquad\enspace\enspace-\frac{d^2\mu_B^2rs}{1-(d-1)\nu-(r-1)d\mu_B}\bigg)\label{lemma6in2}.
	\end{align}
\end{lemma1}

\begin{IEEEproof}
	Based on \textbf{Lemma \ref{lemma2}}, the upper bound in (\ref{lemma6in1}) follows from the following inequality:
	\begin{align}
		\|\ddot{\mathbf{A}}^{\mathbf{\Theta}}_{\mathbf{\Xi}}\mathbf{x}_{\mathbf{\Xi}}\|_2^2&=\|\mathbf{P}^{\bot}_{\mathbf{D}_{\mathbf{\Theta}}}\mathbf{D}_{\mathbf{\Xi}}\mathbf{x}_{\mathbf{\Xi}}\|_2^2\nonumber\\
		&\leq\|\mathbf{D}_{\mathbf{\Xi}}\mathbf{x}_{\mathbf{\Xi}}\|_2^2\nonumber\\
		&\leq(1+(d-1)\nu+(s-1)d\mu_B)\|\mathbf{x}_{\mathbf{\Xi}}\|_2^2.\nonumber
	\end{align}
	The low bound in (\ref{lemma6in1}) is derived by
	\begin{align}
		\|\mathbf{P}^{\bot}_{\mathbf{D}_{\mathbf{\Theta}}}\mathbf{D}_{\mathbf{\Xi}}\mathbf{x}_{\mathbf{\Xi}}\|_2^2&=\|\mathbf{D}_{\mathbf{\Xi}}\mathbf{x}_{\mathbf{\Xi}}\|_2^2-\|\mathbf{P}_{\mathbf{D}_{\mathbf{\Theta}}}\mathbf{D}_{\mathbf{\Xi}}\mathbf{x}_{\mathbf{\Xi}}\|_2^2\nonumber\\
		&=\|\mathbf{D}_{\mathbf{\Xi}}\mathbf{x}_{\mathbf{\Xi}}\|_2^2-\|(\mathbf{D}_{\mathbf{\Theta}}^{\dagger})^{\rm H}\mathbf{\mathbf{D}^{\rm H}_{\mathbf{\Theta}}}\mathbf{D}_{\mathbf{\Xi}}\mathbf{x}_{\mathbf{\Xi}}\|_2^2\nonumber\\
		&\geq(1-(d-1)\nu  -(s-1)d\mu_B)\|\mathbf{x}_{\mathbf{\Xi}}\|_2^2\nonumber\\
		&\quad-\frac{\|\mathbf{\mathbf{D}^{\rm H}_{\mathbf{\Theta}}}\mathbf{D}_{\mathbf{\Xi}}\mathbf{x}_{\mathbf{\Xi}}\|_2^2}{1-(d-1)\nu-(r-1)d\mu_B}\nonumber\\
		&\geq(1-(d-1)\nu  -(s-1)d\mu_B)\|\mathbf{x}_{\mathbf{\Xi}}\|_2^2\nonumber\\
		&\quad-\frac{d^2\mu_B^2rs\|\mathbf{x}_{\mathbf{\Xi}}\|_2^2}{1-(d-1)\nu-(r-1)d\mu_B},\nonumber
	\end{align}
	where the first inequality follows from \textbf{Lemmas \ref{lemma2}} and \textbf{\ref{lemma3}}, and the second inequality is due to \textbf{Lemma \ref{lemma4}}.
	
	We now have the bounds presented in (\ref{lemma6in1}). The following inequality holds:
	\begin{align}
		\mathbf{x}^{\rm H}_{\mathbf{\Xi}}(\ddot{\mathbf{A}}^{\mathbf{\Theta}}_{\mathbf{\Xi}})^{\rm H}\ddot{\mathbf{A}}^{\mathbf{\Theta}}_{\mathbf{\Xi}}\mathbf{x}_{\mathbf{\Xi}}\geq&\bigg(1-(d-1)\nu  -(s-1)d\mu_B\nonumber\\
		&-\frac{d^2\mu_B^2rs}{1-(d-1)\nu-(r-1)d\mu_B}\bigg)\mathbf{x}^{\rm H}_{\mathbf{\Xi}}\mathbf{x}_{\mathbf{\Xi}}.
	\end{align}
Letting $\mathbf{x}_{\mathbf{\Xi}}$ be the eigenvector corresponding to the eigenvalue $\sigma_{\min}((\ddot{\mathbf{A}}^{\mathbf{\Theta}}_{\mathbf{\Xi}})^{\rm H}\ddot{\mathbf{A}}^{\mathbf{\Theta}}_{\mathbf{\Xi}})$, we have
\begin{align}
	&\mathbf{x}^{\rm H}_{\mathbf{\Xi}}(\ddot{\mathbf{A}}^{\mathbf{\Theta}}_{\mathbf{\Xi}})^{\rm H}\ddot{\mathbf{A}}^{\mathbf{\Theta}}_{\mathbf{\Xi}}\mathbf{x}_{\mathbf{\Xi}}\nonumber\\
	&=\sigma_{\min}((\ddot{\mathbf{A}}^{\mathbf{\Theta}}_{\mathbf{\Xi}})^{\rm H}\ddot{\mathbf{A}}^{\mathbf{\Theta}}_{\mathbf{\Xi}})\mathbf{x}^{\rm H}_{\mathbf{\Xi}}\mathbf{x}_{\mathbf{\Xi}}\nonumber\\
	&\geq\bigg((1-(d-1)\nu  -(s-1)d\mu_B)\nonumber\\
	&\quad\enspace-\frac{d^2\mu_B^2rs}{1-(d-1)\nu-(r-1)d\mu_B}\bigg)\mathbf{x}^{\rm H}_{\mathbf{\Xi}}\mathbf{x}_{\mathbf{\Xi}}.
\end{align}
Therefore, the inequality in (\ref{lemma6in2}) holds.
\end{IEEEproof}

\textbf{Lemma \ref{lemma1}} applies to the conventional block-sparse formulation. For hierarchical block-sparse recovery, we give the following corollary, the proof of which is similar to that of \textbf{Lemma \ref{lemma1}}.
\begin{Corollary4}
	Given a matrix $\mathbf{A}$, suppose that $|\mathbf{\Theta}|=r$, $|\mathbf{\Xi}|=s$, $\mathbf{\Theta}\cap\mathbf{\Xi}=\mathbf{\emptyset}$. Let the minimum block length unit is equal to $d$, $d^*$ is an integer multiple of $d$, and $\nu_{d^*}$ and $\mu_{d^*}$ denote the hierarchical sub-coherence and hierarchical block coherence of $\mathbf{A}$. When $(d^*-1)\nu_{d^*}  +(k_t-1)d^*\mu_{d^*}<1$ and $(d^*-1)\nu_{d^*}+(\lceil\frac{rd}{d^*}\rceil-1)d^*\mu_{d^*}<1$, $\forall\mathbf{x}_{\mathbf{\Xi}}$ and for the $t$th hierarchical mode, we have
	\begin{align}
		&\bigg((1-(d^*-1)\nu_{d^*}  -(k_t-1)d^*\mu_{d^*})\nonumber\\
		&-\frac{d^{*^2}\mu_{d^*}^2\lceil\frac{rd}{d^*}\rceil k_t}{1-(d^*-1)\nu_{d^*}-(\lceil\frac{rd}{d^*}\rceil-1)d^*\mu_{d^*}}\bigg)\|\mathbf{x}_{\mathbf{\Xi}}\|_2^2\nonumber\\
		&\leq\|\ddot{\mathbf{A}}^{\mathbf{\Theta}}_{\mathbf{\Xi}}\mathbf{x}_{\mathbf{\Xi}}\|_2^2\leq\bigg(1+(d^*-1)\nu_{d^*}+(k_t-1)d^*\mu_{d^*}\bigg)\|\mathbf{x}_{\mathbf{\Xi}}\|_2^2.\nonumber
	\end{align}
\end{Corollary4}

\begin{lemma3}\label{lemma3}
	Let $|\mathbf{\Xi}|=s$. When $(d-1)\nu+(s-1)d\mu_B<1$, $\forall\mathbf{x}_{\mathbf{\Xi}}$ we have
	\begin{align}
		&\sqrt{1-(d-1)\nu-(s-1)d\mu_B}\|(\mathbf{A}^{\dagger}_{\mathbf{\Xi}})^{\rm H}\mathbf{x}_{\mathbf{\Xi}}\|_2\nonumber\\
		&\leq\|\mathbf{x}_{\mathbf{\Xi}}\|_2\leq\sqrt{1+(d-1)\nu+(s-1)d\mu_B}\|(\mathbf{A}^{\dagger}_{\mathbf{\Xi}})^{\rm H}\mathbf{x}_{\mathbf{\Xi}}\|_2.\nonumber
	\end{align}
\end{lemma3}

\begin{IEEEproof}
	Suppose that $\mathbf{A}_{\mathbf{\Omega}}$ has singular value decomposition $\mathbf{A}_{\mathbf{\Omega}}=\mathbf{U}\mathbf{\Sigma}\mathbf{V}^{H}$. Then, we obtain that the minimum diagonal atom of $\mathbf{\Sigma}$ satisfies $\sigma_{\min}\geq\sqrt{1-(d-1)\nu-(r-1)d\mu_B}$. From \cite[A.1]{jwang2017}, we have $(\mathbf{A}^{\dagger}_{\Omega})^{\rm H}=\mathbf{U}\mathbf{\Sigma}^{-1}\mathbf{V}^{H}$, where $\mathbf{\Sigma}^{-1}$ is the diagonal matrix formed by replacing nonzero diagonal entries of $\mathbf{\Sigma}$ by their reciprocal. Therefore, all singular
	values of $(\mathbf{A}^{\dagger}_{\Omega})^{\rm H}$ are upper bounded by $\frac{1}{\sigma_{\min}}\leq\frac{1}{\sqrt{1-(d-1)\nu-(r-1)d\mu_B}}$. The lower bound in \textbf{Lemma \ref{lemma3}} has now been proved.
	
	Note that from \textbf{Lemma \ref{lemma2}}, we have $\|\mathbf{A}_{\mathbf{\Theta}}\|_2\leq\sqrt{1+(d-1)\nu+(r-1)d\mu_B}$. Then, the upper bound in \textbf{Lemma \ref{lemma3}} can be proved by directly using standard relationships between the singular values of $\mathbf{A}_{\mathbf{\Theta}}$ and the singular values of basic functions of $\mathbf{A}_{\mathbf{\Theta}}$. This completes the proof.
\end{IEEEproof}

Based on \textbf{Lemmas \ref{lemma1}} and \textbf{\ref{lemma3}}, the following corollary holds.

\begin{Corollary2}
	Let $|\mathbf{\Theta}|=r$, $|\mathbf{\Xi}|=s$, and $\mathbf{\Theta}\cap\mathbf{\Xi}=\mathbf{\emptyset}$. When $(d-1)\nu  +(s-1)d\mu_B<1$ and $(d-1)\nu+(r-1)d\mu_B<1$, $\forall\mathbf{x}_{\mathbf{\Xi}}$ we have
	\begin{align}
		&\bigg((1-(d-1)\nu  -(s-1)d\mu_B)\nonumber\\
		&\quad-\frac{d^2\mu_B^2rs}{1-(d-1)\nu-(r-1)d\mu_B}\bigg)^{\frac{1}{2}}\|(\ddot{\mathbf{A}}^{\dagger}_{\mathbf{\Xi}})^{\rm H}\mathbf{x}_{\mathbf{\Xi}}\|_2\nonumber\\
		&\leq\|\mathbf{x}_{\mathbf{\Xi}}\|_2\leq\sqrt{1+(d-1)\nu+(s-1)d\mu_B}\|(\ddot{\mathbf{A}}^{\dagger}_{\mathbf{\Xi}})^{\rm H}\mathbf{x}_{\mathbf{\Xi}}\|_2.\nonumber
	\end{align}
\end{Corollary2}

\subsection{Proof of \textbf{Lemma \ref{lemma8}}}\label{profoflemma8} % Ap-E
\begin{IEEEproof}
Note that 
\begin{align}
	\rho_{c(d_1,d_3)}(\mathbf{A}\mathbf{B}_{[t]})&=\sum_{i}\Bigg\|\sum_{j}\mathbf{A}_{[i,j]}\mathbf{B}_{[j,t]}\Bigg\|_2\nonumber\\
	&\leq\sum_i\sum_j\|\mathbf{A}_{[i,j]}\mathbf{B}_{[j,t]}\|_2\nonumber\\
	&\leq\sum_i\sum_j\|\mathbf{A}_{[i,j]}\|_2\|\mathbf{B}_{[j,t]}\|_2.\label{lemma82}
\end{align}
Thus, we have
\begin{align}
	\sum_i\|\mathbf{A}_{[i,j]}\|_2\leq\max_t\sum_i\|\mathbf{A}_{[i,t]}\|_2 = \rho_{c(d_1,d_2)}(\mathbf{A}).\label{lemma83}
\end{align}
Substituting (\ref{lemma83}) into (\ref{lemma82}) yields
\begin{align}
	\rho_{c(d_1,d_3)}(\mathbf{A}\mathbf{B}_{[t]})&\leq\rho_{c(d_1,d_2)}(\mathbf{A})\sum_j\|\mathbf{B}_{[j,t]}\|_2\nonumber\\
	&=\rho_{c(d_1,d_2)}(\mathbf{A})\rho_{c(d_2,d_3)}(\mathbf{B}_{[t]}).\nonumber
\end{align}
Therefore, the following inequality holds:
\begin{align}
	\rho_{c(d_1,d_3)}(\mathbf{A}\mathbf{B})&=\max_t\rho_{c(d_1,d_3)}(\mathbf{A}\mathbf{B}_{[t]})\nonumber\\
	&\leq\max_t\rho_{c(d_1,d_2)}(\mathbf{A})\rho_{c(d_2,d_3)}(\mathbf{B}_{[t]})\nonumber\\
	&=\rho_{c(d_1,d_2)}(\mathbf{A})\max_t\rho_{c(d_2,d_3)}(\mathbf{B}_{[t]})\nonumber\\
	&=\rho_{c(d_1,d_2)}(\mathbf{A})\rho_{c(d_2,d_3)}(\mathbf{B}).\nonumber
\end{align}
This completes the proof.
\end{IEEEproof}

\subsection{Proof of \textbf{Lemma \ref{lemma6}}}\label{profoflemma6} % Ap-E

\begin{IEEEproof}
	Without loss of generality, suppose that $\max\limits_{1\leq p\leq n}\|\mathbf{x}^p_{[1]}\|_2=\|\mathbf{x}^i_{[1]}\|_2$, and $\max\limits_{1\leq t\leq n} \|\mathbf{x}^t\|_2=\|\mathbf{x}^j\|_2$. Since $\|\mathbf{x}^i_{[1]}\|_2\geq\|\mathbf{x}^j_{[1]}\|_2$, we have $\|\mathbf{x}^j_{[2]}\|_2\geq\|\mathbf{x}^i_{[2]}\|_2\geq\min\limits_{1\leq q\leq n}\|\mathbf{x}^q_{[2]}\|_2$. Thus $\max\limits_{1\leq p\leq n}\|\mathbf{x}^p_{[1]}\|^2_2+\min\limits_{1\leq q\leq n}\|\mathbf{x}^q_{[2]}\|^2_2\leq \max\limits_{1\leq t\leq n} \|\mathbf{x}^t\|^2_2$ has been proved. The bound $\max\limits_{1\leq q\leq n}\|\mathbf{x}^q_{[2]}\|^2_2+\min\limits_{1\leq p\leq n}\|\mathbf{x}^p_{[1]}\|^2_2\leq \max\limits_{1\leq t\leq n} \|\mathbf{x}^t\|^2_2$ can be proved similarly.

As for the upper bound, for $j=\arg\max\limits_{1\leq t\leq n}\|\mathbf{x}^t\|^2_2$, we have
\begin{align}
	 \|\mathbf{x}^j\|^2_2=\|\mathbf{x}^j_{[1]}\|_2^2+\|\mathbf{x}^j_{[2]}\|_2^2\leq\max_{1\leq p\leq n}\|\mathbf{x}^p_{[1]}\|^2_2 + \max_{1\leq q\leq n}\|\mathbf{x}^q_{[2]}\|^2_2.\nonumber
\end{align}
This completes the whole proof.
\end{IEEEproof}

\subsection{Proof of Remark \ref{rmk4}}\label{profofrmk4}

\begin{IEEEproof}
	Observe that
	\begin{align}
		&\|\ddot{\mathbf{A}}^{\rm H}_{\mathbf{\Xi}^*\backslash\mathbf{\Theta}}\mathbf{r}^l\|_{(d^*)2,\infty}\nonumber\\
		&=\|\mathbf{D}^{\rm H}_{\mathbf{\Xi}^*\backslash\mathbf{\Theta}}\mathbf{P}^{\bot}_{\mathbf{D}_{\mathbf{\Theta}}}\mathbf{y}\|_{(d^*)2,\infty}\nonumber\\
		&=\|\mathbf{D}^{\rm H}_{\mathbf{\Xi}^*\backslash\mathbf{\Theta}}\mathbf{P}^{\bot}_{\mathbf{D}_{\mathbf{\Theta}}}\mathbf{D}_{\mathbf{\Xi}^*\backslash\mathbf{\Theta}}\mathbf{x}_{\mathbf{\Xi}^*\backslash\mathbf{\Theta}}\|_{(d^*)2,\infty}\nonumber\\
		&\geq\sigma_{\min}(\mathbf{D}^{\rm H}_{\mathbf{\Xi}^*\backslash\mathbf{\Theta}}\mathbf{P}^{\bot}_{\mathbf{D}_{\mathbf{\Theta}}}\mathbf{D}_{\mathbf{\Xi}^*\backslash\mathbf{\Theta}})\|\mathbf{x}_{\mathbf{\Xi}^*\backslash\mathbf{\Theta}}\|_{(d^*)2,\infty}\nonumber\\
		&\geq\sigma_{\min}(\mathbf{D}^{\rm H}_{\mathbf{\Xi}^*\backslash\mathbf{\Theta}}\mathbf{D}_{\mathbf{\Xi}^*\backslash\mathbf{\Theta}})\|\mathbf{x}_{\mathbf{\Xi}^*\backslash\mathbf{\Theta}}\|_{(d^*)2,\infty}\nonumber\\
		&\geq(1-(d^*-1)\nu_{d^*}-(k_t-\overline{\alpha}-1)d^*\mu_{d^*})\|\mathbf{x}_{\mathbf{\Xi}^*\backslash\mathbf{\Theta}}\|_{(d^*)2,\infty}\nonumber\\
		&=\delta^*\|\mathbf{x}_{\mathbf{\Xi}^*\backslash\mathbf{\Theta}}\|_{(d^*)2,\infty},\label{dayu2}
	\end{align}
	where the first equality is because $\mathbf{P}^{\bot}_{\mathbf{D}_{\mathbf{\Theta}}}=(\mathbf{P}^{\bot}_{\mathbf{D}_{\mathbf{\Theta}}})^{\rm H}\mathbf{P}^{\bot}_{\mathbf{D}_{\mathbf{\Theta}}}$, and the third inequality follows from \cite[Lemma~5]{ttcai2011}.
	
	On the other hand, 
	\begin{align}
		&\|\ddot{\mathbf{A}}^{\rm H}_{\mathbf{\Xi}^{\Delta}\backslash\mathbf{\Theta}}\mathbf{r}^l\|_{(d^{\Delta})2,\infty}\nonumber\\
		&=\|\mathbf{D}^{\rm H}_{\mathbf{\Xi}^{\Delta}\backslash\mathbf{\Theta}}\mathbf{P}^{\bot}_{\mathbf{D}_{\mathbf{\Theta}}}\mathbf{y}\|_{(d^{\Delta})2,\infty}\nonumber\\
		&=\|\mathbf{D}^{\rm H}_{\mathbf{\Xi}^{\Delta}\backslash\mathbf{\Theta}}\mathbf{P}^{\bot}_{\mathbf{D}_{\mathbf{\Theta}}}\mathbf{D}_{\mathbf{\Xi}^*\backslash\mathbf{\Theta}}\mathbf{x}_{\mathbf{\Xi}^*\backslash\mathbf{\Theta}}\|_{(d^{\Delta})2,\infty}\nonumber\\
		&=\|\mathbf{D}^{\rm H}_{\mathbf{\Xi}^{\Delta}\backslash\mathbf{\Theta}}(\mathbf{I}-\mathbf{P}_{\mathbf{D}_{\mathbf{\Theta}}})\mathbf{D}_{\mathbf{\Xi}^*\backslash\mathbf{\Theta}}\mathbf{x}_{\mathbf{\Xi}^*\backslash\mathbf{\Theta}}\|_{(d^{\Delta})2,\infty}\nonumber\\
		&\leq\|\mathbf{D}^{\rm H}_{\mathbf{\Xi}^{\Delta}\backslash\mathbf{\Theta}}\mathbf{D}_{\mathbf{\Xi}^*\backslash\mathbf{\Theta}}\mathbf{x}_{\mathbf{\Xi}^*\backslash\mathbf{\Theta}}\|_{(d^{\Delta})2,\infty}\nonumber\\
		&\quad+\|\mathbf{D}^{\rm H}_{\mathbf{\Xi}^{\Delta}\backslash\mathbf{\Theta}}\mathbf{P}_{\mathbf{D}_{\mathbf{\Theta}}}\mathbf{D}_{\mathbf{\Xi}^*\backslash\mathbf{\Theta}}\mathbf{x}_{\mathbf{\Xi}^*\backslash\mathbf{\Theta}}\|_{(d^{\Delta})2,\infty}.\label{twoterms}
	\end{align}
	Note that the terms in (\ref{twoterms}) satisfy that
	\begin{align}
		&\|\mathbf{D}^{\rm H}_{\mathbf{\Xi}^{\Delta}\backslash\mathbf{\Theta}}\mathbf{D}_{\mathbf{\Xi}^*\backslash\mathbf{\Theta}}\mathbf{x}_{\mathbf{\Xi}^*\backslash\mathbf{\Theta}}\|_{(d^{\Delta})2,\infty}\nonumber\\
		&\leq\rho_{r(d^{\Delta},d^*)}(\mathbf{D}^{\rm H}_{\mathbf{\Xi}^{\Delta}\backslash\mathbf{\Theta}}\mathbf{D}_{\mathbf{\Xi}^*\backslash\mathbf{\Theta}})\|\mathbf{x}_{\mathbf{\Xi}^*\backslash\mathbf{\Theta}}\|_{(d^{*})2,\infty}\nonumber\\
		&=\rho_{c(d^*,d^{\Delta})}(\mathbf{D}^{\rm H}_{\mathbf{\Xi}^*\backslash\mathbf{\Theta}}\mathbf{D}_{\mathbf{\Xi}^{\Delta}\backslash\mathbf{\Theta}})\|\mathbf{x}_{\mathbf{\Xi}^*\backslash\mathbf{\Theta}}\|_{(d^{*})2,\infty}\nonumber\\
		&\leq\bigg\lceil\frac{d^{\Delta}}{d^*}\bigg\rceil (k_t-\overline{\alpha})d^*\mu_{d^*}\|\mathbf{x}_{\mathbf{\Xi}^*\backslash\mathbf{\Theta}}\|_{(d^{*})2,\infty},\nonumber\\
		&\|\mathbf{D}^{\rm H}_{\mathbf{\Xi}^{\Delta}\backslash\mathbf{\Theta}}\mathbf{P}_{\mathbf{D}_{\mathbf{\Theta}}}\mathbf{D}_{\mathbf{\Xi}^*\backslash\mathbf{\Theta}}\mathbf{x}_{\mathbf{\Xi}^*\backslash\mathbf{\Theta}}\|_{(d^{\Delta})2,\infty}\nonumber\\
		&=\|\mathbf{D}^{\rm H}_{\mathbf{\Xi}^{\Delta}\backslash\mathbf{\Theta}}\mathbf{D}_{\mathbf{\Theta}}(\mathbf{D}_{\mathbf{\Theta}}^{\rm H}\mathbf{D}_{\mathbf{\Theta}})^{-1}\mathbf{D}_{\mathbf{\Theta}}^{\rm H}\mathbf{D}_{\mathbf{\Xi}^*\backslash\mathbf{\Theta}}\mathbf{x}_{\mathbf{\Xi}^*\backslash\mathbf{\Theta}}\|_{(d^{\Delta})2,\infty}\nonumber\\
		&\leq\rho_{c(d^*,d^{\Delta})}(\mathbf{D}_{\mathbf{\Theta}}^{\rm H}\mathbf{D}_{\mathbf{\Xi}^{\Delta}\backslash\mathbf{\Theta}})\rho_{c(d^*,d^*)}((\mathbf{D}_{\mathbf{\Theta}}^{\rm H}\mathbf{D}_{\mathbf{\Theta}})^{-1})\nonumber\\
		&\quad\times\rho_{c(d^*,d^*)}(\mathbf{D}^{\rm H}_{\mathbf{\Xi}^*\backslash\mathbf{\Theta}}\mathbf{D}_{\mathbf{\Theta}})\|\mathbf{x}_{\mathbf{\Xi}^*\backslash\mathbf{\Theta}}\|_{(d^{*})2,\infty},\nonumber
	\end{align}
	where 
	\begin{align}
		&\rho_{c(d^*,d^{\Delta})}(\mathbf{D}_{\mathbf{\Theta}}^{\rm H}\mathbf{D}_{\mathbf{\Xi}^{\Delta}\backslash\mathbf{\Theta}})\leq\bigg\lceil\frac{rd}{d^*}\bigg\rceil\bigg\lceil\frac{(k_t-\overline{\alpha})d^{\Delta}}{d^*}\bigg\rceil d^*\mu_{d^*},\nonumber\\
		&\rho_{c(d^*,d^*)}((\mathbf{D}_{\mathbf{\Theta}}^{\rm H}\mathbf{D}_{\mathbf{\Theta}})^{-1})\leq\frac{1}{1-(d^*-1)\nu_{d^*}-(\lceil\frac{rd}{d^*}\rceil-1)d^*\mu_{d^*}},\nonumber\\
		&\rho_{c(d^*,d^*)}(\mathbf{D}^{\rm H}_{\mathbf{\Xi}^*\backslash\mathbf{\Theta}}\mathbf{D}_{\mathbf{\Theta}})\leq \bigg\lceil\frac{rd}{d^*}\bigg\rceil (k_t-\overline{\alpha})d^*\mu_{d^*}.\nonumber
	\end{align}
	
	Therefore, (\ref{twoterms}) becomes that
	\begin{align}
		&\|\ddot{\mathbf{A}}^{\rm H}_{\mathbf{\Xi}^{\Delta}\backslash\mathbf{\Theta}}\mathbf{r}^l\|_{(d^{\Delta})2,\infty}\nonumber\\
		&\leq\bigg\lceil\frac{d^{\Delta}}{d^*}\bigg\rceil (k_t-\overline{\alpha})d^*\mu_{d^*}\|\mathbf{x}_{\mathbf{\Xi}^*\backslash\mathbf{\Theta}}\|_{(d^{*})2,\infty}\nonumber\\
		&\quad+\frac{\lceil\frac{rd}{d^*}\rceil^2\lceil\frac{(k_t-\overline{\alpha})d^{\Delta}}{d^*}\rceil (k_t-\overline{\alpha})d^{*^2}\mu^2_{d^*}}{1-(d^*-1)\nu_{d^*}-(\lceil\frac{rd}{d^*}\rceil-1)d^*\mu_{d^*}}\|\mathbf{x}_{\mathbf{\Xi}^*\backslash\mathbf{\Theta}}\|_{(d^{*})2,\infty}\nonumber\\
		&=\delta^{\Delta}\|\mathbf{x}_{\mathbf{\Xi}^*\backslash\mathbf{\Theta}}\|_{(d^{*})2,\infty}.\label{lastcom}
	\end{align}
	
	%Combining (\ref{dayu1}), (\ref{dayu2}) and (\ref{lastcom}) yields
	Combining (\ref{dayu2}) and (\ref{lastcom}) yields
	\begin{align}
		&(1-(d^*-1)\nu_{d^*}-(k_t-\overline{\alpha}-1)d^*\mu_{d^*})\nonumber\\
		&\geq\bigg\lceil\frac{d^{\Delta}}{d^*}\bigg\rceil (k_t-\overline{\alpha})d^*\mu_{d^*}\nonumber\\
		&\quad
		+\frac{\lceil\frac{rd}{d^*}\rceil^2\lceil\frac{(k_t-\overline{\alpha})d^{\Delta}}{d^*}\rceil (k_t-\overline{\alpha})d^{*^2}\mu^2_{d^*}}{1-(d^*-1)\nu_{d^*}-(\lceil\frac{rd}{d^*}\rceil-1)d^*\mu_{d^*}},\nonumber
	\end{align}
i.e., $\delta^*\geq\delta^{\Delta}$, which is a sufficient condition for the inequality $\|\ddot{\mathbf{A}}^{\rm H}_{\mathbf{\Xi}^*\backslash\mathbf{\Theta}}\mathbf{r}^l\|_{(d^*)2,\infty}\geq\|\ddot{\mathbf{A}}^{\rm H}_{\mathbf{\Xi}^{\Delta}\backslash\mathbf{\Theta}}\mathbf{r}^l\|_{(d^{\Delta})2,\infty}$ to hold. 
\end{IEEEproof}

\ifCLASSOPTIONcaptionsoff
  \newpage
\fi

%\bibliographystyle{ieeetran}
%\bibliography{SBOLS}

\begin{IEEEbiographynophoto}{Liyang Lu}	
(Member, IEEE) received the B.S. degree from the School of Information Engineering, Beijing University of Posts and Telecommunications (BUPT), China, in 2017, and the Ph.D. degree from the School of Artificial Intelligence, BUPT, in 2022. He was a postdoctoral fellow with the Department of Electronic Engineering, Tsinghua University, Beijing, China. He is currently an Assistant Professor with the School of Artificial Intelligence, BUPT. His research interests include compressed sensing, cognitive radios, MIMO communications, and intelligent communications.
\end{IEEEbiographynophoto}

\begin{IEEEbiographynophoto}{Haochen Wu}	
	(Student Member, IEEE) received the B.S. degree (Hons.) from the Department of Electronic Engineering, Tsinghua University, Beijing, China, in 2023, where he is currently pursuing his Ph.D. degree. His research interests include signal processing, compressive sensing, and channel estimation and precoding in near-field or unified near/far-field communication systems.
\end{IEEEbiographynophoto}

\begin{IEEEbiographynophoto}{Wenbo Xu}
	(Senior Member, IEEE) received the
	B.S. degree from the School of Information Engi
	neering, Beijing University of Posts and Telecom
	munications (BUPT), China, in 2005, and the Ph.D.
	degree from the School of Information and Communication Engineering, BUPT, in 2010. Since 2010,
	she has been with BUPT, where she is currently a
	Professor with the School of Artificial Intelligence.
	Her current research interests include sparse signal
	processing, machine learning, and signal processing
	in wireless networks.
\end{IEEEbiographynophoto}

\begin{IEEEbiographynophoto}{Zhaocheng Wang}
	(Fellow, IEEE) received the B.S.,
	M.S., and Ph.D. degrees from Tsinghua University in
	1991, 1993, and 1996, respectively.
	From 1996 to 1997, he was a Post-Doctoral Fellow with Nanyang Technological University, Singapore.
	From 1997 to 1999, he was a Research Engineer/a Senior Engineer with OKI Techno Centre (Singapore) Pte.
	Ltd., Singapore. From 1999 to 2009, he was a Senior
	Engineer/a Principal Engineer with Sony Deutschland
	GmbH, Germany. Since 2009, he has been a Professor
	with the Department of Electronic Engineering, Tsinghua University, where he is currently the Director of the Broadband Communication Key Laboratory, Beijing National Research Center for Information Science
	and Technology (BNRist). He has authored or coauthored two books, which have
	been selected by IEEE Press Series on Digital and Mobile Communication (Wiley
	IEEE Press). He has also authored/coauthored more than 200 peer-reviewed
	journal articles. He holds 60 U.S./EU granted patents (23 of them as the first
	inventor). His research interests include wireless communications, millimeter
	wave communications, and optical wireless communications.
	Prof. Wang is a fellow of the Institution of Engineering and Technology. He was
	a recipient of the ICC2013 Best Paper Award, the OECC2015 Best Student Paper
	Award, the 2016 IEEE Scott Helt Memorial Award, the 2016 IET Premium Award,
	the 2016 National Award for Science and Technology Progress (First Prize), the
	ICC2017 Best Paper Award, the 2018 IEEE ComSoc Asia Pacific Outstanding
	Paper Award, and the 2020 IEEE ComSoc Leonard G. Abraham Prize. 
\end{IEEEbiographynophoto}

\vspace{-11cm}

\begin{IEEEbiographynophoto}{H. Vincent Poor}
	(S'72, M'77, SM'82, F'87) received the Ph.D. degree in EECS from Princeton University in 1977.  From 1977 until 1990, he was on the faculty of the University of Illinois at Urbana-Champaign. Since 1990 he has been on the faculty at Princeton, where he is currently the Michael Henry Strater University Professor. During 2006 to 2016, he served as the dean of Princeton's School of Engineering and Applied Science, and he has also held visiting appointments at several other universities, including most recently at Berkeley and Caltech. His research interests are in the areas of information theory, stochastic analysis and machine learning, and their applications in wireless networks, energy systems and related fields. Among his publications in these areas is the book Machine Learning and Wireless Communications.  (Cambridge University Press, 2022). Dr. Poor is a member of the National Academy of Engineering and the National Academy of Sciences and is a foreign member of the Royal Society and other national and international academies. He received the IEEE Alexander Graham Bell Medal in 2017.
\end{IEEEbiographynophoto}

\end{document}